\documentclass[12pt,dvipsnames]{article}
\usepackage{iftex}
\ifLuaTeX
\usepackage{polyglossia} 
\usepackage{fontspec} 
\setmainlanguage{english}
\else 
\usepackage[utf8]{inputenc}
\usepackage[T1]{fontenc}
\usepackage[english]{babel} 
\fi

\usepackage{multirow}

\usepackage[top=2cm,bottom=2cm,outer=2.5cm,inner=2.5cm,marginparwidth=2.5cm]{geometry}
\usepackage{fvextra} 
\usepackage{authblk} 
\usepackage{mathrsfs}

\usepackage{xcolor}
\usepackage{tabularx}
\usepackage{youngtab}

\usepackage[all]{xy}
\definecolor{pale}{RGB}{255,250,225}
\definecolor{marginal}{RGB}{230,255,245}
\definecolor{marginal}{RGB}{255,230,245}
\usepackage[margin=1cm]{caption}

\ifLuaTeX
\usepackage[unicode,breaklinks]{hyperref}
\usepackage{pdftexcmds} 
\makeatletter
\let\pdfstrcmp\pdf@strcmp
\makeatother
\else
\usepackage[breaklinks]{hyperref}
\fi

\usepackage{amsmath}
\usepackage{amsfonts}
\usepackage{amssymb}
\usepackage{empheq} 
\usepackage{romanbar} 
\usepackage{dsfont} 
\usepackage{slashed}

\usepackage{lmodern} 
\usepackage{caption}
\usepackage{subcaption} 
\usepackage{titlesec} 
\usepackage{fancyhdr} 
\usepackage{abstract} 
\usepackage[super]{nth}
\usepackage{tocloft} 
\usepackage{comment}

\hypersetup{
pdfstartview={FitH},
pdftitle={Crosscap Defects},
pdfauthor={N. Drukker, S. Komatsu and A. Wallberg},
pdfsubject={},
pdfcreator={},
pdfkeywords={},
colorlinks=true,
citecolor=MidnightBlue,
linkcolor=MidnightBlue,
urlcolor=MidnightBlue
}

\numberwithin{equation}{section}
\newcommand\frontmatter{%
\clearpage
\pagenumbering{roman}
}

\newcommand\mainmatter{%
\clearpage
\pagenumbering{arabic}
}

\usepackage{mathtools}

\DeclareMathOperator\arctanh{arctanh}
\DeclareMathOperator{\Tr}{Tr}

\DeclareMathOperator{\Li}{Li}
\newcommand{\vev}[1]{\left\langle #1 \right\rangle}

\DeclareMathOperator{\diag}{diag}

\def\bH{\mathbb{H}}
\def\bR{\mathbb{R}}
\def\RP{{\mathbb{R}\mathbb{P}}}

\def\H{\mathbb{H}}

\def\bZ{\mathbb{Z}}

\def\cM{{\mathcal{M}}}
\def\cN{{\mathcal{N}}}
\def\cO{{\mathcal{O}}}

\def\dd{\text{d}}

\def\ra{\rightarrow}
\def\para{\parallel}

\def\Npm{N_{+-}}

\def \comma{\,,\qquad }

\newcommand{\beq}{\begin{equation}}
\newcommand{\eeq}{\end{equation}}
\newcommand{\bal}{\begin{equation}\begin{aligned}{}}
\newcommand{\eal}{\end{aligned} \end{equation}}
\newcommand{\ip}{\ipl}
\newcommand{\jp}{\jpl}
\newcommand{\im}{\imi}
\newcommand{\jm}{\jmi}

\usepackage{accents}
\newcommand{\ipl}{{\accentset{+}{\imath}}}
\newcommand{\jpl}{{\accentset{+}{\jmath}}}
\newcommand{\imi}{{\accentset{-}{\imath}}}
\newcommand{\jmi}{{\accentset{-}{\jmath}}}
\newcommand{\ipm}{{\accentset{\pm}{\imath}}}
\newcommand{\jpm}{{\accentset{\pm}{\jmath}}}

\newcommand{\capwidth}{332.101pt}
\newcommand{\halfcapwidth}{197.563pt}

\usepackage{tikz}
\usetikzlibrary{shapes.misc,arrows,decorations.markings,decorations.pathmorphing, patterns,calc,shapes.geometric}
\usepackage{pgfplots}
\pgfplotsset{compat=1.18} 

\title{\begin{flushright}
{\normalsize \ttfamily CERN-TH-2026-095}
\end{flushright}
\vspace{1 cm}

Crosscap Defects
\vspace{5mm}}
\author{Nadav~Drukker%
\thanks{\href{mailto:nadav.drukker@gmail.com}{nadav.drukker@gmail.com}}}
\author[2]{Shota Komatsu%
\thanks{\href{mailto:shota.komatsu@cern.ch}{shota.komatsu@cern.ch}}}
\author[2,3]{Anders Wallberg%
\thanks{\href{mailto:anders.heide.wallberg@cern.ch}{anders.heide.wallberg@cern.ch}}}
\affil[1]{\it Department of Mathematics, King's College London,\protect\\London, 
WC2R 2LS, United Kingdom\vspace{4pt}}
\affil[2]{\it Department of Theoretical Physics, CERN\protect\\ 1211 Meyrin, Switzerland\vspace{4pt}}
\affil[3]{\it  Laboratory for Theoretical Fundamental Physics, EPFL\protect\\1015 Lausanne, Switzerland}
\date{}

\begin{document}

\frontmatter
\maketitle

\begin{abstract}
We introduce a novel class of defects, termed {\it crosscap defects}, in conformal field theory (CFT) in general dimensions. These arise from quotienting the spacetime by a $\mathbb{Z}_2$ automorphism, and provide higher-codimension generalisations of CFTs on real projective space ($\mathbb{RP}^{d}$). 
Crosscap defects extend along a $p$-dimensional fixed locus of the $\bZ_2$ action and preserve an $SO(p+1,1)\times PO(d-p)$ subgroup of the conformal group. The two-point functions of operators in this setup exhibit three operator product expansion channels: bulk, image, and defect. These lead to several {\it crosscap crossing equations}, which we present. We analyse conformal block decompositions and show that the blocks are identical to defect CFT blocks up to a redefinition of cross ratios. As concrete examples, we study crosscap defects in the $O(N)$ model at the Gaussian and Wilson--Fisher fixed points in the $\varepsilon$-expansion. We compute explicitly the associated CFT data as a function of $p$ and find that, unlike standard defects, displacement and tilt operators are absent for generic $p$. They provide examples of defect conformal manifolds without exactly marginal operators.
\end{abstract}

\thispagestyle{empty}

\mainmatter

\tableofcontents

\section{Introduction}
\label{sec:intro}
Conformal field theories (CFTs) are fundamental both in high-energy and condensed matter physics, appearing as ultraviolet (UV) and infrared (IR) fixed points of renormalisation group (RG) flows and describing various critical phenomena. While CFTs in 
$1+1$ dimensions have been extensively studied in the past due to their infinite-dimensional Virasoro symmetry, higher-dimensional CFTs have gained more attention recently owing to the successful application of the conformal bootstrap and its numerical implementations \cite{Poland:2018epd}.

One way to enrich the study of CFTs is by introducing boundaries and defects (impurities). Boundaries and impurities are common in real-world materials and possess new critical phenomena that do not exist in their absence. Specifically, RG fixed points with boundaries or defects are described by boundary CFT (BCFT) \cite{Liendo:2012hy,Gliozzi:2015qsa} or defect CFT (DCFT) \cite{Billo:2016cpy}, which can be analysed by extensions of the conformal bootstrap approach.

Another way to probe detailed structures of CFT is by placing them on non-trivial manifolds. A key example is CFTs on $S^{1}\times\mathbb{R}^{d-1}$, corresponding to finite-temperature CFTs (TCFT) \cite{Iliesiu:2018fao}. However, studying CFTs at finite temperature is more challenging than in flat space because of the lack of positivity properties needed for numerical bootstrap and the restrictions on the use of operator product expansion (OPE) due to the thermal circle. 
A simpler yet informative setting is CFTs on real projective space $\RP^{d}$ (XCFT), which has been explored in the literature e.g.~\cite{Nakayama:2016cim,Hasegawa:2016piv,Hogervorst:2017kbj,Hasegawa:2018yqg,Giombi:2020xah}. Unlike finite-temperature CFTs, the topology of $\RP^{d}$ does not impose strong constraints on the applicability of OPE though it still lacks positivity. For instance, the two-point function on $\RP^{d}$ always admits convergent OPE expansions in Euclidean kinematics while the two-point function in TCFTs does not have any convergent OPE expansion whenever the distance between the operators is greater than the size of the thermal circle.
Notably, both setups arise from discrete quotients of flat space; $\mathbb{Z}$ for finite-temperature CFT and $\mathbb{Z}_2$ for $\RP^{d}$. 

The purpose of this paper is to study CFTs on non-trivial manifolds that combine virtues of both defect CFTs and CFTs at finite temperature or on $\RP^{d}$. These arise from $\mathbb{Z}_2$ quotients of $\mathbb{R}^{d}$, distinct from the one that leads to $\RP^{d}$, and can be viewed as higher codimension versions of $\RP^{d}$. We call them {\it crosscap defects}, or {\it crosscap defect CFT} (XDCFT). Their key features include:
\begin{itemize}
\item Similar to CFTs at finite temperature or on $\RP^{d}$, there are two ways to perform the OPE of two bulk local operators: directly or via their $\mathbb{Z}_2$ images. This results in a crossing equation for the two-point function, relating two different expansions, both involving bulk operators only.
\item Unlike CFTs at finite temperature or on $\RP^{d}$, $\mathbb{Z}_2$ quotients generally introduce a fixed locus in spacetime, which plays a role similar to defects (hence the name, crosscap defects). Consequently, the two-point function admits yet another expansion in terms of operators localised at the fixed locus, resembling the defect channel OPE in DCFT and inheriting its positivity properties.

\item 
Crosscap defects with a dimension-$p$ fixed locus break the Euclidean conformal group $SO(d+1,1)$ to $SO(p+1,1)\times PO(q)$, where we use $q=d-p$ and $PO(q)=O(q)/\{I,-I\}$ is the \textit{projective orthogonal group}, which coincides with the special orthogonal group $SO(q)$ for odd $q$. This closely resembles the symmetry-breaking pattern of 
$p$-dimensional conformal defects. However, there is a small difference: the transverse symmetry of conformal defects is typically $O(q)$, which includes reflections (although this is rarely stated explicitly in the literature). In contrast, crosscap defects lack reflections because the $\mathbb{Z}_2$ quotient transforms the space around the fixed locus from $S^{q-1}$ to $\RP^{q-1}$, and the symmetry is reduced to $PO(q)$.
\item CFTs with crosscap defects can be mapped via Weyl transformation to CFTs on $\bH_{p+1}\times \RP^{d-p-1}$, where $\bH_{p+1}$ is a hyperbolic space (also known as Euclidean AdS) in $p+1$ dimensions.
\item Crosscap defects can be enriched with $\mathbb{Z}_2$ quotients of other global symmetries of the theory.\footnote{In two dimensions, crosscaps can be defined even with non-$\mathbb{Z}_2$ symmetry actions \cite{Huiszoon:1999xq, Fuchs:2000cm, Brunner:2002em}, including non-invertible symmetries \cite{Harada:2025uhh}. It would be interesting to explore such possibilities for crosscap defects.} 
In superconformal field theories, this enables the construction of BPS crosscap defects that preserve a fraction of supersymmetry. Some of these BPS crosscap defects have holographic duals corresponding to AdS spaces with orientifolds, offering a setup to explore the non-perturbative physics of orientifolds. These topics will be addressed in our upcoming paper \cite{toappear}. In the examples in Sections~\ref{sec:free} and~\ref{sec:interacting} of this paper, we allow breaking of $O(N)$ to $O(N_+)\times O(N_-)$. 
\end{itemize}
In this paper, we focus on basic properties of crosscap defects, such as the crossing equation and the conformal block decomposition, laying the foundation for future studies. To illustrate the construction, we analyse crosscap defects in the $O(N)$ model at both the Gaussian and Wilson--Fisher fixed points. We compute the associated conformal data as analytic functions of the fixed-locus dimension 
$p$, allowing continuous interpolation between distinct physical regimes. We also find that XDCFTs in general do not possess displacement or tilt operators, in marked contrast to standard DCFTs.

\section{Review of CFTs on projective space and embedding space}
\label{sec:RPd}
Our new crosscap defect CFTs are in many ways a generalisation of CFTs on $\mathbb{RP}^d$, discussed in \cite{Nakayama:2016cim,Hasegawa:2016piv,Hogervorst:2017kbj,Hasegawa:2018yqg,Giombi:2020xah}. We start by reviewing some known results in this context and introduce the embedding space formalism which is useful for our construction.

\subsection{CFTs on \texorpdfstring{$\RP^d$}{RPd}}\label{sec:CFTsonRPd}
One starts with a CFT with internal global symmetry group $G$ on $\bR^d$ with coordinates $x^\mu\,,\,\mu\in\{1,\dots,d\}$. Assuming the theory is parity invariant, the symmetry group of the CFT is $O^+(d+1,1)\times G$.%
\footnote{Recall that $O(d+1,1)$ has four connected components and $O^+(d+1,1)$, the group of \textit{orthochronous} orthogonal transformations, is a $\bZ_2$ quotient of it.}
One then takes the quotient of this space by the conformal inversion 
\begin{equation}
\label{eq:inversion}
\iota:x^\mu\mapsto x'^{\mu}\equiv-\frac{x^\mu}{x^2}\,,
\end{equation}
which squares to the identity, $\iota^2=\mathbb{I}$, and breaks the conformal symmetry down to $O(d+1)$. Using the stereographic projection, the inversion \eqref{eq:inversion} is equivalent to the antipodal map on $S^d$, and hence this gives a CFT on $\RP^d$. We alternate between using the description on $\bR^d/\iota$ and $\RP^d$ below and refer to such theories as \textit{XCFTs}. 

The symmetry group of the XCFT is given by $PO(d+1)\times G$, which for even $d$ is simply $SO(d+1)\times G$. In general, one can further include the action of a $\mathbb{Z}_2$ subgroup of the $G$ in the definition of $\iota$. This breaks the internal symmetry down to a normaliser, $G\rightarrow K$. Combined with the symmetry breaking of the conformal group, the XCFT has symmetry group
\begin{equation}
    \frac{O(d+1)\times K}{\iota}\,.
\end{equation}
Operators transform under $\iota$ as
\begin{equation}
    \iota \left(\mathcal{O}_i(x)\right)=\mathcal{O}'_i(x')\,,
    \label{eq:RPiO}
\end{equation}
and since $\iota$ squares to the identity, $\mathcal{O}_i''=\mathcal{O}_i$. The space of operators can then be divided into $\iota$-even and $\iota$-odd sectors. 

Crucially, the conformal symmetry breaking implies that there are fewer restrictions on correlators in the XCFT compared to a usual CFT. Indeed, a scalar operator\footnote{One-point functions of spinning operators still vanish by the remaining conformal symmetry \cite{Giombi:2020xah}.\label{1ptvanish}} $\mathcal{O}_i$ of conformal dimension $\Delta_i$ can have a non-vanishing one-point function of the form
\begin{equation}
\langle\mathcal{O}_i\rangle_{\text{sphere}}=A_i\comma \langle\mathcal{O}_i(x)\rangle_{\text{flat}}=\frac{A_i}{\left(1+x^2\right)^{\Delta_i}}\,,
\label{eq:RP1pt}
\end{equation}
provided that $\mathcal{O}_i$ is $\iota$-even. Similarly, the two-point function of scalar operators can be non-vanishing if both operators are in the same sector under $\iota$, and
takes the form
\begin{equation}
\langle\mathcal{O}_i(x)\mathcal{O}_j(y)\rangle_{\text{flat}}=\frac{G_{ij}(\eta(x,y))}{\left(1+x^2\right)^{\Delta_i}\left(1+y^2\right)^{\Delta_j}}\label{eq:RP2pt}\,,
\end{equation}
where
\begin{equation}
\eta(x,y)=\frac{(x-y)^2}{(1+x^2)(1+y^2)}\label{eq:etadef}
\end{equation}
is an invariant under the action of $O(d+1)$ which satisfies $\eta\left(x',y\right)=1-\eta(x,y)$. Furthermore, $G_{ij}$ must satisfy a crossing relation $G_{ij}(\eta)=\pm G_{ij}(1-\eta)$ where the sign depends on whether $\mathcal{O}_i$ is even or odd \cite{Nakayama:2016cim,Hasegawa:2016piv,Hogervorst:2017kbj,Hasegawa:2018yqg,Giombi:2020xah}.

\subsection{Embedding space formalism}
\label{sec:embed}

A useful way to analyse the kinematics of CFTs is by embedding the $d$-dimensional spacetime in a $d+2$-dimensional space $\mathbb{R}^{d+1,1}$ with coordinates $X^M$, $M\in\{0,1,\dots,d+1\}$ and metric $\eta_{MN}=\diag(-1,1,\dots,1)$. The advantage is that the conformal group $O^+(d+1,1)$ is linearly realised as Lorentzian transformations on this space. Points in the $ d$-dimensional spacetime are represented by light-rays in embedding space, i.e.\ subsets satisfying
\begin{equation}
X\cdot X=0\,,\qquad
X\sim \lambda X\quad \text{for}\quad \lambda\in \bR\,.
\label{eq:lightray}
\end{equation}
Imposing these two constraints precisely accounts for the additional two dimensions.

One can then define operators directly in embedding space. Restricting to primary symmetric traceless operators $\mathcal{O}_i^{\mu_1\dots \mu_J}(x)$ of spin $J$, we promote them to $\mathcal{O}_i^{M_1\dots M_J}(X)$. In order for an operator to have a well-defined pull-back to the light-ray, it must be homogeneous
\begin{equation}
    \mathcal{O}_i^{M_1\dots M_N}(\lambda X)=\lambda^{-\Delta_i}\mathcal{O}_i^{M_1\dots M_J}(X)\label{homo}\,,
\end{equation}
with $\Delta_i\in\mathbb{R}_+$ its conformal dimension. For spinning operators, the additional tensor components are projected away by requiring transversality
\begin{equation}
    X_{M}\mathcal{O}_i^{MM_2\dots M_J}(X)=0\,,
    \label{spinray1}
\end{equation}
and the extra gauge freedom
\begin{equation}
\mathcal{O}_i^{MM_2\dots M_J}(X)\sim \mathcal{O}_i^{MM_2\dots M_J}(X)+X^{M}\Lambda^{M_2\dots M_J}(X)\label{spinray2}\,,
\end{equation}
for any $\Lambda^{M_2\dots M_J}$. To see that this results in tensors with the right number of degrees of freedom, note that for $J=1$ the vector primary operator should have $d$ components while the embedding space operator has $d+2$ and indeed \eqref{spinray1} and \eqref{spinray2} precisely projects away $2$ components.

To simplify computations involving spinning operators, it is convenient to further introduce the $d+2$-dimensional complex polarisation vector $Z_M$ such that the symmetric spinning operator $\mathcal{O}_i^{M_1\dots M_J}(X)$ is replaced by the $J$-th degree polynomial
\begin{equation}
    \mathcal{O}_i(X,Z)\equiv Z_{M_1}\dots Z_{M_J}\mathcal{O}_i^{M_1\dots M_J}(X)\,.
\end{equation}
Given such a polynomial, one can always go back to the physical spinning operator by using the formula
\begin{equation}
    \mathcal{O}_i^{\mu_1\dots \mu_J}(X)=\frac{1}{J!\left(\frac{D-2}{2}\right)_J}\frac{\partial X_{M_1}}{\partial x_{\mu_1}}\dots\frac{\partial X_{M_J}}{\partial x_{\mu_J}} D_{M_1}\dots D_{M_J}\mathcal{O}_i(X,Z)\,.
\end{equation}
where $D_M$ is the Todorov operator (see e.g.~\cite{Costa:2011scc,dobrev_dynamical_1976})
\begin{align}
    D_M\equiv\left(\frac{D-2}{2}+Z\cdot \partial_Z\right)\partial_{Z_M}-\frac{1}{2}Z_M\partial_Z\cdot\partial_Z\,,\label{eq:Todorov}
\end{align}
Tracelessness, homogeneity \eqref{homo}, transversality \eqref{spinray1} and the gauge freedom \eqref{spinray2} translate into the conditions
\begin{equation}
    X\cdot X=Z\cdot Z=Z\cdot X=0\comma \mathcal{O}(\lambda X,Z+\alpha X)=\lambda^{-\Delta}\mathcal{O}(X,Z)\,,\label{eq:polycond}
\end{equation}
which must be kept in mind when constructing correlators. It is easy to see that, due to the constraints \eqref{eq:polycond}, $Z$ can only appear through the antisymmetric tensor
\begin{equation}
\label{CMN}
    C^{MN}\equiv X^MZ^N-Z^MX^N\,,
\end{equation}
which satisfies $C^M_{\phantom{N}M}=C^{MN}C_{NM}=C^{MN}C_{NK}C^{KL}=C^{MN}X_N=0$.

Requiring compatibility of correlators with \eqref{eq:polycond} is then sufficient to prove the usual CFT constraints on $n$-point functions. For the one-point function $\langle\mathcal{O}(X,Z)\rangle$, there are no non-vanishing possible contractions of $X$ and $C$ and thus the one-point function must vanish (unless $\Delta=0$, i.e.\ $\mathcal{O}$ is the identity operator). For two-point functions, the structure is fixed to be
\begin{equation}
    \langle\mathcal{O}_i(X,Z)\mathcal{O}_j(Y,W)\rangle=\delta_{ij}\frac{(C\cdot D)^{J_i}}{(-2X\cdot Y)^{\Delta_i+J_i}}\label{Or2pt}\,,
\end{equation}
where $D^{MN}=Y^MW^N-W^MY^N$ and the factor of $-2$ in the denominator is conventional \cite{Costa:2011scc}. Similarly, three-point functions are fixed up to a constant, four-point functions can depend on cross ratios like $(X_1\cdot X_3)(X_2\cdot X_4)/((X_1\cdot X_2)(X_3\cdot X_4))$ and $(X_1\cdot X_4)(X_2\cdot X_3)/((X_1\cdot X_2)(X_3\cdot X_4))$ and so on.

\paragraph{Different conformal frames:}
One strength of the embedding space formalism is that it allows us to describe different conformally-flat spacetimes in a uniform manner. Different spacetimes like $\bR^d$ or $S^d$ simply amount to different choices of sections of the light-rays \eqref{eq:lightray}. To obtain $\bR^d$, we identify $x_\mu=X_\mu$ for $\mu\in\{1,\dots,d\}$, and fix the ray condition by 
\begin{equation}
X_0=\frac{1+x_\mu x^\mu}{2}\,,
\qquad
X_{d+1}=\frac{1-x_\mu x^\mu}{2}\,.
\label{eq:Rray}
\end{equation}
In what follows, we call it the {\it flat frame}. Note that the denominator of \eqref{Or2pt} becomes $-2X\cdot Y=(x-y)^2$ in this frame.
To get a CFT on $S^d$, one instead takes the {\it sphere frame}, given by
\begin{equation}
X_0=1\comma\sum_{M=1}^{d+1}(X_M)^2=1\,.
\label{eq:Sd}
\end{equation}
Another natural section that we use is
\begin{equation}
      X_0>0\comma -X_0^2+X_{d+1}^2+\sum_{M=1}^{p}\left(X_M\right)^2=-1\comma
    \sum_{M=p+1}^d\left(X_M\right)^2=1\,,
\label{eq:LSray}
\end{equation}
which we call the {\it hyperbolic frame}.
The first $p+1$ directions define Euclidean AdS space in $p+1$ dimensions, $\bH_{p+1}$, while the remaining ones parametrise $S^{q-1}$, where $q=d-p$. The hyperbolic frame is the most relevant for studying defect CFTs, with the properties of the defect encoded on the boundary of $\bH_{p+1}$.

\section{Crosscap defects in CFT}
\label{sec:crosscap}

\subsection{Definition}
\label{subsec:ccross}
We are now ready to describe the crosscap defects. First, for CFTs on $\RP^d$ discussed in Section~\ref{sec:CFTsonRPd}, the embedding space description is obtained by identifying $X^M\sim -X^M$ for $M\in\{1,\dots,d+1\}$ and leaving $X^0$ invariant.
This change of signs cannot be reabsorbed by the identification
\eqref{eq:lightray} and in the sphere frame \eqref{eq:Sd} it is merely the antipodal map.

Building on this, we define {\it crosscap defects} by the identification on a subset of the coordinates 
\begin{equation}
X^M\sim\iota_p(X^M)\,,
\qquad
\iota_p(X^M)=\begin{cases}
X^M\,,&M\in \{0,1,\dots,p,d+1\}\,,\\
-X^M\,,&M\in \{p+1,\dots,d\}\,.
\label{eq:embediota}
\end{cases}
\end{equation}
We collectively denote the two sets of coordinates as $X_+$ and $X_-$, respectively, such that $\iota(X_\pm)=\pm X_\pm$.
This action is non-trivial for any $0\leq p<d$ and leaves invariant the
flat frame \eqref{eq:Rray}, where it induces the action
\begin{equation}
\iota_p(x^\mu)=x'^{\mu}\equiv \begin{cases}
x^\mu \,,&\mu\in\{1,\dots p\}\,,
\\
-x^\mu\,,& \mu\in\{p+1,\dots d\}\,.
\end{cases}
\label{eq:flatiota}
\end{equation}
We denote the resulting quotient $\bR^d/\iota_p$ and mirroring the two sets $X_+$ and $X_-$, we denote the sets of coordinates in flat space by $x^\alpha_\para$ and $x^a_\perp$ with $\alpha\in\{1,\dots,p\}$ and $a\in\{p+1,\dots,d\}$ respectively. Note that $\iota_p$ has an $\mathbb{R}^p$ fixed locus spanned by $x_\para^\alpha$ at $x^a_\perp=0$. It is straightforward to get instead an $S^p$ fixed locus, in which case one simply has to define $\iota_p$ by $\iota_p(X^M)=-X^M$ for $M\in\{p+2,\dots,d,d+1\}$, i.e.\ include a sign-flip of $X^{d+1}$ \eqref{eq:Rray}. This only leads to minor modifications to the discussion below. Note that a sign flip of $X^0$ could always be undone at the cost of sign flips of other coordinates by utilising the light-ray condition \eqref{eq:lightray} with $\lambda=-1$. Thus, without loss of generality, we only consider $\iota_p$ which leaves $X^0$ invariant.

Another useful description of this space comes from the hyperbolic frame defined in \eqref{eq:LSray}. In this case, $\iota_p$ \eqref{eq:embediota} leaves $\mathbb{H}_{p+1}$ invariant, and it acts as the antipodal map on $S^{q-1}$, resulting in a quotient spacetime of the form $\H_{p+1}\times\RP^{q-1}$. Due to the similarities with both crosscap CFTs and defect CFTs, we call these new theories \textit{crosscap defect CFTs (XDCFT)}.

Although the dimension of the fixed locus simply appears as a parameter in our calculations, physics looks quite different for different values of $p$. For integer values, it breaks down as follows
\begin{itemize}
\item $\boldsymbol{p=d:}$ In this case the action of $\iota_d$ is trivial and the resulting theory in flat space is simply a \textit{CFT} on $\bR^d$. This case is excluded from the discussion below.

\item $\boldsymbol{p=d-1:}$ In the hyperbolic frame, we get a theory on $\mathbb{H}_d\times\mathbb{RP}^0$. The former space has a boundary while the latter is simply a point, and the resulting theory is thus a \textit{BCFT}.

\item $\boldsymbol{p=d-2}:$ In the hyperbolic frame, $\iota_{d-2}$ acts by mapping $S^1\ra\mathbb{RP}^1$ which is itself a circle but with half the radius. As such, we get a theory on $\mathbb{H}_{d-1}\times S^1$ which looks like a \textit{TCFT} on $\mathbb{H}_{d}$ with $\beta=1/2$. Note that a straightforward generalisation would lead to other temperatures, see the conclusions (Section~\ref{sec:conclusion}) for more context. This codimension 2 defect additionally shares many similarities with 
monodromy defects, studied for example in~\cite{Billo:2013jda,Gaiotto:2013nva,Yamaguchi:2016pbj,Giombi:2021uae}. 

\item $\boldsymbol{0<p\leq d-2:}$ In the flat frame, we get a theory with a $p$-dimensional fixed locus. This is similar to a \textit{DCFT}, though with novel features discussed below. Indeed, as we see later, we get a $p$-dimensional CFT living on the fixed locus coupled to the bulk CFT. 
\item $\boldsymbol{p=0:}$ In the flat frame, $\iota_0$ acts on all the coordinates of $\bR^d$ and the fixed points are just $x=0$ and $x=\infty$. This looks like a \textit{CFT with two operator insertions}\footnote{A similar viewpoint of treating two operator insertions as point-like defects was advocated recently in \cite{Chen:2026ium} in the context of heavy-heavy-light-light four-point functions.}, similar to twist fields in 2d CFTs \cite{Calabrese:2009qy}. Going to the hyperbolic frame is the usual state-operator map replacing the radial coordinate in $\bR^d/\iota_0$ with a Euclidean time along $\bH_1\simeq\bR$ and spatial slices that are $\RP^{d-1}$.

\item $\boldsymbol{p=-1:}$ $\iota_{-1}$ acts non-trivially on $d+1$ coordinates. 
In this case, in embedding space we need to slightly modify the definition in \eqref{eq:embediota} and take it to flip the signs of $X_1,\dots,X_{d+1}$. This results in a \textit{XCFT} on $\RP^d$ discussed in Section~\ref{sec:RPd}. 
Note that the action $X_{d+1}\mapsto -X_{d+1}$ alone is incompatible with the flat frame \eqref{eq:Rray}, and one has to combine it with the rescaling $X_M \to \frac{1}{x_\mu x^{\mu}}X_M$. The combined action corresponds to the inversion \eqref{eq:inversion}.
\end{itemize}
Although the cases above appear quite different, we stress that all the calculations in this paper are essentially continuous in $p\in(-1,d-1)$ and we can thus smoothly connect theories with different physics. For the free $O(N)$ model studied in Section~\ref{sec:free}, we find that the conformal data is independent of the dimension of the defect. For the interacting theory in Section~\ref{sec:interacting}, the conformal data does display an analytic dependence on $p$.

When constructing a CFT on the crosscap defect, as on $\RP^d$, one can include the action of a $\mathbb{Z}_2$ subgroup of the internal symmetry $G$ in the definition of $\iota_p$. This induces a further action on operators $\mathcal{O}(x)\mapsto \mathcal{O}'(x')$. We again classify operators by this additional $\bZ_2$ action and refer to their $\bZ_2$ charge as $\iota_p$ chirality. 
Denoting by $K$ the subgroup of $G$ that is invariant under the non-trivial $\bZ_2$ element (so its commutant or normaliser), the combination of \eqref{eq:flatiota} and the action on the fields implies that the XDCFT has the symmetry group
\begin{equation}
    O^+(p+1,1)\times \frac{O(q)\times K}{\iota_p}\,,\label{eq:symdown}
\end{equation}
where $O^+(p+1,1)$ acts as Lorentz transformations of the $X_+$ coordinates that leave $X_0$ positive and $O(q)$ as orthogonal transformations of the $X_-$ coordinates. Note that unless $\iota_p$ acts trivially on the operators, the group $K$ still includes the $\bZ_2$ subgroup, as does $O(q)$, which is the reason for modding out by their diagonal combination $\iota_p$. If the internal symmetry is unbroken, the preserved symmetry is $O^+(p+1,1)\times PO(q)\times G$.

\paragraph{Comparison with DCFT:}
Crosscap defects share many properties with conformal defects (see e.g.~\cite{Billo:2016cpy}). In both cases, there is a bulk CFT with the standard CFT operators and a $p$-dimensional subspace with preserved $p$-dimensional conformal symmetry on which different operators can reside and one can define the bulk and defect correlation functions. Furthermore, as in defect CFTs, one can enrich the crosscap defect by coupling it to localised degrees of freedom on the fixed locus or by turning on operators on the fixed locus. The main distinction is that the space surrounding the crosscap defect is $\RP^{q-1}$ while for a regular defect it is $S^{q-1}$. Consequently, in the latter case, there always is a ``trivial defect'' which is really just the bulk CFT vacuum with the full $O^+(d+1,1)$ symmetry in a special coordinate frame \eqref{eq:LSray}. By contrast, the crosscap defect is never trivial, even if we do not turn on operators or introduce new degrees of freedom at the fixed locus.
In this work, we mostly restrict to the simplest crosscap defects defined solely in terms of the geometry plus the extra possible action of $\iota_p$ on operators. However, as we show in Section~\ref{sec:1ptint}, in the analysis of the $p=2$ XDCFT in the $O(N)$ model, divergences require turning on a counterterm, which is an operator localised at the fixed locus.

One further important distinction from DCFT is the absence of displacement and tilt operators. In DCFT, the breaking of the spacetime symmetry by the defect implies the existence of a special operator on the defect with protected conformal dimension $\hat{\Delta}=p+1$. It is known as the \emph{displacement} operator. Similarly, if the defect breaks the internal symmetry $G$ of the bulk CFT down to $K$, there are operators with dimension $\hat{\Delta}=p$, called {\it tilt} operators, which generate a {\it defect conformal manifold} $G/K$. These operators are generically absent in XDCFT. Roughly speaking, this is because the fixed locus of XDCFTs and the choice of $K$ are determined globally by \eqref{eq:flatiota} and one cannot deform it locally.

Despite the absence of tilt operators, one can still define a defect conformal manifold, as there is a global choice in the location of the defect and choice of subgroups $K$ parametrised by $G/K$. Defect conformal manifolds without tilt operators are known to arise also in monodromy defects \cite{Billo:2013jda,Gaiotto:2013nva,Yamaguchi:2016pbj,Giombi:2021uae}. Both are examples of conformal manifolds without marginal operators. Exactly marginal operators imply a conformal manifold and in certain settings the converse is also true~\cite{Komatsu:2025cai}. One of the conditions in~\cite{Komatsu:2025cai} is a local stress tensor, which does not exist in DCFT and XDCFTs, so there is no contradiction. The absence of tilt and displacement operators is discussed in the case of the free $O(N)$ model in Section~\ref{sec:disp}, but a more general and rigorous analysis would be interesting.

\subsection{Correlation functions}\label{sec:corfunc}
We analyse now the possible structures of one-point and two-point correlation functions. For this we need the full set of invariants one can make out of $X_\pm$. 
The dot products $X_{+}\cdot X_{+}$ and $X_{-}\cdot X_{-}$ are basically identical to the ones used in DCFT, denoted $X\circ X$ and $X\bullet X$ in \cite{Billo:2016cpy}, so the structure of the correlation functions discussed below closely parallels the ones for DCFT explained in \cite{Billo:2016cpy}. 

For symmetric traceless operators, the polarisation vector $Z^M$ additionally splits into $Z_+^M$ and $Z_-^M$. The tensor $C^{NM}$ \eqref{CMN} then splits into three, $C_{++}$, $C_{+-}$ and $C_{--}$, depending on whether two, one or zero of its indices belong to the positive directions in \eqref{eq:embediota}. 
It is easy to see from the definition that
\begin{equation}
\begin{aligned}
C_{++}^{MN}&=-X_+^MC_{+-}^{NK}X_{-,K}+C_{+-}^{MK}X_{-,K}X_{+,N}\,,
\\
C_{--}^{MN}&=-X_-^MX_{+,K}C_{+-}^{KN}-X_{+,K}C_{+-}^{KM}X_{-,N}\,,
\end{aligned}
\end{equation}
so all correlation functions can be written in terms of $C_{+-}$ and $X_\pm$ only. Furthermore, one can check the additional three tensorial relations: $C_{+-}^{NM}C_{+-,MK}C_{+-}^{KL}\propto(C_{+-}\cdot C_{+-})C_{+-}^{NL}$, $X_{+,N}C^{NM}_{+-}X_+^KC_{+-,KM}\propto(C_{+-}\cdot C_{+-})(X_+\cdot X_+)$ and $C_{+-}^{NM}X_{+,N}X_{-,M}=0$. These further cut down the number of independent structures.

\subsubsection{Bulk one-point functions}\label{sec:Bulk1pt}
\paragraph{Scalar operators:}
The one-point function of scalar operators can depend only on $ X_-\cdot X_- (=-X_{+}\cdot X_{+})$. In the hyperbolic frame, equation \eqref{eq:LSray} sets this to unity, while in the flat frame $X_-\cdot X_-=|x_\perp|^2$. We hence have
\begin{equation}
\langle\mathcal{O}_i(X)\rangle=\frac{A_i}{(X_-\cdot X_-)^{\Delta_i/2}}
\quad\Rightarrow\,\quad
\langle\mathcal{O}_i(x)\rangle_{\text{hyperbolic}}=A_i\,,\quad
\langle\mathcal{O}_i(x)\rangle_{\text{flat}}=\frac{A_i}{|x_\perp|^{\Delta_i}}\,,
\label{eq:1pt}
\end{equation}
where the constants $\{A_i\}$ constitute \textit{new data} of the crosscap theory. This one-point function is manifestly $\iota_p$-even, which implies that just as in $\RP^d$, $\iota_p$-odd scalars generically cannot have non-vanishing one-point functions, though an exception to this rule is shown in \eqref{eq:1ptcodd}.

For parity-odd operators, the one-point function always vanishes. But there is one extra structure that can appear for an $\iota_p$-odd scalar, but only in the special case of $p=d-1$, i.e.\ $q=1$ or BCFT. In this case, $C_{+-}\cdot C_{+-}=0$, and thus only scalar operators can have non-vanishing $A_i$. However, since $X_-=x_\perp$ is one-dimensional, it is also possible to have non-vanishing \textit{parity-even, $\iota_p$-odd, scalar} one-point functions, which take the form 
\begin{equation}
    \langle\mathcal{O}_i(X)\rangle=\frac{A_i}{(X_-^2)^{(\Delta_i-1)/{2}}X_-}\qquad\Rightarrow\qquad
\langle\mathcal{O}_i(x)\rangle_{\text{flat}}=\frac{A_i}{|x_\perp|^{\Delta_i-1}x_\perp}\label{eq:1ptcodd}\,.
\end{equation}

\paragraph{Spinning operators:}
Since the one-point function of a spin-$J$ symmetric traceless tensor must be a $J$'th order polynomial in $Z$ and homogeneous with weight $\Delta$ in $X$, this leaves only one possible structure for the one-point function of an even parity spinning operator, namely
\begin{align}
    \langle\mathcal{O}_i(X,Z)\rangle=A_i\frac{\left(C_{+-}\cdot C_{+-}\right)^{J_i/2}}{(X_-\cdot X_-)^{(\Delta_i+J_i)/2}}\,.\label{eq:spin1pt}
\end{align}
This is only a polynomial in $Z$ if $J$ is even and is manifestly $\iota_p$-even. For parity-odd or $\iota_p$-odd spinning operators, their one-point function must vanish.

\subsubsection{Defect operators and correlation functions}\label{sec:bd2pt}
The $p$-dimensional fixed locus hosts operators that are different from those in the bulk and we denote them by $\hat{\mathcal{O}}$. Such operators carry conformal dimension $\hat{\Delta}$ and spin $\hat{J}$ for the defect conformal group $O^+(p+1,1)$, as well as quantum number $\hat{s}$ under the transverse rotation group $PO(q)$. Throughout this paper, we focus on operators $\hat{\mathcal{O}}$ that are symmetric traceless tensors under both $O^+(p+1,1)$ and $PO(q)$, which thus carry a single conformal spin $\hat{J}$ and a single transverse spin $\hat{s}$. 

Similarly to the bulk operators, they can be promoted to operators in a $p+2$-dimensional embedding space with coordinates $\hat{X}_+$, polarisation vector $\hat{Z}_+$ and transverse polarisation vector $\hat{Z}_-$, the latter of which keeps track of the $PO(q)$ symmetry. These vectors must satisfy
\begin{equation}
    \hat{Z}_-\cdot\hat{Z}_-=\hat{Z}_+\cdot\hat{Z}_+=\hat{X}_+\cdot\hat{X}_+=\hat{X}_+\cdot \hat{Z}_+=0\,,\label{eq:deftrans}
\end{equation}
which implies that the dependence of correlation functions on $\hat{Z}_+$ must be through the structure $\hat{D}_{++}=\hat{X}_+\otimes \hat{Z}_+-\hat{Z}_+\otimes\hat{X}_+$, while $\hat{Z}_-$ can appear freely. 

\paragraph{Defect correlators:}
Since the defect theory has the symmetries of a $p$-dimensional CFT, it has vanishing one-point functions, while the two-point functions take the form
\begin{equation}
\langle\hat{\mathcal{O}}_i(\hat{X}_+,\hat{Z}_+,\hat{Z}_-)\hat{\mathcal{O}}_j(\hat{Y}_+,\hat{W}_+,\hat{W}_-)\rangle
=\delta_{ij}\frac{(\hat{C}_{++}\cdot\hat{D}_{++})^{\hat{J}_i}(\hat{Z}_-\cdot \hat{W}_-)^{\hat{s}_i}}{(-2 \hat{X}_+\cdot\hat{Y}_+)^{\hat{\Delta}_i+\hat{J}_i}}\,,\label{eq:dd}
\end{equation}
where $\hat{D}^{MN}_{++}=\hat{Y}^M_+\hat{W}^N_+-\hat{W}^M_+\hat{Y}^N_+$. This is the natural generalisation of \eqref{Or2pt} to a CFT with a global $PO(q)$ symmetry. As discussed in Section~\ref{sec:embed}, one can likewise construct the three-point functions that depend on the defect structure constants $\hat C_{ijk}$.

\paragraph{Two-point function with a bulk scalar:}
Using the embedding coordinates, we can constrain the bulk--defect $2$-point function. In particular, when the bulk operator is a scalar, the two-point function is fully fixed to take the form
\begin{equation}
\langle\mathcal{O}_{i}(X)\hat{\mathcal{O}}_{j}(\hat{Y}_+,\hat{Z}_+,\hat{Z}_-)\rangle
=\delta_{\hat{J}_j0}B_{ij}\frac{(X_-\cdot \hat{Z}_-)^{\hat{s}_j}}{(-2X_+\cdot\hat{Y}_+)^{\hat{\Delta}_j}(X_-\cdot X_-)^{\frac{\Delta_i-\hat{\Delta}_j+\hat{s}_j}{2}}}\,.\label{bdscalar}
\end{equation}
The coefficients $B_{ij}$ are new data for the XDCFT. Together with the one-point functions $A_i$ and the dimensions $\hat\Delta_j$ and structure constants $\hat C_{ijk}$, this completes the XDCFT data. 

Notice that the right-hand side of \eqref{bdscalar} is even under $\iota_p$ if $\hat{s}_j$ is even and odd if $\hat{s}_j$ is odd, so for generic $q$ the two-point function of a $\iota_p$-even bulk operator and an odd transverse-spin defect operator must vanish and conversely for an $\iota_p$-odd bulk operator. As shown in Appendix~\ref{sec:bdspin}, this also holds for spinning bulk operators.

The exception is for $q=1$ (BCFT), for which there is an additional possible structure, where one power of $|X_-|$ in the denominator of \eqref{bdscalar} is replaced by $X_-$, as in the one-point function \eqref{eq:1ptcodd}. This is the appropriate form for $\iota_p$-even bulk operator and odd transverse-spin defect operator and vice versa.

\paragraph{Two-point function with bulk spinning operators:}
For spinning bulk operators there are seven independent $SO(p+1,1)\times PO(q)$ covariant contractions one can write down. In odd $q=d-p$ dimensions there are additional structures involving the Levi--Civita tensor. Those allow for two-point functions of operators of unmatched $\iota_p$ chirality or spin, generalising the $q=1$ case above. This is because for odd $q$, $PO(q)=SO(q)$, so the antisymmetric tensor preserves the symmetry. Due to the theory being parity invariant, the $\epsilon$ tensor in the defect directions cannot be used. All these structures are given in Appendix~\ref{sec:bdspin}.

\subsubsection{Bulk two-point functions} \label{eq:bb2pt}
For the two-point functions of bulk operators at $X$ and $Y$, there are combinations of the inner products that are invariant under conformal transformations, namely two independent cross ratios\footnote{Using the identification $X_{+}\cdot X_{+} \leftrightarrow X \circ X$ and $X_{-}\cdot X_{-} \leftrightarrow X \bullet X$, these cross ratios are related to the ones in \cite{Billo:2016cpy} by $2(\kappa_{-}-\kappa_{+})\leftrightarrow \xi$ and $\kappa_{+}\leftrightarrow\cos\phi$.}
\begin{equation}
    \kappa_\pm\equiv\mp\frac{X_\pm\cdot Y_\pm}{\sqrt{(X_-\cdot X_-)\,(Y_-\cdot Y_-)}}\,.\label{eq:kappas}
\end{equation}
In the flat frame, the cross ratios take the form
\begin{align}
\begin{split}
&\kappa_{+}=\frac{x^2+y^2}{2|x_\perp||y_\perp|}-\frac{x_\parallel\cdot y_\parallel}{|x_\perp||y_\perp|}=\frac{(x-y)^2+(x'-y)^2}{4|x_\perp||y_\perp|}\,,
\label{eq:flatkappa}\\
&\kappa_{-}=\frac{x_\perp\cdot y_\perp}{|x_\perp||y_\perp|}=-\frac{(x-y)^2-(x'-y)^2}{4|x_\perp||y_\perp|}\,.
\end{split}
\end{align}

We also find it useful to introduce the light-cone coordinates 
\begin{align}
(z,\bar{z})=(rt,r/t) 
\end{align}
which are defined by
\begin{equation}
    \kappa_+=\frac{1}{2}\left(r+\frac{1}{r}\right)\comma \kappa_-=\frac{1}{2}\left(t+\frac{1}{t}\right)\,.\label{eq:LCcoord}
\end{equation}
Their behaviour under the $\mathbb{Z}_2$ quotient is $\iota_p(z,\bar{z})=(-z,-\bar{z})$ corresponding to $\iota_p(\kappa_\pm)=\pm\kappa_\pm$.
To visualise these coordinates in the flat frame, one can use the residual conformal transformations to set $x_\parallel=0$, $x_\perp=(1,0,\dots)$ and $y_\parallel=(1,0,\dots)$, $y_\perp=(\text{Re}\,z,\text{Im}\, z,0,\dots)$. In the hyperbolic frame, the only invariants of two points are their geodesic distances on $\bH_{p+1}$ and $\RP^{q-1}$ 
denoted $\theta_\mathbb{H}$ and $\theta_{\RP}$ respectively. Accordingly, the light-cone coordinates are expressed in this frame as
\begin{equation}
    z=\exp(\theta_\bH+i\theta_{\RP})\,,
    \qquad
    r\geq0\,.
    \label{light-cone}
\end{equation}

\paragraph{Scalar two-point functions:}
The two-point function of bulk primary scalar operators can now be written in the form
\begin{equation}
\langle\mathcal{O}_i(X)\mathcal{O}_j(Y)\rangle
=\frac{G_{ij}\left(\kappa_+,\kappa_-\right)}{(X_-\cdot X_-)^{\Delta_i/2}(Y_-\cdot Y_-)^{\Delta_j/2}}\,,
\label{eq:2pt}
\end{equation}
where $G_{ij}$ is a function whose properties we study below. Note that this structure is non-zero only when both operators have the same $\iota_p$ chirality.

Extra possible structures for $q=1$ have one power of $X_-$ instead of $|X_-|$ or one power of $Y_-$ instead of $|Y_-|$. Those allow for non-vanishing two-point functions of operators with unmatched $\iota_p$ chirality (see \eqref{eq:1ptcodd}).

Comparing the crosscap defect two-point function \eqref{eq:2pt} to that on $\RP^d$ \eqref{eq:RP2pt}, we see that they take a very similar form, with the appropriate change in the denominator to powers of $x_\perp$ and $y_\perp$. One notable difference is that generally the two-point function for the crosscap defect has two cross ratios while for $\mathbb{R}\mathbb{P}^{d}$ there is a single one, since $\kappa_+=1$ is fixed. Likewise for the $p=d-1$ crosscap (a boundary), $\kappa_-=1$ and only $\kappa_+$ remains as a cross ratio.

\paragraph{Spinning two-point functions:}
For spinning operators $\mathcal{O}_1(X,Z)$, $\mathcal{O}_2(Y,W)$, 
there are twelve possible independent combinations of $X$, $Y$, $C_{+-}$ and $D_{+-}=Y_+\otimes W_--W_+\otimes Y_-$. For odd $q$ there are further structures involving the $\epsilon$ symbol of $SO(q)$. The enumeration of all such structures is shown in Appendix~\ref{sec:bbspin}. 

In principle, this analysis can be repeated for higher-point functions, which are expressed in terms of a larger and larger set of cross ratios which can systematically be constructed from inner products of the coordinates $X_\pm$ and $C_{+-}$. We do not pursue that here.

\subsection{Conformal blocks and crossing}
\label{sec:blocks}
\begin{figure}[t]
\centering
\begin{tikzpicture}[scale=0.5, baseline=(vert_cent.base)]
\node[inner sep=0pt,outer sep=0pt] (vert_cent) at (0,0) {$\phantom{\cdot}$};
\draw[very thick,black] (-3.5,0)--(3.5,0);
\draw[very thick, dashed] (-2,0.06)--(2,0.06);
\fill[black] (-2,2) circle (4pt);
\node[above left] at (-2,2) {$\cO_i$};
\fill[black] (2,2) circle (4pt);
\node[above right] at (2,2) {$\cO_j$};
\draw[dashed] (-2,2) to (-2,-2);
\draw[dashed] (2,2) to (2,-2);
\filldraw[color=black,fill=white,line width=0.35mm] (-2,-2) circle (4pt);
\node[below left] at (-2,-2) {$\cO_i'$};
\filldraw[color=black,fill=white,line width=0.35mm] (2,-2) circle (4pt);
\node[below right] at (2,-2) {$\cO_j'$};
\node[above] at (0,0) {$\hat\cO_k$};
\node[below] at (0,-5) {Defect channel};
\end{tikzpicture}
\qquad
\begin{tikzpicture}[scale=0.5, baseline=(vert_cent.base)]
\node[inner sep=0pt,outer sep=0pt] (vert_cent) at (0,0) {$\phantom{\cdot}$};
\draw[very thick,black] (-3.5,0)--(3.5,0);
\fill[black] (-1,3) circle (4pt);
\node[above left] at (-1,3) {$\cO_i$};
\fill[black] (1,3) circle (4pt);
\node[above right] at (1,3) {$\cO_j$};
\draw[dashed] (-1,3) to (0,2);
\draw[dashed] (1,3) to (0,2);
\draw[dashed] (0,-2) to (0,2);
\draw[dashed] (-1,-3) to (0,-2);
\draw[dashed] (1,-3) to (0,-2);
\filldraw[color=black,fill=white,line width=0.35mm] (-1,-3) circle (4pt);
\node[below left] at (-1,-3) {$\cO_i'$};
\filldraw[color=black,fill=white,line width=0.35mm] (1,-3) circle (4pt);
\node[below right] at (1,-3) {$\cO_j'$};
\node[below right] at (0,2) {$\cO_k$};
\node[below] at (0,-5) {Bulk channel};
\end{tikzpicture}
\qquad
\begin{tikzpicture}[scale=0.5, baseline=(vert_cent.base)]
\node[inner sep=0pt,outer sep=0pt] (vert_cent) at (0,0) {$\phantom{\cdot}$};
\draw[very thick,black] (-3.5,0)--(3.5,0);
\fill[black] (-1,3) circle (4pt);
\node[above left] at (-1,3) {$\cO_i$};
\fill[black] (1,3) circle (4pt);
\node[above right] at (1,3) {$\cO_j$};
\draw[dashed] (-1,3) to (0,1.8);
\draw[dashed] (-1,-3) to (0,-1.8);
\draw[dashed] (0,-1.8) to (0,1.8);
\draw[dashed] (1,3)[out=-95,in=60] to (0,-1.8);
\draw[dashed] (1,-3)[out=95,in=-60]to (0,1.8);
\filldraw[color=black,fill=white,line width=0.35mm] (-1,-3) circle (4pt);
\node[below left] at (-1,-3) {$\cO_i'$};
\filldraw[color=black,fill=white,line width=0.35mm] (1,-3) circle (4pt);
\node[below right] at (1,-3) {$\cO_j'$};
\node[below left] at (0,1.8) {$\cO_k$};
\node[below] at (0,-5) {Image channel};
\end{tikzpicture}
\caption{The three different operator product expansion channels. We indicate the fixed locus of $\iota_p$ in the covering space $\bR^d$ with the thick horizontal line, and for every operator include its mirror image. In the defect channel, the OPE is between $\cO_i\times\cO_i'$, exchanging the defect operator $\hat\cO_k$. In the bulk channel, the OPE is $\cO_i\times\cO_j$, exchanging the operator $\cO_k$ with their images. Lastly, in the image channel the roles of $\cO_j$ and $\cO_j'$ are 
interchanged. }
\label{fig:channels}
\end{figure}
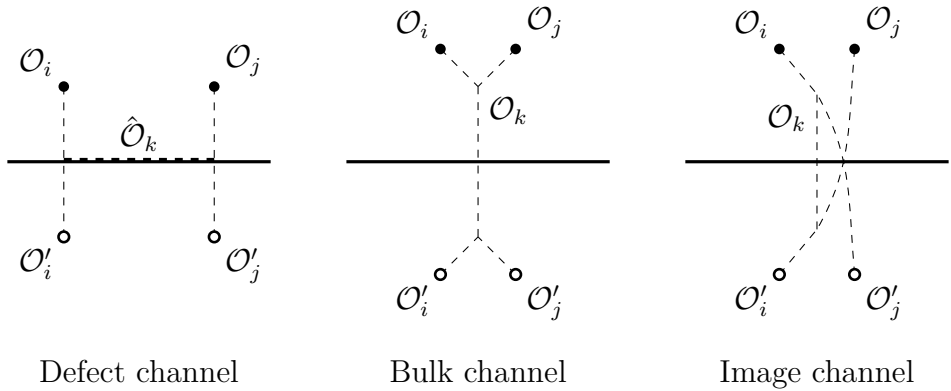

Examining the two-point functions of scalar operators \eqref{eq:2pt}, there are three limits where the operator $\cO_i(x)$ collides with another operator or the image of an operator under $\iota_p$: $x\to x'$, $x\to y$ and $x\to y'$. These lead to three operator product expansion (OPE) channels, which we respectively denote as the \emph{defect, bulk and image channels}, see Figure~\ref{fig:channels}. The analysis is again very similar to the case of defect CFT \cite{Billo:2016cpy} with the addition of the image channel (familiar from $\mathbb{R}\mathbb{P}^{d}$ \cite{Giombi:2020xah}).

We present now the three different expansion channels and their conformal blocks. The conformal blocks are obtained by enforcing that they are eigenfunctions of certain conformal Casimir equations, with eigenvalue
\begin{equation}
    C_{d,\Delta,J}=\Delta(\Delta-d)+J(J+d-2)\,.\label{eq:Cas}
\end{equation}

\subsubsection{Defect channel}
\label{sec:defdefchan}
In the defect channel, the operator $\cO_i$ collides with its own image $\mathcal{O}_i'$ which produces operators localised on the fixed locus, whose two-point function can then be computed using \eqref{eq:dd}. In this channel, it is convenient to work with the light-cone coordinates \eqref{eq:LCcoord} and the expansion is in small $r$ at fixed $t$.

The conformal block expansion allows for exchange of the defect identity and other defect primaries $\hat{\mathcal{O}}_{k}$ with the appropriate bulk--defect couplings $A_i$ and $B_{ik}$
\begin{equation}
G_{ij}\left(\kappa_+,\kappa_-\right)=A_iA_j+\sum_{k} B_{ik}B_{jk}g_\text{defect}\big(\hat{\Delta}_{k},\hat{s}_{k}\big|r,t\big)\,.
\label{eq:defectchannel}
\end{equation}
This block depends on the transverse spin $\hat{s}_{k}$ and conformal dimensions $\hat{\Delta}_k$ of the defect operator $\hat\cO_{\hat k}$, but not on the conformal dimensions of $\mathcal{O}_i$ and $\mathcal{O}_j$.

The conformal block is a solution to an appropriate Casimir equation. Since the symmetry group, up to discrete factors, is the same as the one for defect CFTs, the equation is precisely the same as in \cite{Billo:2016cpy}
\begin{align}
\begin{split}
\left(\frac{1}{2}\left(L^x_+\right)^2+C_{p,\hat{\Delta},0}\right)g_{\text{defect}}(\hat{\Delta},\hat{s}|r,t)=0\,,
\\
\left(\frac{1}{2}\left(L^x_-\right)^2+C_{q,0,\hat{s}}\right)g_{\text{defect}}(\hat{\Delta},\hat{s}|r,t)=0\,,
\end{split}\label{eq:defectCasimir}
\end{align}
where $L^x_+$ and $L^x_-$ are the Casimir operators for $SO(p+1,1)$ and $PO(q)$, respectively, acting only on the point $x$ and the $C$'s are defined in \eqref{eq:Cas}. Explicitly, using embedding space coordinate, $(L^x_\pm)^2=(X_\pm^M\partial_{X_\pm^N}-X_\pm^N\partial_{X_\pm^M})^2$.

Since $(L_+)^2$ acts only on $\kappa_+$ (or $r$) and $(L_-)^2$ only on $\kappa_-$ (or $t$), the equations are solved by separation of variables as \cite{Billo:2016cpy,Gimenez-Grau:2022ebb}
\begin{equation}
    g_{\text{defect}}(\hat{\Delta},\hat{s}|
        r,t)=r^{\hat{\Delta}}{}_2F_1\left(\hat{\Delta},\frac{p}{2};\hat{\Delta}+1-\frac{p}{2};r^2\right)\times (2t)^{-\hat{s}}{}_2F_1\left(-\hat{s},\frac{q}{2}-1;2-\frac{q}{2}-\hat{s};t^2\right).\label{eq:gdefect}
\end{equation}
The factor of $2^{-\hat{s}}$ is conventional and ensures a nice expansion in the defect channel \cite{Billo:2016cpy}.

One easily observes that the blocks for dimensions $p$ and $d-p-2$ are related by
\begin{equation}
    2^{\hat{s}}g_{\text{defect}}(\hat{\Delta},\hat{s}|r,t)\Big|_{p\to p'}=2^{\hat{\Delta}}g_{\text{defect}}(\hat{s},\hat{\Delta}|t,r)\Big|_{p\to d-p'-2}\,,\label{eq:gdefsym}
\end{equation}
which holds for arbitrary $p'$. For standard defects, this relation is known already (cf.~\cite{Hamilton:2006fh,Czech:2016xec}) and the geometric interpretation is given in \cite{Fukuda:2017cup}. Further relations among defect conformal blocks in different dimensions are discussed in \cite{Ghosh:2025lzb}. We use this relation below.

For integer $\hat{s}$, the second hypergeometric function in \eqref{eq:gdefect} is expressible as a Gegenbauer polynomial
\begin{equation}
\label{gegenbauer}
t^{-\hat{s}}{}_2F_1\left(-\hat{s},\frac{q}{2}-1;2-\frac{q}{2}-\hat{s};t^2\right)
=\frac{\Gamma(\hat{s}+1)\Gamma(q/2-1)}{\Gamma(q/2+\hat{s}-1)}C_{\hat{s}}^{(\frac{q}{2}-1)}(\kappa_-)\,.
\end{equation}
Recall from \eqref{light-cone} that $\kappa_-=\cos\theta_\RP$ which is not single valued on $\RP^{q-1}$ but only on the double cover $S^{q-1}$. For even $\hat{s}$, the Gegenbauer polynomial is even and thus a proper function on $\RP^{q-1}$ while for odd $\hat s$ it is not. This guarantees that only even $\hat s$ operators appear in the defect channel OPE of operators even under $\iota_p$ and only odd $\hat s$ appear for operators odd under $\iota_p$, which is consistent with \eqref{bdscalar}. 

Note that this channel disappears as $p\ra -1$ (i.e.\ $\mathbb{RP}^d$), since there is no fixed locus.

\subsubsection{Bulk channel}
\label{sec:defbulkchan}
In the bulk channel, one uses the regular bulk OPE 
$\cO_i\times\cO_j\ni\cO_k$ with the usual structure constants $C_{ijk}$ and then computes the one-point function of $\cO_k$ in terms of its corresponding constant $A_k$ \eqref{eq:1pt}. For scalar operators, the expansion then takes the form
\begin{equation}
G_{ij}(\kappa_+,\kappa_-)=(2(\kappa_+-\kappa_-))^{-\frac{\Delta_i+\Delta_j}{2}}\sum_k C_{ijk}A_kg_\text{bulk}(\Delta_{ij};\tau_k,J_k|\kappa_+,\kappa_-)\,,
\label{eq:bulkchannel-kappa}
\end{equation}
where the factor of $(2(\kappa_+-\kappa_-))^{-\frac{\Delta_i+\Delta_j}{2}}$ is included such that the block only depends on $\Delta_{ij}=\Delta_i-\Delta_j$ and the twist $\tau_k=\Delta_k-J_k$.

It is actually more convenient to use the light-cone coordinates \eqref{eq:LCcoord}, where
\begin{equation}
G_{ij}(z,\bar z)
=\frac{|z|^{\frac{\Delta_i+\Delta_j}{2}}}{|1-z|^{\Delta_i+\Delta_j}}
\sum_k C_{ijk}A_kg_\text{bulk}(\Delta_{ij};\tau_k,J_k|z,\bar z)\,,
\label{eq:bulkchannel}
\end{equation}
and this is the notation we follow below.

As shown in \cite{Billo:2016cpy}, the bulk blocks satisfy the conformal Casimir equation 
\begin{equation}
\left(\frac{1}{2}(L^x+L^y)^2+C_{d,\Delta_k,J_k}\right)\frac{|z|^{\frac{\Delta_i+\Delta_j}{2}}}{|1-z|^{\Delta_i+\Delta_j}|x_\perp|^{\Delta_i}|y_\perp|^{\Delta_j}}g_{\text{bulk}}(\Delta_{ij};\tau_k,J_k|z,\bar z)=0\,,\label{eq:bulkCasimir}
\end{equation}
where $(L^x)^2=(X^M\partial_{X^N}-X^N\partial_{X^M})^2$ is the quadratic Casimir for the full $SO(d+1,1)$ conformal group, the operator $L^x+L^y$ acts on both coordinates and the eigenvalue $C_{d,\Delta_k,J_k}$ is defined in \eqref{eq:Cas}. This equation is hard to solve, so following \cite{Gimenez-Grau:2022ebb}, we present the perturbative solution for $g_{\text{bulk}}$ in an expansion around $(z,\bar{z})=(1,1)$ or $\kappa_+=\kappa_-=1$, as
\begin{align}
\label{eq:gbulkexp}
g_{\text{bulk}}(\Delta_{ij};\tau_k,J_k|z,\bar{z})
&=\bar{z}^{\Delta_{ij}/4}\sum_{n=0}^\infty\sum_{m=-n}^nc_{n,m}(\tau_k,J_k)(1-z)^{\frac{\tau_k}{2}+n}(1-\bar{z})^{\frac{\tau_k}{2}+J_k+m}
\\\nonumber
&\quad\times{}_2F_1\left(\frac{\tau_k}{2}+J_k+m,\frac{\tau_k+\Delta_{ij}}{2}+J_k+m;\tau_k+2(J_k+m),1-\bar{z}\right).
\end{align}
The coefficients $c_{n,m}$ are obtained by solving the Casimir equation order by order in $1-z$ and $1-\bar{z}$, with the boundary condition $c_{0,0}(\tau_k,J_k)=2^{-J_k}$.

We note that the bulk blocks for dimensions $p$ and $d-p-2$ also satisfy a relation similar to the one for the defect block \eqref{eq:gdefsym}, namely
\begin{equation}
    g_{\text{bulk}}(\Delta_{ij};\tau_k,J_k|z,\bar z)\Big|_{p\to p'}
=e^{-i\frac{\pi}{2}(\tau_k+2J_k)}g_{\text{bulk}}(\Delta_{ij};\tau_k,J_k|z,1/\bar z)\Big|_{p\to d-p'-2}\,,\label{eq:gbulksym}
\end{equation}
as can be obtained by writing out the differential equation \eqref{eq:bulkCasimir} and identifying the correct asymptotics. In terms of $\kappa_\pm$, the change is $\kappa_+\leftrightarrow\kappa_-$.

In the particular cases of integer $p$ or $q$ discussed in Section~\ref{subsec:ccross}, this relation leads to the following results:
\begin{itemize}
\item $\boldsymbol{p=d-1}$: The crosscap defect is a boundary. As mentioned previously, $\kappa_-=1$ is fixed, so $\bar z=z$ is real and positive and only scalar operators can have non-vanishing one-point functions. We therefore only need to compute the bulk block with $\tau_k=\Delta_k,J_k=0$ which allows for the exact computation of the bulk conformal block~\cite{McAvity:1995zd}
\begin{equation}
g_{\text{bulk}}(\Delta_{ij},\Delta_k,0|z,z)
=\frac{(z-1)^{\Delta_k}}{z^{\frac{\Delta_k}{2}}2^{J_k}}
\,{}_2F_1\left(\frac{\Delta_k+\Delta_{12}}{2},\frac{\Delta_k-\Delta_{12}}{2};\Delta_k+1-\frac{d}{2};\frac{(z-1)^2}{z}\right).
\label{eq:gbulkpdm1}
\end{equation}
\item $\boldsymbol{p=d-2}$: It was noted in \cite{Billo:2016cpy} that for $p=d-2$, the differential equation \eqref{eq:bulkCasimir} simplifies and is related to the regular conformal block for four-point functions in the bulk theory. Specifically,
\begin{equation}
g_{\text{bulk}}(\Delta_{ij},\tau_k,J_k|z,\bar z)
=e^{-i\frac{\pi}{2}(\tau_k+2J_k)}g_{\text{4pt}}\left(\Delta_{ij},0;\tau_k,J_k\Big|-\frac{|z-1|^2}{\bar z},\frac{z}{\bar z}\right),
\label{eq:q=2}
\end{equation}
where on the right-hand side $g_{\rm 4pt}(\Delta_{12},\Delta_{34};\tau,J|U,V)$ is the standard four-point conformal block in $d$ dimensions for scalar external operators of dimensions $\Delta_{1,2,3,4}$, an exchanged operator with twist $\tau$ and spin $J$ and complexified $U$, $V$ cross ratios \cite{Dolan:2000ut}. 
While we don't have a direct interpretation of this relation in this case, it is related via \eqref{eq:gbulksym} to the $p=0$ case, where the connection to the four-point block is straightforward, as discussed below.

\item $\boldsymbol{p=0}$: 
Per the discussion in Section~\ref{subsec:ccross}, the two-point function in the presence of the 0-dimensional crosscap defect can be viewed as a four-point function in the theory without a defect but with two extra twist fields. If we focus on the sphere containing the four operators and use conformal transformations to place the twist fields at $0$ and $\infty$ and one of the two bulk operators at $1$, then in our light-cone coordinates \eqref{eq:LCcoord}, the second operator is at $z$. Thus, the bulk channel is nothing but the usual $t$-channel of the bulk CFT, and the corresponding block is then given by
\begin{equation}\label{eq:p=0rel4pt}
g_{\text{bulk}}(\Delta_{ij},\tau_k,J_k|z,\bar z)
=g_\text{4pt}(\Delta_{ij},0;\tau_k,J_k||1-z|^2,|z|^2)\,.
\end{equation}
The function on the right-hand side is the same bulk conformal block as in \eqref{eq:q=2}. Note that the cross ratios $U$, $V$ seem to be exchanged here, compared to their standard definitions. This simply reflects the fact that it is a $t$-channel block.

\item $\boldsymbol{p=-1}$: This is the usual crosscap, whose conformal block was computed in \cite{Nakayama:2016cim}. It can alternatively be derived from the BCFT conformal block \eqref{eq:gbulkpdm1} by using \eqref{eq:gbulksym}. Again, only bulk scalars can have non-vanishing one-point functions, while it is now $\kappa_+$ (or $r$) that is held fixed, meaning $\bar z=1/z$, i.e.\ $|z|=1$. The block is given by
\begin{equation}
g_\text{bulk}(\Delta_{ij},\Delta_k,0|z,1/z)
=\frac{(z-1)^{\Delta_k}}{z^\frac{\Delta_k}{2}2^{J_k}}
\,{}_2F_1\left(\frac{\Delta_k+\Delta_{12}}{2},\frac{\Delta_k-\Delta_{12}}{2};\Delta_k+1-\frac{d}{2};\frac{(z-1)^2}{z}\right).
\end{equation}
\item $\boldsymbol{p=(d-2)/2}$: For the special case of $p=(d-2)/2$, the relation \eqref{eq:gbulksym} becomes a non-trivial identity for the bulk block, $g_{\text{bulk}}(z,\bar z)=g_{\text{bulk}}(z,1/\bar z)$. It would be interesting to explore the implications of this identity on the bootstrap analysis.
\end{itemize}
For other values of $p$, we find neither simplified dynamics nor control of either side of \eqref{eq:gbulksym}, and can therefore not compute the block explicitly. Note that since we mostly treat $p$ as a continuous parameter, we use the general expression \eqref{eq:gbulkexp} in the computations in Sections~\ref{sec:freebulk} and~\ref{sec:Bulk}, truncating the series at finite $n$ and extrapolating the result.

\subsubsection{Image channel}
\label{sec:defimchan}
In the image channel the OPE is between $\mathcal{O}_i$ and the image of $\cO_j$ under $\iota_p$, denoted $\cO'_{j}$ (see Figure \ref{fig:channels}),
and is valid for $z,\bar z\sim1$. The block expansion is
\begin{equation}
G_{ij}(z,\bar z)=\frac{|z|^{\frac{\Delta_i+\Delta_j}{2}}}{|z+1|^{\Delta_i+\Delta_j}}
\sum_k C_{ijk}A_kg_{\text{image}}(\Delta_{ij};\tau_k,J_k|z,\bar z)\,.
\label{eq:imagechannel}
\end{equation}

This channel is very similar to the bulk channel. The replacement of the operator at $y$ with its image at $y'$ amounts to replacing $(z,\bar z)\to(-z,-\bar z)$, or $\kappa_-\to-\kappa_-$, and if the operator $\mathcal{O}_{j}$ is $\iota_p$-odd, also an overall change in sign. Therefore the conformal block for the image channel is\footnote{We can also define the conformal block without an overall sign ($g_{\rm image}(z,\bar z)=g_{\rm bulk}(-z,-\bar z)$). In this case, we have to replace $C_{ijk}$ with the structure constant with the image operator $C_{ij'k}$, which is related to $C_{ijk}$ by an overall sign $C_{ijk}=\pm C_{ij'k}$.}
\begin{equation}
g_{\text{image}}(z,\bar z)
=\pm g_{\text{bulk}}(-z,-\bar z)\,,
\label{eq:crossing}
\end{equation}
where the sign depends on whether the operators are even or odd (recalling that the two-point function vanishes if they have different parities).
This is very similar to the case of a CFT on real projective space \cite{Giombi:2020xah}. 

Note that this channel does not exist for $p= d-1$ where $z=\bar z\geq0$ (or because $\kappa_{-}$ is fixed to be $1$). Physically, this is because one cannot go around the fixed locus to make an operator $\cO_i$ and the image of $\cO_j$ collide in the bulk. In other words, one cannot draw a sphere which contains $\mathcal{O}_i$ and $\mathcal{O}^{\prime}_{j^{\prime}}$ that does not contain $\mathcal{O}_j$. In a technical sense, this implies that the image operator is always outside the radius of convergence of the OPE when $p=d-1$.

\subsubsection{Crosscap crossing equations}
\label{sec:crossing}
As in DCFT, equating the different expansions \eqref{eq:bulkchannel}, \eqref{eq:imagechannel}, and \eqref{eq:defectchannel} leads to non-trivial constraints on the XDCFT data, which we call {\it the crosscap crossing equations}. The simplest of these comes from matching the bulk and image channels using \eqref{eq:crossing} and is equivalent to imposing\footnote{Or in terms of $\kappa_\pm$, this is $G_{ij}(\kappa_{+},\kappa_{-})=\pm G_{ij}(\kappa_{+},-\kappa_{-})$}
$G_{ij}(z,\bar z)=\pm G_{ij}(-z,-\bar z)$. Despite its simple form, it constrains the XDCFT data once it is expanded into conformal blocks, since the blocks are not simple power series in $z$, $\bar z$ but rather in $z\mp1$, $\bar z\mp1$. It would be interesting to perform the bootstrap analyses of these crossing equations both numerically and analytically.

\section{Crosscap defects in the free \texorpdfstring{$O(N)$}{O(N)} model}\label{sec:free}
Let us now examine a simple example of an XDCFT to see how these things play out in practice. One of the simplest CFTs is the theory of $N$ free scalars $\left\{\phi^i\right\}_{i=1}^N$ with conformal dimension $\Delta=d/2-1$ governed by the action
\begin{equation}
S^{F}=\int\dd^{d} x\,\frac{1}{2}\partial_\mu \phi^i\partial^\mu \phi^i\,,
\label{eq:freeaction}
\end{equation}
where the $F$ superscript denotes that we are working with a free theory. There is an $O(N)$ symmetry rotating the different scalars. 

Upon introducing the crosscap defect, the $\iota_p$ charges of the scalars can be freely chosen. We take $N_+$ $\iota_p$-even scalars, denoted $\phi^{\overset{+}{\imath}}$, and $N_-$ $\iota_p$-odd scalars, denoted $\phi^{\overset{-}{\imath}}$, such that $N=N_++N_-$. It is also convenient to introduce the opposite combination $\Npm=N_+-N_- $. Thus, in the presence of the crosscap, the $O(N)$ global internal symmetry is broken to $K=O(N_+)\times O(N_-)$ and the total symmetry group is
\begin{equation}
    O^+(p+1,1)\times O(N_+)\times\frac{O(q)\times O(N_-)}{\iota_p}\,,
\end{equation}
where $\iota_p$ acts by flipping the sign of all directions in the fundamental representation of $O(q)\times O(N_-)$.
A seemingly possible enrichment of the free crosscap CFT is the addition of terms localised on the fixed locus. For any of the $\iota_p$-even fields, we can include a defect action term
\begin{equation}
\exp\left[-S^{D}\right]\,,
\qquad
S^{D}=
\frac{1}{2}h_0\int\dd^{p}x_{\para}\,\phi^{\overset{+}{\imath}}\phi^{\overset{+}{\imath}}\,.
\label{counter}
\end{equation}
This is a relevant deformation for $p>d-2$, i.e.\ $q<2$, triggering an RG flow. In the absence of a crosscap, this defect was studied for $p=2$ in \cite{Trepanier:2023tvb, Giombi:2023dqs, Raviv-Moshe:2023yvq} and for other values of $p$ in \cite{deSabbata:2024xwn}. In our context, a closer analogue is the case of boundary CFT or $q=1$, where $h_0$ flows to infinity and the boundary conditions are changed from Neumann to Dirichlet (see e.g.~\cite{McAvity:1995zd}), in which case the deformation is not really new and simply amounts to increasing $N_-$. We do not include such a term in the present section. This term plays a more important role for the interacting theory; in Section~\ref{sec:interacting} it is required as a counterterm around $p=2$.

\subsection{Correlation functions}
\label{sec:freecor}
Since the theory is free, computing correlators of $\phi$'s in the presence of the crosscap simply amounts to Wick contractions with massless propagators. The propagator of a scalar field on $\bR^d/\iota_p$ is the sum of two propagators in $\bR^d$, each of which is given by
\begin{equation}
K(x,y)\equiv\frac{\mathcal{N}_\phi^2}{|x-y|^{d-2}}
=\frac{\cN_\phi^2}{(2|x_\perp||y_\perp|(\kappa_+-\kappa_-))^{d/2-1}}\,,
\qquad
    \mathcal{N}_\phi^2=\frac{\Gamma(d/2-1)}{4\pi^{d/2}}\,.\label{free-prop}
\end{equation}
Written in terms of the cross ratios \eqref{eq:flatkappa}, the correlator in the XDCFT has indeed the structure of bulk scalar two-point functions \eqref{eq:2pt}
\begin{align}
\langle\phi^{\ipm}(x)\phi^{\jpm}(y)\rangle
&=\frac{\mathcal{N}_\phi^2\,G^{F}_{\ipm\jpm}(\kappa_+,\kappa_-)}{|x_\perp|^{d/2-1}|y_\perp|^{d/2-1}}\,,\label{eq:free2ptform}
\\G^{F}_{\ipm\jpm}(\kappa_+,\kappa_-)
&=\delta_{\ipm\jpm}\left[\frac{1} {(2(\kappa_+-\kappa_-))^{d/2-1}}\pm 
\frac{1}{(2(\kappa_++\kappa_-))^{d/2-1}}\right].
\label{free2pt}
\end{align}
From this, by taking the limit $y\to x$ and removing the divergence by appropriate normal-orderings, we immediately read off the $1$-point functions of an $O(N)$ scalar and symmetric traceless $2$-tensor (both of which are $SO(p+1,1)$ scalars with dimension $\Delta=d-2$)
\beq\label{eq:SandT}
S=\frac{1}{\cN_S}\phi^i\phi^i\,,\qquad
T^{ij}=\frac{1}{\cN_T}
\left(\phi^i\phi^j-\frac{\delta^{ij}}{N}\phi^k\phi^k\right),
\qquad
\cN_S=\sqrt{2N}\mathcal{N}_\phi^2\,,\quad \mathcal{N}_T=\sqrt{2}\mathcal{N}_\phi^2\,.
\eeq
The normalisation constants $\cN_S$ and $\cN_T$ are chosen such that two-point functions in $\mathbb{R}^{d}$ are normalised as in \eqref{Or2pt}
\begin{equation}
\langle S(x)S(y)\rangle_{\mathbb{R}^{d}}=\frac{1}{|x-y|^{2(d-2)}}\,,\quad 
\langle T^{ij}(x)T^{k\ell}(y)\rangle_{\mathbb{R}^{d}}
=\frac{\frac{1}{2}(\delta^{ik}\delta^{j\ell}+\delta^{i\ell}\delta^{jk})-\frac{1}{N}\delta^{ij}\delta^{k\ell}}{|x-y|^{2(d-2)}}\,.
\label{eq:normalisations}
\end{equation}
The one-point functions then evaluate to
\begin{equation}
\begin{aligned}
\label{1pt-free-new}
\langle S(x)\rangle_{}
&=\frac{\Npm}{\sqrt{2N}}
\frac{1}{2^{d-2}|x_\perp|^{d-2}}\,,
\qquad&
\langle T^{\ipl\jmi}(x)\rangle_{}
&=\langle T^{\imi\jpl}(x)\rangle_{}=0\,,
\\
\langle T^{\ipl\jpl}(x)\rangle_{}
&=\frac{\sqrt{2}\,N_-}{N}\frac{\delta^{\ipl\jpl}}{2^{d-2}|x_\perp|^{d-2}}\,,
\qquad&
\langle T^{\imi\jmi}(x)\rangle_{}
&=-\frac{\sqrt{2}\,N_+}{N}
\frac{\delta^{\imi\jmi}}{2^{d-2}|x_\perp|^{d-2}}\,.
\end{aligned}
\end{equation}
We reproduce this expression (and the generalisation to operators carrying spin) by using the bulk channel expansion in Section~\ref{sec:freebulk}.

\subsection{Defect channel expansion}\label{sec:freedefect}
We now want to expand the bulk two-point function \eqref{free2pt} in order to read off further conformal data, namely bulk--defect two-point functions and bulk one-point functions. We start with the defect channel expansion. Let us recall that due to \eqref{bdscalar}, only defect scalars $\hat{J}=0$ can appear. We further note that only operators with even transverse spin should contribute for the $\iota_p$-even fields $\phi^{\ipl}$ while only odd spin operators should contribute for $\phi^{\imi}$. The relevant defect operators, denoted $\hat{\mathcal{O}}_{\hat{s}}^{\ipl}$ and $\hat{\mathcal{O}}_{\hat{s}}^{\imi}$ respectively, are
\begin{equation}
\hat{\mathcal{O}}_{\hat{s}}^{\ipm,a_1\dots a_{\hat{s}}}(x_\para)\sim
\lim_{x_\perp\to0}\partial^{a_1}_\perp\cdots \partial_\perp^{a_{\hat{s}}}\phi^{\ipm}(x_\para,x_\perp)\comma (-1)^{\hat{s}}=\pm1\,.
\label{eq:defectprims}
\end{equation}
The constants of proportionality are chosen so that the two-point functions of these operators take the form \eqref{eq:dd}. They are $SO(p+1,1)$ scalars as required and have defect conformal dimensions $ \hat{\Delta}_{\hat{\mathcal{O}}^{\ipm}_{\hat{s}}}=\hat{s}+d/2-1$. Note that any contracted derivatives $\partial_\perp^2$ can be traded for $\partial^2_\para$ due to the equations of motion $(\partial_\perp^2+\partial_\para^2)\phi=0$, making such operators defect descendants. Therefore the operators in \eqref{eq:defectprims} are in symmetric traceless representations of $PO(q)$. Furthermore, since any operators surviving in the limit $x_\perp\ra 0$ must be invariant under $\iota_p$, $\hat{s}$ is even for $\ipl$ while it is odd for $\imi$.

Expanding \eqref{free2pt} in the defect channel conformal blocks \eqref{eq:gdefect}, we find the following bulk--defect two-point function coefficients, independent of $p$,
\begin{equation}
\Big(B^{F}_{\phi^{\ipm}\hat{\mathcal{O}}^{\jpm}_{\hat{s}}}\Big)^2=\delta^{\ipm \jpm}
(1\pm(-1)^{\hat s})\frac{2^{\hat{s}}\Gamma\big(\frac{d}{2}-1+\hat{s}\big)}{\Gamma(\hat{s}+1)\Gamma\big(\frac{d}{2}-1\big)}\,.
\label{eq:freespectrum}
\end{equation}
It is instructive to compare this (for $p=2$) with the result for surface defects in free scalar in \cite{Lauria:2020emq} (see also \cite{Behan:2020nsf}). As shown there, the contact term $\langle \Box \phi (x) \phi(y)\rangle\propto \delta^{d}(x-y)$ strongly constrains the bulk--defect OPE data. In particular, in the absence of $\psi^{(-)}$ modes in \cite{Lauria:2020emq}, they are uniquely fixed to be
\footnote{As compared to the expression in \cite{Lauria:2020emq}, \eqref{eq:surfacebulkdefect} contains an extra factor of $2^{\hat{s}}$. This comes purely from the different definition of the bulk--defect two-point function \eqref{bdscalar}.}
\begin{align}
\left(B^\text{DCFT}_{\phi \hat{\mathcal{O}}_{\hat{s}}}\right)^2=\frac{2^{\hat{s}}\Gamma\big(\frac{d}{2}-1+\hat{s}\big)}{\Gamma(\hat{s}+1)\Gamma\big(\frac{d}{2}-1\big)}\,.\label{eq:surfacebulkdefect}
\end{align}
Physically, these OPE data are realised in the ``trivial defect'', a defect corresponding to inserting nothing on $\mathbb{R}^{d}$.
The two results, \eqref{eq:freespectrum} and \eqref{eq:surfacebulkdefect}, differ by a factor of $(1\pm (-1)^{\hat{s}})$. These differences come from the difference of the contact term, which for the crosscap defects is modified to $\langle\phi^{\ipm}(x)\phi^{\jpm}(y)\rangle\propto \delta^{\ipm \jpm}\left(\delta^{d}(x-y)\pm \delta^{d}(x-y^{\prime})\right)$, due to the presence of the image operators. 
The differences can also be attributed to the ratio of the volume of the transverse $\RP^q$ compared to $S^q$ for a regular defect. Similarly, the above expressions for $d=4$ and $p=1$ agree with the result for the magnetic line defect \cite{Gimenez-Grau:2022ebb,
Cuomo:2021kfm,
Nishioka:2022qmj,Bianchi:2023gkk} 
in the free $O(N)$ model, up to the same overall factor $(1\pm (-1)^{\hat{s}})$.

\subsection{Bulk channel expansion}
\label{sec:freebulk}
\begin{figure}[t]
\centering
\includegraphics[width=\capwidth]{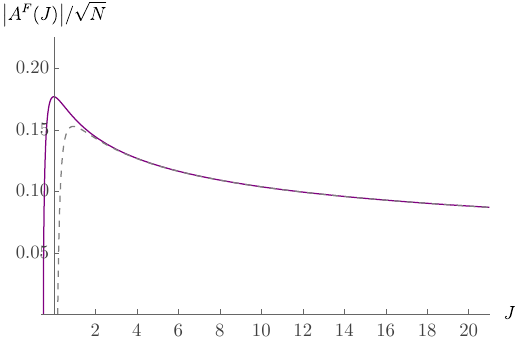}
    \caption{Plot of the one-point function of $S_J$ \eqref{eq:alphaF} in $d=4$ as a function of $J$, in the limit $N\ra\infty$ with $N_-$ held fixed. The exact expression is pictured in purple, while the large $J$ asymptotics \eqref{eq:alpha0asym} is given by the dashed grey line.}
    \label{fig:Bafree}
\end{figure}

We now similarly expand \eqref{free2pt} in the bulk channel, finding contributions from the identity and operators with twist $d-2$ and even spin $J$. They are spinning generalisations of \eqref{eq:SandT} and are thus either singlets or symmetric traceless two-tensors of $O(N)$, and we denote them $S_{J}$ and $T^{ij}_{J}$, respectively. Schematically, they are of the form
\begin{equation}
S_{J,\mu_1\dots\mu_J}
\sim\phi^{i}\partial_{\mu_1}\dots\partial_{\mu_J}\phi^{i}
\comma 
T^{ij}_{J,\mu_1\dots \mu_J}
\sim\phi^{(i}\partial_{\mu_1}\dots\partial_{\mu_J}\phi^{j)}
-\frac{\delta^{ij}}{N}S_{J,\mu_1\dots\mu_J}\,,
\label{bulkops}
\end{equation}
with normalisation chosen in agreement with \eqref{Or2pt}, i.e.\ in embedding space coordinates the two-point functions in $\mathbb{R}^{d}$ are given by
\begin{equation}
\begin{split}
&\langle S_J(X,Z)S_J(Y,W)\rangle_{\mathbb{R}^d}=\frac{(C\cdot D)^J}{(-2 X\cdot Y)^{d-2}}\,,
    \\
&\langle T^{ij}_J(X,Z)T^{k\ell}_J(Y,W)\rangle_{\mathbb{R}^d}=\frac{(C\cdot D)^J}{(-2 X\cdot Y)^{d-2}}\left[\frac{1}{2}\left(\delta^{ik}\delta^{j\ell}+\delta^{i\ell}\delta^{jk}\right)-\frac{1}{N}\delta^{ij}\delta^{k\ell}\right]\,,
\end{split}
\end{equation}
where we recall that $C_{MN}=X_MZ_N-Z_MX_N$, $D_{MN}=Y_MW_N-W_MY_N$. These operators are the unique primaries built from two $\phi$'s and $J$ derivatives since any other arrangements of derivatives differ only by descendants of lower spin operators. See more explicit expressions in \cite{Henriksson:2022rnm}, including the corresponding OPE coefficients which are summarised in Appendix~\ref{ap:bulk}.
Notice in particular that the operators in \eqref{eq:SandT} are $S=S_{0}$ and $T=T_{0}$. 
Expanding the bulk $2$-point function into conformal blocks and dividing by the OPE coefficients, we can extract the $1$-point functions of these operators explicitly. For the $O(N)$ singlets, they read
\begin{align}\label{eq:alphaF}
A^{F}_{S_{J}}=\frac{\Npm}{\sqrt{N}}\frac{(-1)^{J/2}}{\pi^{1/4}2^{d-3/2}}\sqrt{\frac{\Gamma\big(\frac{J+1}{2})\Gamma\big(\frac{J+d-3}{2}\big)\Gamma\big(\frac{2J+d-2}{2}\big)}{\Gamma\big(\frac{J+2}{2}\big)\Gamma\big(\frac{J+d-2}{2}\big)\Gamma\big(\frac{2J+d-3}{2}\big)}}\,,
\end{align}
which is independent of $p$ and we recall $N_{+-}=N_+-N_-$. For the symmetric traceless $2$-tensor $T_{J}^{ij}$, we find that $1$-point functions form a diagonal matrix with respect to the indices $i$ and $j$,
\begin{equation}
\begin{split}
\langle T_{J}^{ij}(X,Z)\rangle
&=A^{F}_{T_J}
\frac{\left(C_{+-}\cdot C_{+-}\right)^{J/2}}{(X_-\cdot X_-)^{(d-2)/2}}
        \begin{pmatrix}
        N_- \,\mathbb{I}_{N_{+}\times N_{+}}&0_{N_{+}\times N_{-}}
        \\0_{N_{-}\times N_{+}}&-N_+\,\mathbb{I}_{N_{-}\times N_{-}}
        \end{pmatrix}^{ij}\,,
        \\
A^{F}_{T_{J}}
&=\frac{2}{\Npm\sqrt{N}}A^F_{S_J}
=\frac{2}{N}\frac{(-1)^{J/2}}{\pi^{1/4}2^{d-3/2}}\sqrt{\frac{\Gamma\big(\frac{J+1}{2})\Gamma\big(\frac{J+d-3}{2}\big)\Gamma\big(\frac{2J+d-2}{2}\big)}{\Gamma\big(\frac{J+2}{2}\big)\Gamma\big(\frac{J+d-2}{2}\big)\Gamma\big(\frac{2J+d-3}{2}\big)}}\,.
        \label{eq:tensornorm}
    \end{split}
\end{equation}
Indeed, these agree with the computation of the one-point functions of $S$ and $T$ in \eqref{1pt-free-new} for $J=0$. We note that at large $J$, the one-point functions go like
\begin{equation}
\begin{split}
    &A^F_{S_J}\underset{J\ra+\infty}{\sim}\frac{\Npm}{\sqrt{N}}\frac{(-1)^{J/2}}{2^{d-2}(\pi J)^{1/4}}\left(1-\frac{2d-5}{16J}+O(J^{-2})\right)\,,
    \\
    &A^F_{T_J}\underset{J\ra+\infty}{\sim}\frac{2}{N}\frac{(-1)^{J/2}}{2^{d-2}(\pi J)^{1/4}}\left(1-\frac{2d-5}{16J}+O(J^{-2})\right)\,.
    \end{split}
    \label{eq:alpha0asym}
\end{equation}
The exact expression for $A_{S_J}$ as well as its large $J$ expansion are plotted in Figure~\ref{fig:Bafree}. Note that, as in the defect channel, the results are independent of the dimension $p$ of the defect.

The expressions in \eqref{eq:alphaF} and \eqref{eq:tensornorm} are rather similar to those for the line defect in the free $O(N)$ model \cite{Liendo:2019jpu, Bianchi:2022sbz}. This is despite the fact that the structure of the two-point function is different. In addition to the direct propagator that gives the identity contribution in the bulk OPE, the line defect has direct interaction with the line and the crosscap defect has the interaction with the image of the second operator. It just happens that in the free field kinematics, both are represented by geometric series in $z$ or $z^2$ and the factors in \eqref{eq:alphaF} and \eqref{eq:tensornorm} come from translating those to the free conformal block basis.

\subsection{Displacement and tilt operators}\label{sec:disp}
In standard defect CFTs, the defect operator spectrum contains the displacement operator with $\hat{\Delta} =p+1$, which generates a local deformation of the defect. In addition, if the defect breaks the global symmetry, there exist tilt operators with dimension $\hat{\Delta}=p$, parametrizing the defect conformal manifold. Under the defect conformal group $SO(p+1,1)$, both operators transform as scalar while under the transverse $SO(q)$, the displacement operator is a vector while the tilt operators are scalar. In addition, the displacement operator is a singlet under the global symmetry preserved by the defect. 

It is rather straightforward to see that no such operators exist for crosscap defects in the free $O(N)$ model for generic $p$. As discussed above, in free theories, the defect operators are obtained by simply restricting operators in the bulk to the fixed locus. However, it is impossible in general to construct operators that have the correct scaling dimensions and quantum numbers out of 
$\phi^{\ipl}$, $\partial_{\perp}\phi^{\imi}$, and derivatives. The only exception is $p=d-1$, namely BCFT, for which the displacement operator is proportional to $ \partial_{\perp} \phi^{\imi} \partial_{\perp} \phi^{\imi}-\partial_{\parallel}\phi^{\ipl}\partial_{\parallel}\phi^{\ipl}$ and the tilt operator to $\phi^{\ipl} \partial_{\perp} \phi^{\jmi}$.

This statement also holds for the interacting $O(N)$ model at all orders in the $\varepsilon$ expansion. We expect this to be true more generally, beyond the examples studied in this paper. Namely, we expect that the crosscap defects with $p\neq d-1$ do not have the displacement or the tilt operators. The basic reason for this expectation is that, unlike the standard defects, the fixed locus of the crosscap defects are defined by the geometrical quotient and it is impossible to deform the fixed locus locally without affecting the bulk. It would be interesting to study crosscap defects in other theories and verify this statement.
\section{Crosscap defects in the interacting \texorpdfstring{$O(N)$}{O(N)} model}
\label{sec:interacting}
Having discussed the free theory, we now turn on interactions. We consider 
the $O(N)$ symmetric version of the Wilson--Fisher (WF) 
theory~\cite{Wilson:1971dc}, deforming \eqref{eq:freeaction} to
\begin{equation}
S^{\text{WF}}=
    \int_{\bR^d}\dd^{d} x\left(\frac{1}{2}\partial_\mu \phi^i\partial^\mu \phi^i+\frac{\lambda}{4!}(\phi^i\phi^i)^2\right).
\label{eq:intaction}
\end{equation}
In dimension $d=4-\varepsilon$, the $\beta$-function is perturbative 
in $\varepsilon$ with a non-trivial fixed point at 
\begin{align}
\lambda_*=(4\pi)^2\frac{3\varepsilon}{N+8}+O(\varepsilon^2)\,.\label{eq:lambdastar}
\end{align}
We work to first order in $\varepsilon\sim\lambda$ throughout. This means that there are two contributions to correlators: free correlators evaluated in $d=4-\varepsilon$ to linear order in $\varepsilon$ and the first interacting graphs evaluated in $d=4$. The main Feynman integral we need to compute to express the one-point and two-point functions is (see Figure~\ref{fig:Feynman}a)
\begin{align}
    I(d,p|x,y)=-\lambda\int\dd^p u_\para\,\dd^{q}u_\perp\,K(x,u)K(y,u)K(u,\iota_p(u))\,,\label{IpInt}
\end{align}
where the propagators are free and given in \eqref{free-prop}. A second graph with $K(y,\iota_p(u))$ (Figure~\ref{fig:Feynman}b) is the same as $I(d,p|x,\iota_p(y))$. For general values of $d,p,x$ and $y$ we reduce this to a single integral over a hypergeometric function, see \eqref{eq:genf}. When either $y=x$ \textit{or} $d=4$ and $p\in\mathbb{Z}$, the integrand simplifies and we can find a closed form expression for the correlator. For other values, the integral can be evaluated perturbatively when expanded in one channel or another.

\subsection{One-point functions}\label{sec:1ptint}

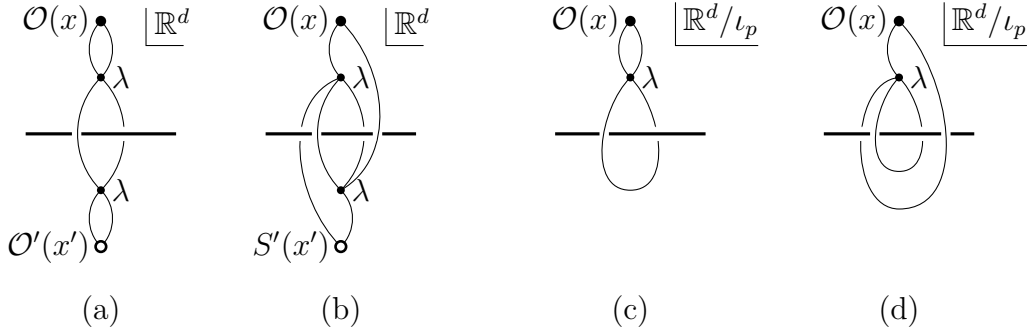
\begin{figure}[t]
\centering
\begin{tikzpicture}[scale=0.5, baseline=(vert_cent.base)]
\node[inner sep=0pt,outer sep=0pt] (vert_cent) at (0,0) {$\phantom{\cdot}$};
\fill[black] (0,3) circle (4pt);
\node[left] at (0,3) {$\cO(x)$};
\draw (0,3) to [out=-135,in=135](0,1.5)to [out=45,in=-45](0,3);
\draw (0,-1.5) to [out=-135,in=135](0,-3)to [out=45,in=-45](0,-1.5);
\draw (0,1.5) to [out=-135,in=135](0,-1.5);
\draw (0,1.5) to [out=-45,in=45](0,-1.5);
\filldraw[color=black,fill=white,line width=0.35mm] (0,-3) circle (4pt);
\node[left] at (0,-3) {$\cO'(x')$};
\draw[color=white,fill=white] (0.75,0) +(-5pt,-5pt) rectangle +(5pt,5pt);
\draw[very thick,black] (-2,0)--(-0.77,0);
\draw[very thick,black] (-.52,0)--(2,0);
\fill[black] (0,1.5) circle (3pt);
\fill[black] (0,-1.5) circle (3pt);
\node[right] at (0,1.5) {$\lambda$};
\node[right] at (0,-1.5) {$\lambda$};
\node[below] at (0,-4) {(a)};
\node at (1.9,3) {$\mathbb{R}^d$};
\draw[black] (1.2,3.35)--(1.2,2.35)--(2.2,2.35);
\end{tikzpicture}
\quad
\begin{tikzpicture}[scale=0.5, baseline=(vert_cent.base)]
\node[inner sep=0pt,outer sep=0pt] (vert_cent) at (0,0) {$\phantom{\cdot}$};
\fill[black] (0,3) circle (4pt);
\node[left] at (0,3) {$\cO(x)$};
\draw (0,3) to [out=-135,in=135](0,1.5)to [out=-160,in=135](0,-3);
\draw (0,-3)to [out=45,in=-45](0,-1.5)to [out=30,in=-45](0,3);
\draw (0,1.5) to [out=-135,in=135](0,-1.5);
\draw (0,1.5) to [out=-45,in=45](0,-1.5);
\filldraw[color=black,fill=white,line width=0.35mm] (0,-3) circle (4pt);
\node[left] at (0,-3) {$S'(x')$};
\draw[color=white,fill=white] (-1.1,0) +(-5pt,-5pt) rectangle +(5pt,5pt);
\draw[color=white,fill=white] (0.75,0) +(-5pt,-5pt) rectangle +(5pt,5pt);
\draw[very thick,black] (-2,0)--(-0.77,0);
\draw[very thick,black] (-.52,0)--(.85,0);
\draw[very thick,black] (1.1,0)--(2,0);
\fill[black] (0,1.5) circle (3pt);
\fill[black] (0,-1.5) circle (3pt);
\node[right] at (0,1.5) {$\lambda$};
\node[right] at (0,-1.5) {$\lambda$};
\node[below] at (0,-4) {(b)};
\node at (1.9,3) {$\mathbb{R}^d$};
\draw[black] (1.2,3.35)--(1.2,2.35)--(2.2,2.35);
\end{tikzpicture}
\hskip1.2cm
\begin{tikzpicture}[scale=0.5, baseline=(vert_cent.base)]
\node[inner sep=0pt,outer sep=0pt] (vert_cent) at (0,0) {$\phantom{\cdot}$};
\fill[black] (0,3) circle (4pt);
\node[left] at (0,3) {$\cO(x)$};
\draw (0,3) to [out=-135,in=135](0,1.5)to [out=45,in=-45](0,3);
\draw (0,1.5) to [out=-135,in=180](0,-1.5)to [out=0,in=-45](0,1.5);
\draw[color=white,fill=white] (0.8,0) +(-5pt,-5pt) rectangle +(5pt,5pt);
\draw[very thick,black] (-2,0)--(-0.82,0);
\draw[very thick,black] (-.57,0)--(2,0);
\fill[black] (0,1.5) circle (3pt);
\node[right] at (0,1.5) {$\lambda$};
\node[below] at (0,-4) {(c)};
\node at (2.4,3) {$\mathbb{R}^d/\iota_p$};
\draw[black] (1.2,3.5)--(1.2,2.35)--(3.4,2.35);
\end{tikzpicture}
\quad
\begin{tikzpicture}[scale=0.5, baseline=(vert_cent.base)]
\node[inner sep=0pt,outer sep=0pt] (vert_cent) at (0,0) {$\phantom{\cdot}$};
\fill[black] (0,3) circle (4pt);
\node[left] at (0,3) {$\cO(x)$};
\draw (0,3) to [out=-135,in=135](0,1.5)to [out=-175,in=180](0,-2)to [out=0,in=-45](0,3);
\draw (0,1.5) to [out=-135,in=180](0,-1)to [out=0,in=-45](0,1.5);
\draw[color=white,fill=white] (-0.9,0) +(-5pt,-5pt) rectangle +(5pt,5pt);
\draw[color=white,fill=white] (0.75,0) +(-5pt,-5pt) rectangle +(5pt,5pt);
\draw[very thick,black] (-2,0)--(-0.77,0);
\draw[very thick,black] (-.52,0)--(1.12,0);
\draw[very thick,black] (1.37,0)--(2,0);
\fill[black] (0,1.5) circle (3pt);
\node[right] at (0,1.5) {$\lambda$};
\node[below] at (0,-4) {(d)};
\node at (2.4,3) {$\mathbb{R}^d/\iota_p$};
\draw[black] (1.2,3.5)--(1.2,2.35)--(3.4,2.35);
\end{tikzpicture}
\caption{The interacting Feynman diagrams contributing to the one 
point function of $S$. (a) and (b) on the left are respectively $I(d,p|x,x)$ and $I(d,p|x,x')$ represented in the covering space $\bR^d$. 
(c) and (d) on the right are again $I(d,p|x,x)$ and $I(d,p|x,x')$ but now represented without the images in $\bR^d/\iota_p$.}
\label{fig:onept}
\end{figure}

Before studying the two-point function, let us look first at the one-point functions of the scalars $S$ and $T$, the interacting versions of the operators \eqref{eq:SandT} studied in Section~\ref{sec:freecor}.

\subsubsection{$O(N)$ singlet $S$}
\label{sec:WF-S}
As we turn on the coupling, the operator $S$ gets renormalised. in the MS prescription, the bare operator $S_\text{b}$ is related to the renormalised one by $S_{\text{b}}\equiv\phi^i\phi^i=\cN_SZ_S S$ 
(c.f.~\cite{Cuomo:2021kfm}), where
\begin{equation}
\label{NSZS}
\cN_S=\sqrt{2N}\mathcal{N}_\phi^2
\left(1-\varepsilon\frac{N+2}{2(N+8)}(1+\log\pi+\gamma_E) 
+O(\varepsilon^2) 
\right),
\qquad
Z_S=1-\frac{\lambda(N+2)}{3(4\pi)^2\varepsilon}\,,
\end{equation}
where we include the one-loop correction to $\cN_S$ in \eqref{eq:SandT} and $\gamma_E$ the Euler--Mascheroni constant. The operator's scaling dimension is now 
$\Delta_S=2-\frac{6}{N+8}\varepsilon+O(\varepsilon^2)$ (see, e.g.~\cite{Henriksson:2022rnm}).

In the computation of the one-point function, the relevant diagrams at order $\varepsilon$ are shown in Figure~\ref{fig:onept} 
and are computed in Appendix~\ref{app:1pt}. 
Evaluating the flavour factors using the interaction vertex in Figure~\ref{fig:onept}, the first graph is weighted by 
$\frac{\lambda}{6}\Npm(N+2)$ and the second one by $\frac{\lambda}{6}(\Npm^2+2N)$. Then there is an extra factor of two accounting for
\bal
(K(x,u)+K(x,u'))^2K(u,u')
&=K(x,u)^2K(u,u')+K(x,u')^2K(u,u')
\\&\quad+2K(x,u)K(x,u')K(u,u')\,,
\eal
where given that $u$ is integrated over all of $\bR^d$, the first two terms are equal. This gives the unrenormalised one-point function
\bal
\langle S_{\text{b}}(x)\rangle&=
\frac{\Npm\Gamma(d/2-1)}{(4\pi)^{d/2}|x_\perp|^{d-2}}
+\frac{\Npm(N+2)}{3}I(4-\varepsilon,p|x,x)
+\frac{\Npm^2+2N}{3}I(4,p|x,x')
\\
\label{S1pt}
&=
\frac{\Npm}{(4\pi)^2x_\perp^2}\left(1
+\frac{\varepsilon}{2}\big(\log(4\pi x_\perp^2)+\gamma_E\big)\right)
-\frac{\lambda\Npm(N+2)}{3(4\pi)^4x_\perp^2}
\left(\frac{1}{\varepsilon}+1+\log(4\pi x_\perp^2)
+\gamma_E\right)
\\&\quad
-\frac{\lambda}{6(4\pi)^4x_\perp^2}
\left(\Npm(N+2)+\frac{\Npm^2+2N}{p-1}\right)
\left(H_{-1/2}-H_{-p/2}\right)\,.
\eal
Here, the first term is the one-point function in the free theory \eqref{1pt-free-new} (before unit-normalising \eqref{eq:SandT}), and $H$ is the harmonic number.

The divergence as $\varepsilon\ra 0$ cancels between the interacting terms and the free one-point function \eqref{1pt-free-new} using the normalisation \eqref{NSZS}. The resulting one-point function for the renormalised field is then
\begin{align}
\label{eq:S-WF}
\langle S(x)\rangle
&=\frac{\Npm}{8\sqrt{2N}|x_\perp|^{\Delta_S}}
\\&\quad\nonumber
\times\bigg(2
-\frac{\varepsilon}{N+8}
\bigg(N+2-6\log4
+\left(N+2+\frac{\Npm^2+2N}{\Npm(p-1)}\right)
(H_{-1/2}-H_{-p/2})
\bigg)\bigg).
\end{align}
Notice that the naive pole at $p=1$ is cancelled by the harmonic numbers, but there is an actual pole at $p=2$ where $H_{-p/2}$ blows up. Expanding around $p=2-\delta$ for small $\delta$ yields
\begin{align}
\langle S(x)\rangle&\simeq
\frac{1}{4\sqrt{2N}|x_\perp|^{\Delta_S}}
\bigg(\frac{\varepsilon}{\delta}
\frac{(N+\Npm)(\Npm+2)}{N+8}
+\Npm-\frac{\varepsilon}{2(N+8)}\bigg(
N\Npm+2\Npm^2
\nonumber\\&\quad
+4N+2\Npm
-(N\Npm+\Npm^2+2N+8\Npm)\log4
\bigg)\bigg)
+O(\delta)\,.
\label{eq:Sdefect}
\end{align}
Unlike the $1/\varepsilon$ pole, this $1/\delta$ pole is not regulated by bulk renormalisation. Instead, this divergence needs to be cancelled by introducing a defect localised counterterm \eqref{counter} for all $\phi^{\ipl}$. The tree-level Feynman integral contribution of this term to $\vev{S(x)}$ is
\begin{equation}
N_+ D(d,p|x,x)\equiv -h_0N_+\int\dd^pu_\parallel K(x,u)^2\,.
\end{equation}
This diagram was computed in~\cite{Trepanier:2023tvb, Giombi:2023dqs, Raviv-Moshe:2023yvq,deSabbata:2024xwn} and reproduced in Appendix~\ref{sec:counter}. In our context, it evaluates to
\begin{equation}
\label{h-tree}
N_+ D(d,p|x,x)=
-\frac{h_0N_+\Gamma(d/2-1)^2}{16\pi^{d-p/2}|x_\perp|^{2d-4-p}}
\simeq-\frac{h_0N_+}{16\pi^3x_\perp^2}\,,
\end{equation}
where the last expression is at leading order in $\varepsilon$ and 
$\delta$. One can then use $h_0$ to cancel the divergence in 
\eqref{eq:Sdefect}, namely by choosing
\begin{equation}
h_0=\frac{\lambda}{\delta}\frac{(\Npm+2)}{24\pi}\,.
\label{h0}
\end{equation}
The self-interactions of the counterterm \eqref{counter} lead to a further series of divergences which were calculated to order $O(\lambda h^{10})$ for a defect in the absence of the crosscap in \cite{deSabbata:2024xwn}. This is not too hard to adapt to our case and then at any given order one would need to write the appropriate $\beta$-function for the coupling $h$ and find the fixed point. We do not pursue this here.

\subsubsection{$O(N)$ tensor $T^{ij}$}
The story is very similar for $T_\text{b}^{ij}\equiv\frac{1}{2}(\phi^i\phi^j+\phi^j\phi^i)-\frac{1}{N}\delta^{ij}\phi^k\phi^k=\cN_TZ_T T^{ij}$. In this case, the first graph in Figure~\ref{fig:onept} comes with a factor of $\delta^{ij}\frac{\lambda}{6}(\Npm\pm2)-\Tr$ and the second graph with $\pm\delta^{ij}\frac{\lambda}{6}(\Npm\pm2)-\Tr$, with the signs depending on the charge of $\phi^{\ipm}$ under $\iota_p$. Taken together, we conclude
\begin{align}
\label{eq:Tint}
\big<T_{\text{b}}^{\ip\jp}(x)\big>
&=
\delta^{\ip\jp}\frac{N_-}{N(4\pi)^2x_\perp^2}
\bigg[\left(1
+\frac{\varepsilon}{2}\big(\log(4\pi x_\perp^2)+\gamma_E\big)\right)
\nonumber\\
&\quad+
\frac{\lambda}{3(4\pi)^2}
\bigg(4\left(\frac{1}{\varepsilon}+1+\log(4\pi x^2_\perp)+\gamma_E\right)
+\left(2+\frac{\Npm}{p-1}\right)\left(H_{-1/2}-H_{-p/2}\right)\bigg)\bigg]\,,
\nonumber\\
\big<T_{\text{b}}^{\ip\jm}(x)\big>
&=\big<T_{\text{b}}^{\im\jp}(x)\big>=0\,,
\\
\big<T_{\text{b}}^{\im\jm}(x)\big>
&=
-\delta^{\im\jm}\frac{N_+}{N(4\pi)^2x_\perp^2}
\bigg[\left(1
+\frac{\varepsilon}{2}\big(\log(4\pi x_\perp^2)+\gamma_E\big)\right)
\nonumber\\&\quad
+\frac{\lambda}{3(4\pi)^2}\bigg(4\left(\frac{1}{\varepsilon}+1+\log(4\pi x^2_\perp)+\gamma_E\right)
+\left(2+\frac{\Npm}{p-1}\right)\left(H_{-1/2}-H_{-p/2}\right)\bigg)\bigg]\,.
\nonumber
\end{align}
Observe that the one-point function of $T^{\ip\jp}(x)$ and $T^{\im\jm}$ have both a $1/\varepsilon$ pole and an extra $1/\delta$ pole for $p=2-\delta$. The $1/\varepsilon$ divergence is again cancelled by the usual bulk one-loop renormalisation with
\beq
\label{NTZT}
\cN_T= 
\frac{\Gamma(d/2-1)}{2\sqrt2\,\pi^{d/2}} 
\left(1-\varepsilon\frac{1+\log\pi+\gamma_E}{N+8} 
\right),
\qquad
Z_T=1-\frac{2\lambda}{3(4\pi)^2\varepsilon}\,,
\eeq
while the $1/\delta$ divergence is cancelled by the counterterm \eqref{counter}. In fact, there is no further renormalisation freedom and it has to be cancelled by the same $h_0$ in \eqref{h0}. Indeed, this works perfectly because the residues at $p=2$ in \eqref{eq:Tint} are proportional to $\Npm+2$.

The divergence of the one-point functions at $p=2$ carry over to the two-point functions evaluated below and therefore also appear in the conformal block expansions in Section~\ref{sec:Defect} and~\ref{sec:Bulk}. These divergences are again cancelled by the counterterm \eqref{counter}, but would require the computation of higher order terms in $h_0$ in order to fix the conformal data. We do not pursue such an analysis in this paper.

\subsection{Two-point functions}\label{sec:int2pt}
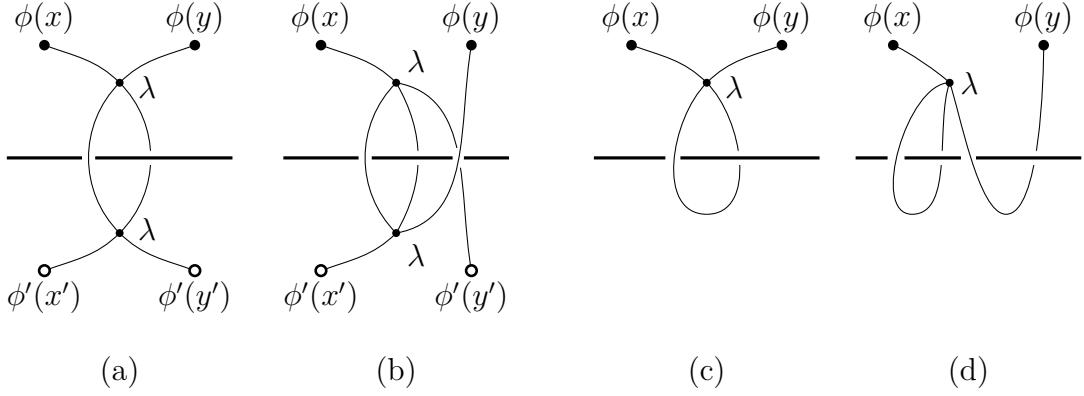
\begin{figure}[t]
\centering
\begin{tikzpicture}[scale=0.5, baseline=(vert_cent.base)]
\node[inner sep=0pt,outer sep=0pt] (vert_cent) at (0,0) {$\phantom{\cdot}$};
\draw[very thick,black] (-3,0)--(-1,0);
\fill[black] (-2,3) circle (4pt);
\node[above] at (-2,3) {$\phi(x)$};
\fill[black] (2,3) circle (4pt);
\node[above] at (2,3) {$\phi(y)$};
\draw (-2,3)to[out=-20,in=135](0,2)to[out=45,in=-160](2,3);
\draw (-2,-3)to[out=20,in=-135](0,-2)to[out=-45,in=160](2,-3);
\draw (0,2) to [out=-135,in=135](0,-2);
\draw (0,2) to [out=-45,in=45](0,-2);
\filldraw[color=black,fill=white,line width=0.35mm] (-2,-3) circle (4pt);
\node[below] at (-2,-3) {$\phi'(x')$};
\filldraw[color=black,fill=white,line width=0.35mm] (2,-3) circle (4pt);
\node[below] at (2,-3) {$\phi'(y')$};
\draw[color=white,fill=white] (0.825,0) +(-5pt,-5pt) rectangle +(5pt,5pt);
\draw[very thick,black] (-.65,0)--(3,0);
\fill[black] (0,2) circle (3pt);
\fill[black] (0,-2) circle (3pt);
\node[right] at (0.2,1.9) {$\lambda$};
\node[right] at (0.2,-1.9) {$\lambda$};
\node[below] at (0,-5) {(a)};
\end{tikzpicture}
\ \ 
\begin{tikzpicture}[scale=0.5, baseline=(vert_cent.base)]
\node[inner sep=0pt,outer sep=0pt] (vert_cent) at (0,0) {$\phantom{\cdot}$};
\draw[very thick,black] (-3,0)--(-1,0);
\fill[black] (-2,3) circle (4pt);
\node[above] at (-2,3) {$\phi(x)$};
\fill[black] (2,3) circle (4pt);
\node[above] at (2,3) {$\phi(y)$};
\draw (-2,3)to[out=-20,in=135](0,2)to[out=-10,in=100](2,-3);
\draw[color=white,fill=white] (1.675,0) +(-7pt,-7pt) rectangle +(7pt,7pt);
\draw (-2,-3)to[out=20,in=-135](0,-2)to[out=10,in=-100](2,3);
\draw (0,2) to [out=-135,in=135](0,-2);
\draw (0,2) to [out=-60,in=60](0,-2);
\filldraw[color=black,fill=white,line width=0.35mm] (-2,-3) circle (4pt);
\node[below] at (-2,-3) {$\phi'(x')$};
\filldraw[color=black,fill=white,line width=0.35mm] (2,-3) circle (4pt);
\node[below] at (2,-3) {$\phi'(y')$};
\draw[color=white,fill=white] (0.575,0) +(-5pt,-5pt) rectangle +(5pt,5pt);
\draw[very thick,black] (-.65,0)--(1.5,0);
\draw[very thick,black] (1.8,0)--(3,0);
\fill[black] (0,2) circle (3pt);
\fill[black] (0,-2) circle (3pt);
\node[above right] at (0,2) {$\lambda$};
\node[below right] at (0,-2) {$\lambda$};
\node[below] at (0,-5) {(b)};
\end{tikzpicture}
\qquad
\begin{tikzpicture}[scale=0.5, baseline=(vert_cent.base)]
\node[inner sep=0pt,outer sep=0pt] (vert_cent) at (0,0) {$\phantom{\cdot}$};
\draw[very thick,black] (-3,0)--(-1.1,0);
\fill[black] (-2,3) circle (4pt);
\node[above] at (-2,3) {$\phi(x)$};
\fill[black] (2,3) circle (4pt);
\node[above] at (2,3) {$\phi(y)$};
\draw (-2,3)to[out=-20,in=135](0,2)to[out=45,in=-160](2,3);
\draw (0,2) to [out=-135,in=180](0,-1.5);
\draw (0,2) to [out=-45,in=0](0,-1.5);
\draw[color=white,fill=white] (0.9,0) +(-5pt,-5pt) rectangle +(5pt,5pt);
\draw[very thick,black] (-.7,0)--(3,0);
\fill[black] (0,2) circle (3pt);
\node[right] at (0.2,1.9) {$\lambda$};
\node[below] at (0,-5) {(c)};
\end{tikzpicture}
\ \ 
\begin{tikzpicture}[scale=0.5, baseline=(vert_cent.base)]
\node[inner sep=0pt,outer sep=0pt] (vert_cent) at (0,0) {$\phantom{\cdot}$};
\draw[very thick,black] (-3,0)--(-2.15,0);
\fill[black] (-2,3) circle (4pt);
\node[above] at (-2,3) {$\phi(x)$};
\fill[black] (2,3) circle (4pt);
\node[above] at (2,3) {$\phi(y)$};
\draw[looseness=.5] (-2,3)to[out=-20,in=135](-.5,2)to[out=-70,in=180](1,-1.5) to [out=0,in=-90] (2,3);
\draw[looseness=.8] (-.5,2) to [out=-180,in=180](-1.5,-1.5) to [out=0,in=-110](-.5,2);
\draw[color=white,fill=white] (-.7,0) +(-5pt,-5pt) rectangle +(5pt,5pt);
\draw[color=white,fill=white] (1.8,0) +(-5pt,-5pt) rectangle +(5pt,5pt);
\draw[very thick,black] (-1.7,0)--(-.2,0);
\draw[very thick,black] (.2,0)--(3,0);
\fill[black] (-.5,2) circle (3pt);
\node[right] at (-.5,2) {$\lambda$};
\node[below] at (0,-5) {(d)};
\end{tikzpicture}
\caption{The interacting Feynman diagram can be represented (on the left) in the covering space $\bR^d$, with the two operators and their images, as well as the interacting vertex and its image. (a) and (b) show different contractions to the operator on the right from the two terms in the propagator \eqref{free-prop}. 
The right figures (c), (d) try to represent the 
same graphs in $\bR^d/\iota_p$, now with the ``tadpole'' propagator 
transversing the non-contractible cycle around the fixed locus.}
\label{fig:Feynman}
\end{figure}

Turning to the two-point functions of the fundamental fields, we write them as in the free theory \eqref{eq:free2ptform} in the form
\begin{equation}
    \langle\phi^{\ipm}(x)\phi^{\jpm}(y)\rangle=\frac{\mathcal{N}_\phi^2\,G_{\ipm\jpm}(z,\bar{z})}{|x_\perp|^{d/2-1}|y_\perp|^{d/2-1}}\,.
\end{equation}
Note that $\phi$ does not get renormalised at one-loop order and therefore $\mathcal{N}_\phi$ is as in the free theory \eqref{free-prop}. The two-point function has an expansion in $\varepsilon$
\begin{equation}
G_{\ipm\jpm}(z,\bar{z})
=\sum_{n=0}^\infty \varepsilon^n\,G^{(n)}_{\ipm\jpm}(z,\bar{z})\,,
\end{equation}
where $G^{(0)}=G^F|_{\varepsilon=0}$ and we split the linear order as
\begin{equation}
G^{(1)}_{\ipm\jpm}(z,\bar{z})=
G^{(1,F)}_{\ipm\jpm}(z,\bar{z})+
G^{(1,I)}_{\ipm\jpm}(z,\bar{z})\pm G^{(1,I)}_{\ipm\jpm}(-z,-\bar{z})\,.
\label{eq:genf}
\end{equation}
Here $G^{(1,F)}$ is the linear piece in $\varepsilon$ in the free two-point function \eqref{free2pt} and the other two terms match the two interacting diagrams in Figure~\ref{fig:Feynman} and are related to the integral \eqref{IpInt}. This separation is natural as the latter two are related by bulk--image crossing $z\to-z$.

Explicitly, $G^{(0)}$ and $G^{(1,F)}$ are given by
\bal
G^{(0)}_{\ipm\jpm}(z,\bar{z})
&=\delta_{\ipm\jpm}\left(\frac{|z|}{|1-z|^2}
\pm\frac{|z|}{|1+z|^2}\right),
\\
G^{(1,F)}_{\ipm\jpm}(z,\bar{z})
&=-\delta_{\ipm\jpm}\left(\frac{|z|}{2|1-z|^2}\log\frac{|z|}{|1-z|^2}
\pm\frac{|z|}{2|1+z|^2}\log\frac{|z|}{|1+z|^2}\right),
\eal
while an expression for $G^{(1,I)}$ in obtained in Appendix~\ref{app:2pt} 
\begin{equation}\label{eq:Glambda}
\begin{split}
G_{\ipm\jpm}^{(1,I)}(z,\bar{z})&=\delta_{\ipm\jpm}\,\frac{\Npm\pm 2}{3}\frac{|x_\perp||y_\perp|}{\varepsilon\,\mathcal{N}^2_\phi}I(4,p|x,y)
\\
&=\delta_{\ipm\jpm}\,\frac{\Npm\pm2}{N+8}\frac{1}{4(p-2)}\int_0^\infty \dd w\frac{{}_2F_1\left(1,1-p/2;2-p/2;\frac{(w+z)(w+\bar{z})}{(w+1)(w+z\bar{z})}\right)}{(w+1)(w+z\bar{z})}\,,
\end{split}
\end{equation}
where we set $d=4$ since $\lambda$ itself is of order $\varepsilon$. For generic values of $p$ we extract in Sections~\ref{sec:Defect} and~\ref{sec:Bulk} the conformal data by performing various expansions of the integrand. However, for integer $p$ the hypergeometric function simplifies and the integral can be computed exactly with the results presented below.

\subsubsection{\texorpdfstring{$\boldsymbol{p=-1}$ (XCFT)}{p=-1}}
For completeness, we also include the result for $\RP^{4-\varepsilon}$, which was previously presented in \cite{Giombi:2020xah}. In this case, there is only one real cross ratio since $\kappa_+=1$ is fixed, which implies $|z|=1$. Using
\begin{equation}
{}_2F_1\big(1,\tfrac{3}{2};\tfrac{5}{2};\alpha\big)
=3\frac{\sqrt{\alpha}\arctanh(\sqrt{\alpha})-\alpha}{\alpha^2}\,,
\end{equation}
we find the two-point function
\begin{equation}
\begin{split}
G_{\ipm\jpm}^{(1)}(z)
&=
G_{\ipm\jpm}^{(1,F)}(z,1/z)
-\delta_{\ipm\jpm}\,\frac{\Npm\pm 2}{2(N+8)}
\Bigg[\frac{z}{(1+z)^2}\log\frac{-4z}{(1-z)^2}\mp\frac{z}{(1-z)^2}\log\frac{4z}{(1+z)^2}\Bigg]\,.\label{p=-12pt}
\end{split}
\end{equation}
Indeed, this results matches those of \cite{Giombi:2020xah} once rewritten in terms of $\eta=(1-z)(1-1/z)/4$. The entire expression to order $\varepsilon$ can also be written as
\begin{align}
&G_{\ipm\jpm}(z)=-\delta_{\ipm\jpm}\left(\frac{z}{(1-z)^2}\mp\frac{z}{(1+z)^2}\right)
\left(1+
\varepsilon\log2\right)
\\&\nonumber
\pm\delta_{\ipm\jpm}\,\frac{\varepsilon}{2}\left(
\frac{\frac{\Npm\pm 2}{N+8}z}{(1-z)^2}
-\frac{z}{(1+z)^2}\right)\log\frac{4z}{(1+z)^2}
+\delta_{\ipm\jpm}\,\frac{\varepsilon}{2}
\left(\frac{z}{(1-z)^2}
-\frac{\frac{\Npm\pm 2}{N+8}z}{(1+z)^2}\right)
\log\frac{-4z}{(1-z)^2}\,.
\end{align}
 By taking $N\ra\infty$ while keeping $N_-$ fixed, this simplifies to
\bal
&-\frac{4z^2}{(1-z^2)^2}
\left(1-\frac{\varepsilon}{2}\log\frac{-4z^2}{(1-z^2)^2}\right)
\sim\left(\frac{2iz}{1-z^2}\right)^{2-\varepsilon}\,.
\eal
Similarly nice expressions exist for $N\ra\infty$ with $N_+$ fixed.

\subsubsection{\texorpdfstring{$\boldsymbol{p=0}$ (point-like defect)}{p=0}}
Using
\begin{equation}
{}_2F_1(1,1;2;\alpha)
=-\frac{\log(1-\alpha)}{\alpha}\,,
\end{equation}
the integral \eqref{eq:genf} yields
\begin{equation}
\begin{split}
G_{\ipm\jpm}^{(1)}(z,\bar{z})
&=G^{(1,F)}_{\ipm\jpm}(z,\bar{z})
-\delta_{\ipm\jpm}\,\frac{(\Npm\pm 2)}{8(N+8)}\frac{|z|}{z-\bar{z}}\\
&\quad\left(\Li_2(z)-\Li_2(\bar{z})-\log|z|\log\frac{1-\bar{z}}{1-z}
\mp\left(\Li_2(-z)-\Li_2(-\bar{z})-\log|z|\log\frac{1+\bar{z}}{1+z}\right)\right),
\end{split}
\end{equation}
where we have assumed without loss of generality that $|z|\leq 1$. This is extremely reminiscent of a four-point box diagram \cite{Usyukina:1992jd}. This is of course no accident since, as mentioned in Section~\ref{subsec:ccross}, the $p=0$ defect consists of only two points, namely the origin and infinity, and the two-point function in the presence of the defect has a similar kinematical structure as the four-point function without a defect.

One can rewrite this expression using dilog identities and also take large $N$ limits, but it does not simplify significantly for generic $z$, $\bar z$.

\subsubsection{\texorpdfstring{$\boldsymbol{p=1}$ (line defect)}{p=1}}
\label{sec:p=1}
In this case
\begin{equation}
{}_2F_1\big(1,\tfrac{1}{2};\tfrac{3}{2};\alpha\big)
=\frac{\arctanh\sqrt{\alpha}}{\sqrt{\alpha}}\,.
\end{equation}
While this expression is simple, the remaining integral \eqref{eq:genf} is not easy to compute. We note, however, that the same integral arises in the case of bulk two-point functions in the presence of a magnetic defect \cite{Gimenez-Grau:2022ebb,
Bianchi:2022sbz}. It was evaluated in \cite{Bianchi:2022sbz} using the defect channel dispersion relation and a defect inversion formula. Using their results, we obtain
\begin{equation}\label{eq:p=12pt}
G^{(1)}_{\ipm\jpm}(z,\bar{z})=G_{\ipm\jpm}^{(1,F)}(z,\bar{z})+\delta_{\ipm\jpm}\,\frac{(\Npm\pm 2)}{8(N+8)}
\Big[\partial_k\mathcal{F}(k,z,\bar{z})
\pm \partial_k\mathcal{F}(k,-z,-\bar{z})\Big]_{k=0}\,,
\end{equation} 
where 
\begin{equation}
\begin{aligned}
\mathcal{F}(k,z,\bar{z})&=\left(\frac{|1-z|^2}{(1+|z|)^2}\right)^k\frac{4|z|}{(1+|z|)^2(2k+1)}
\\&\hskip1cm
F^{112}_{101}\left(\begin{matrix}1+k:1/2;1/2+k,1;\\3/2+k:-;1+k;\end{matrix}\left(\frac{1-|z|}{1+|z|}\right)^2,\frac{|1-z|^2}{(1+|z|)^2}\right),
\end{aligned}
\end{equation}
and $F^{112}_{101}$ is a Kamp\'e de F\'eriet function. We refer to \cite{Bianchi:2022sbz} for more details on the evaluation and the Kamp\'e de F\'eriet function.

The fact that the answer for the crosscap defect is closely related to that of the magnetic line is not a coincidence. In fact, as explained in Appendix~\ref{app:comparison}, at leading order in $\varepsilon$, the Feynman integrals for crosscap defects and defects made of operator insertions are essentially identical.

\begin{figure}[t]
\centering
\begin{tikzpicture}[scale=0.5, baseline=(vert_cent.base)]
\node[inner sep=0pt,outer sep=0pt] (vert_cent) at (0,0) {$\phantom{\cdot}$};
\draw[very thick,black] (-4,0)--(4,0);
\fill[black] (-2,3) circle (4pt);
\node[above left] at (-1,3) {$\phi(x)$};
\fill[black] (2,3) circle (4pt);
\node[above right] at (2,3) {$\phi(y)$};
\draw (-2,3)--(2,-3);
\draw (2,3)--(-2,-3);
\filldraw[color=black,fill=white,line width=0.35mm] (-2,-3) circle (4pt);
\node[below left] at (-2,-3) {$\phi'(x')$};
\filldraw[color=black,fill=white,line width=0.35mm] (2,-3) circle (4pt);
\node[below right] at (2,-3) {$\phi'(y')$};
\fill[black] (0,0) circle (3pt);
\node[above] at (0,.2) {$h$};
\node[below] at (0,-5) {With image};
\end{tikzpicture}
\qquad
\begin{tikzpicture}[scale=0.5, baseline=(vert_cent.base)]
\node[inner sep=0pt,outer sep=0pt] (vert_cent) at (0,0) {$\phantom{\cdot}$};
\draw[very thick,black] (-4,0)--(4,0);
\fill[black] (-2,3) circle (4pt);
\node[above left] at (-1,3) {$\phi(x)$};
\fill[black] (2,3) circle (4pt);
\node[above right] at (2,3) {$\phi(y)$};
\draw (-2,3)--(0,0)--(2,3);
\fill[black] (0,0) circle (3pt);
\node[above] at (0,.2) {$h$};
\node[below] at (0,-5) {Without image};
\end{tikzpicture}
\caption{The counterterm \eqref{counter} for $p=2$ 
gives the additional Feynman diagrams represented on the left 
in the covering space and on the right in the quotient.
}
\label{fig:counterterm}
\end{figure}
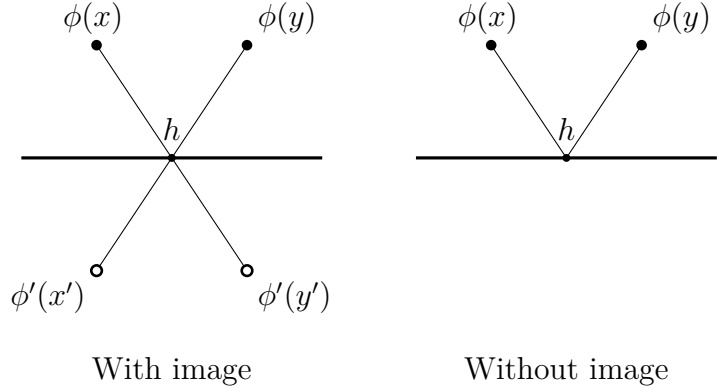

\subsubsection{\texorpdfstring{$\boldsymbol{p=2}$ (surface defect)}{p=2}}

As in the case of the one-point functions \eqref{eq:Sdefect} and \eqref{eq:Tint}, the prefactor to the integral in \eqref{eq:genf} has a pole at $p=2$, so we regularise all the expressions with $p=2-\delta$, where, including the first subleading correction,
\begin{equation}
{}_2F_1\big(1,\tfrac{\delta}{2};1-\tfrac{\delta}{2};\alpha\big)
=1-\delta\,\frac{\log(1-\alpha)}{2}\,.
\end{equation}
The divergences should be cancelled by the same defect-localised counterterm that regularises the divergent one-point functions of $S$ and $T$ in Section~\ref{sec:1ptint}, giving the extra Feynman diagram in Figure~\ref{fig:counterterm}. The computation of this diagram is presented in Appendix~\ref{sec:counter}, where it is found to contribute
\begin{equation}
    D(4,2|x,y)=\frac{4h_0}{(4\pi)^3|x_\perp||y_\perp|}\frac{|z|\log|z|}{1-|z|^2}\,,
\end{equation}
and we recall that $h_0$ is given by \eqref{h0}.

Combining these contributions, we find a finite two-point function
\begin{equation}
\begin{split}
G^{(1)}_{\ipm\jpm}(z,\bar{z})
=G^{(1,F)}_{\ipm\jpm}(z,\bar{z})
&-\delta_{\ipm\jpm}\,\frac{(\Npm\pm 2)}{4(N+8)}\frac{|z|}{1-|z|^2}
\\
&\quad\times\Big[\Li_2(1-|z|^2)-\log|z|\big(\log|1-z|\pm\log|1+z|\big)\Big]\,.
\end{split}
\end{equation}
As already noted, strictly speaking, the result above is incomplete since there are further divergences at higher orders in $h_0$ coming from the self-interaction of the counterterm, and one has to compute and renormalise them in order to obtain a full result. We leave this for the future.
\subsubsection{\texorpdfstring{$\boldsymbol{p=3}$ (BCFT)}{p=3}}
Here
\begin{equation}
{}_2F_1\big(1,-\tfrac{1}{2};\tfrac{1}{2};\alpha\big)
=1-\sqrt{\alpha}\arctanh{\sqrt{\alpha}}\,,
\end{equation}
and we once again have restricted kinematics, this time implying that $z=\bar{z}\in\mathbb{R}$. For $x$ and $y$ on the same side of the boundary, $z>0$ and the two-point function becomes
\begin{align}
    \begin{split}G^{(1)}(z)&=G_{\ipm\jpm}^{(1,F)}(z,z)
-\delta_{\ipm\jpm}\,\frac{(\Npm\pm 2)}{2(N+8)}
    \Bigg[\frac{z}{(1+z)^2}\log\frac{4z}{(1-z)^2}
\mp \frac{z}{(1-z)^2}\log\frac{4z}{(1+z)^2}\Bigg]\,.
    \end{split}
\end{align}

The expression has striking similarity to the $p=-1$ case \eqref{p=-12pt}. Again, the entire expression to order $\varepsilon$ can also be written rather compactly as
\begin{align}
&G_{\ipm\jpm}(z)=\delta_{\ipm\jpm}\left(\frac{z}{(1-z)^2}\pm\frac{z}{(1+z)^2}\right)
\left(1+
\varepsilon\log2\right)
\\&\nonumber
\pm\delta_{\ipm\jpm}\,\frac{\varepsilon}{2}\left(
\frac{\frac{\Npm\pm 2}{N+8}z}{(1-z)^2}
-\frac{z}{(1+z)^2}\right)\log\frac{4z}{(1+z)^2}
-\delta_{\ipm\jpm}\,\frac{\varepsilon}{2}
\left(\frac{z}{(1-z)^2}
+\frac{\frac{\Npm\pm 2}{N+8}z}{(1+z)^2}\right)
\log\frac{4z}{(1-z)^2}\,.
\end{align}

\subsection{Defect channel expansion}\label{sec:Defect}
We proceed now to expand the bulk two-point function in the defect channel. We do this for generic $p$ using the integral representation of $G^{(1,I)}$ in \eqref{eq:Glambda}, by expanding the integrand around $r=0$. Recall the block expansion in the absence of bulk one-point functions for $\phi^{\ipm}$ \eqref{eq:defectchannel}
\begin{equation}
G^{\text{WF}}_{\ipm\jpm}(r,t)=\delta_{\ipm\jpm}\sum_{\hat{s}}\Big(B_{\phi^{\ipm},\hat{\mathcal{O}}^{\ipm}_{\hat{s}}}\Big)^2
g_{\text{defect}}\big(\hat{\Delta}_{\hat{\mathcal{O}}^{\ipm}_{\hat{s}}},\hat{s}\big|r,t\big)\,,
\label{eq:GBulkExpWF}
\end{equation}
where the sum is over the appropriate defect operators. They are in fact in one-to-one correspondence with the free-theory operators in \eqref{eq:defectprims}, since the spectrum is the same up to changes of order $\varepsilon$. These changes are twofold: corrections to the conformal dimension of $\hat{\mathcal{O}}_{\hat{s}}^{\ipm}$ and corrections to the bulk--defect two-point function.

Concretely, starting with the free theory expressions at order $\varepsilon^0$ \eqref{eq:freespectrum}
\begin{equation}
    \Big(B^{2}_{\phi^{\ipm}\hat{\mathcal{O}}^{\ipm}_{\hat{s}}}\Big)^{(0)}=(1\pm(-1)^{\hat s})2^{\hat{s}}\comma \hat{\Delta}^{(0)}_{\hat{\mathcal{O}}^{\ipm}_{\hat{s}}}=\hat{s}+1\,,
\end{equation}
and denoting by $\Big(B^{2}_{\phi^{\ipm}\hat{\mathcal{O}}^{\ipm}_{\hat{s}}}\Big)^{(1)}$ and $\hat{\Delta}^{(1)}_{\hat{\mathcal{O}}^{\ipm}_{\hat{s}}}$ the corrections to this data to first order in $\varepsilon$, the first order term in \eqref{eq:GBulkExpWF} is
\begin{equation}
G^{(1)}_{\ipm\jpm}=\delta_{\ipm\jpm}\sum_{\hat{s}}2^{\hat{s}+1}\hat{\Delta}^{(1)}_{\hat{\mathcal{O}}^{\ipm}_{\hat{s}}}\partial_{\hat{\Delta}}\big[g_{\text{defect}}(\hat{\Delta},\hat{s}|r,t)\big]_{\hat{\Delta}=\hat{s}+1}+\Big(B^{2}_{\phi^{\ipm},\hat{\mathcal{O}}^{\ipm}_{\hat{s}}}\Big)^{(1)}g_{\text{defect}}(\hat{s}+1,\hat{s}|r,t)\,.\label{Gdifexp}
\end{equation}
These two contributions can be disentangled easily since the defect block \eqref{eq:gdefect}
\begin{equation}
    g_{\text{defect}}(\hat{\Delta},\hat{s}|
        r,t)=r^{\hat{\Delta}}{}_2F_1\left(\hat{\Delta},\frac{p}{2};\hat{\Delta}+1-\frac{p}{2};r^2\right) (2t)^{-\hat{s}}{}_2F_1\left(-\hat{s},\frac{q}{2}-1;2-\frac{q}{2}-\hat{s};t^2\right)
\end{equation}
is analytic around $r=0$, while its derivative $ \partial_{\hat{\Delta}}$ contains a non-analytic term proportional to $\log r$. Thus, any non-analytic parts of the left hand side of \eqref{Gdifexp} must come from corrections to the conformal dimensions.

In order to extract the conformal data from \eqref{eq:genf}, we must first calculate $G^{(1,I)}(z,\bar{z})$ as given in \eqref{eq:Glambda}. In the absence of closed-form expressions for the integrals, we perform a series expansion and permute the integral and the sum. For the defect channel, the relevant expansion is the series representation of the hypergeometric function ${}_2F_1(a,b;c;\zeta)$ around $\zeta=0$
\begin{equation}
    {}_2F_1(a,b;c;\zeta)=\sum_{n=0}^\infty\frac{\Gamma(a+n)\Gamma(b+n)\Gamma(c)}{\Gamma(a)\Gamma(b)\Gamma(c+n)\Gamma(1+n)}\zeta^n\label{eq:2F1}\,.
\end{equation}
This allows us to write
\begin{equation}
     G_{\ipm\jpm}^{(1,I)}(r,t)=\delta_{\ipm\jpm}\frac{\Npm\pm2}{4(N+8)}\sum_{n=0}^\infty\frac{r}{p-2(n+1)}\int_0^\infty\dd w\frac{(w^2+r^2+w(rt+r/t))^{n}}{(w^2+r^2+w(1+r^2))^{n+1}}
\end{equation}
We then integrate this for low values of $n$, guess the general result and sum over $n$. By partial fraction decomposition, for each $n$ there is a term in the integrand that looks like $1/(w+r^2+w(1+r^2))$, which integrates to $2\log\,r/(r^2+1)$ and is the source of the non-analytic behaviour around $r=0$ mentioned above. For $G^{(1,I)}(-z,-\bar{z})$ we get the same expansion, but with the change $t\ra -t$.

We are then ready to read off the defect conformal dimensions and bulk--defect two-point functions. We find contributions from operators $\hat{\mathcal{O}}^{\ipm}_{\hat{s}}$, which are the counterparts of the operators \eqref{eq:defectprims} of the free theory, which we recall only exist if $(-1)^{\hat{s}}=\pm1$. Focusing first on the piece of $G^{(1)}$ that is not analytic around $r=0$, the calculation yields
\begin{equation}
    \hat{\Delta}_{\hat{\mathcal{O}}^{\ipm}_{\hat{s}}}=\hat{s}+1-\frac{\varepsilon}{2}\left[1-\frac{\Npm\pm 2}{N+8}\frac{1}{2\hat{s}+(2-p)}\right]+O(\varepsilon^2)\label{eq:defDelta}\,.
\end{equation}
We can then similarly evaluate the bulk--defect two-point functions by analysing the analytic part of $G^{(1)}$. They are
\begin{equation}
    \Big(B_{\phi^{\ipm},\hat{\mathcal{O}}^{\jpm}_{\hat{s}}}\Big)^2=\delta_{\ipm\jpm}\,(1\pm(-1)^{\hat s}){2^{\hat{s}}}\left(1-\frac{\varepsilon}{2}\left[H_{\hat{s}}-\frac{\Npm\pm 2}{N+8}\frac{H_{\hat{s}}-H_{\hat{s}+(2-p)/2}}{2\hat{s}+(2-p)}\right]+O(\varepsilon^2)\right).
\label{eq:defB}
\end{equation}
The first term in the square brackets of \eqref{eq:defDelta} and \eqref{eq:defB} comes from $G^{(1,F)}$, while the second term comes from $G^{(1,I)}(r,t)\pm G^{(1,I)}(r,-t)$, the latter of which vanishes for $p=2$ when $\hat{s}\neq 0$.

\begin{figure}[t]
\centering
    \includegraphics[width=\capwidth]{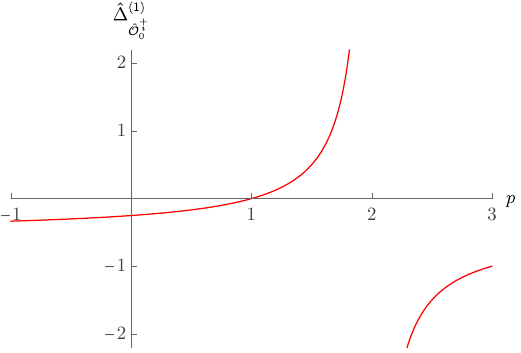}
\caption{The first order correction in $\varepsilon$ to the conformal dimension \eqref{eq:defDelta} of the transverse spin-0 operator $\hat{\mathcal{O}}^{\overset{+}{\imath}}_{0}$, as defined in \eqref{Gdifexp}. This is a function of $p$ and shown for $N\ra\infty$ with $N_-$ fixed. We clearly observe the pole in the anomalous dimension at $p=2$.}
     \label{fig:D0}
\end{figure}

\begin{figure}[t]
\centering
    \includegraphics[width=\halfcapwidth]{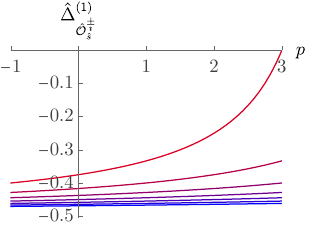}\qquad\includegraphics[width=\halfcapwidth]{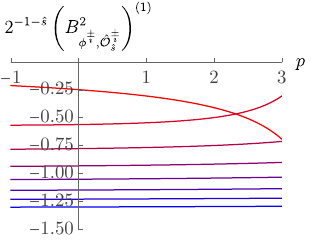} 
    \vspace{0.5 em}
\caption{The first order correction in $\varepsilon$ to the conformal dimensions $\Delta$ \eqref{eq:defDelta} and bulk--defect couplings $B^2$ \eqref{eq:defB}, as defined in \eqref{Gdifexp}. Those are evaluated for $N\to\infty$ with $N_-$ fixed. The left plot is made for $\hat{s}=\{1,\dots,7\}$ (the case of $\hat{s}=0$ is in Figure~\ref{fig:D0}), while the right plot also includes $\hat s=0$. Higher transverse spin corresponds to more blue colours and at $p=-1$ are further down the vertical axis. Note that although the anomalous dimension of $\hat{\mathcal{O}}_0^{\overset{+}{\imath}}$ blows up at $p=2$, its bulk--defect coupling remains finite.}\label{fig:Drest}
\end{figure}

For the operator $\hat{\mathcal{O}}^{\ipm}_{0}$, the $O(\varepsilon)$ correction to the conformal dimension is plotted in Figure~\ref{fig:D0}, and it blows up at $p=2$. This is a reflection of the need for the extra counterterm \eqref{counter}, which indeed cancels the divergence. 
In contrast, when the transverse spin $\hat{s}>0$ the $O(\varepsilon)$ correction to the conformal dimension is finite and depicted in Figure~\ref{fig:Drest}, where the bulk--defect couplings are also plotted. Those remain finite for all operators at any $p$.

Upon taking the limit $N=N_{\pm}\to \infty$, our result \eqref{eq:defB} for $p=1$ coincides with the result for the magnetic line defect \cite{Gimenez-Grau:2022ebb,
Nishioka:2022qmj,Bianchi:2023gkk} in the WF $O(N)$ model up to the overall factor $(1\pm (-1)^{\hat{s}})$ (see e.g.~(138) of \cite{Gimenez-Grau:2022ebb}), generalizing the agreement for the free $O(N)$ model observed in section \ref{sec:freedefect}. This is not surprising since, as mentioned above and discussed further in Appendix \ref{app:comparison}, the one-loop Feynman integrals that appear in crosscap defects and magnetic defects are identical. This also implies that our expansion \eqref{eq:defB} for $p=3$ should apply\footnote{In principle, we can also consider the surface defect ($p=2$), but as mentioned already, one has to renormalise the divergence properly in order to extract the finite OPE data.} also to the boundary in the WF $O(N)$ model with cubic interactions \cite{Harribey:2023xyv,Harribey:2024gjn}, for which the defect CFT data have not been computed explicitly.

\subsection{Bulk channel expansion}\label{sec:Bulk}
We now expand \eqref{eq:genf} in the bulk channel, using \eqref{eq:bulkchannel}
\begin{equation}
    G^{\text{WF}}_{\ipm\jpm}(z,\bar{z})=\delta_{\ipm\jpm}\frac{|1-z|^2}{|z|}\sum_{\tau=2,4}\sum_{J}C_{\phi^{\ipm}\phi^{\ipm}{\mathcal{O}_{J,\tau}}}A_{\mathcal{O}_{J,\tau}}g_{\text{bulk}}(0;\tau,J|z,\bar{z})\,.
\end{equation}
An expansion for $g_{\text{bulk}}$ is in \eqref{eq:gbulkexp}, and we use that only twist $2$ and $4$ operators have non-vanishing three-point functions with two $\phi^{\ipm}$'s to first order in $\varepsilon$. For the twist $2$ operators, we get contributions from the correction to the bulk three-point functions (which are known \cite{Henriksson:2022rnm} and summarised in Appendix~\ref{ap:bulk}), 
the bulk one-point functions and the dimensions of the operators $\mathcal{O}_{J,\tau}$. For the twist $4$ operators, the bulk three-point functions are themselves of order $\varepsilon$. Thus, from the expansion, we can in principle extract the information on the free bulk one-point functions. However, for $\tau>2$, there are multiple classically degenerate operators and the conformal block expansion only provides the average over their OPE data. Because of this, we focus below on the twist-2 operators.

To separate the contribution from anomalous dimensions and the contribution from $1$-point functions, we use the same idea as in \eqref{Gdifexp}, namely that only corrections to the conformal dimensions can be non-analytic. We are thus again tasked with expanding $G^{(1,I)}(z,\bar{z})$ and $G^{(1,I)}(-z,-\bar{z})$, this time around $(z,\bar{z})=(1,1)$. For $G^{(1,I)}(-z,-\bar{z})$, this is straightforward and again utilises the standard series representation of the hypergeometric function \eqref{eq:2F1}, in which case there are no non-analytic components. For $G^{(1,I)}(z,\bar{z})$, the argument $\frac{(w+z)(w+\bar{z})}{(w+1)(w+z\bar{z})}$ of the hypergeometric function in \eqref{eq:Glambda} goes to one in the limit $z\ra 1$ and a different expansion is necessary. Using

\begin{equation}
    {}_2F_1(a,b;a+b;\zeta)=\sum_{n=0}^\infty\frac{\Gamma(a+n)\Gamma(b+n)\Gamma(a+b)}{\Gamma(a)^2\Gamma(b)^2\Gamma(1+n)^2}(1-\zeta)^n\big[2H_n-H_{a-1+n}-H_{b-1+n}-\log(1-\zeta)\big]\label{eq:2f1alt}
\end{equation}
allows to write $G^{(1,I)}(z,\bar{z})$ as
\begin{equation}
\begin{split}
	G^{(1,I)}_{\ipm\jpm}(z,\bar{z})=\delta_{\ipm\jpm}\,\frac{(\Npm\pm 2)\varepsilon}{8(N+8)}\sum_{n=0}^\infty \frac{\Gamma(n+1-p/2)|z|}{\Gamma(n+1)\Gamma(1-p/2)}\int_0^\infty\frac{(w|1-z|^2)^{n}}{((w+1)(w+z\bar{z}))^{n+1}}
    \\
	\left[\log\frac{w|1-z|^2}{(w+1)(w+z\bar{z})}+H_{n-p/2}-H_{n}\right]\dd w\,.
    \end{split}
\end{equation}
We see explicitly the emergence of a $\log|1-z|^2$, from where we can read off corrections to the bulk conformal dimensions. We find that $G^{(1,I)}(z,\bar{z})$ only corrects the conformal data of the $SO(d+1,1)$ scalar operators $S$ and $T$, an observation that was already made at $p=1$ in \cite{Gimenez-Grau:2022ebb}.

We can then evaluate the XDCFT data. For the conformal dimensions, we find
\begin{equation}
\begin{aligned}
\Delta_{S}&=2-\frac{6\varepsilon}{N+8}\comma 
\Delta_{T}=2-\frac{(N+6)\varepsilon}{N+8}\,, 
\\
\Delta_{S_J}&=\Delta_{T_J}=J+2-\varepsilon\,,\qquad
J>0\,.
\end{aligned}
\end{equation}
Thus, at this order, the dimensions of the operators with non-zero spin ($S_{J}$ and $T_J$) are their engineering dimensions while the operators without $SO(d+1,1)$ spin ($S$ and $T$) do receive quantum corrections. These match the known results in the literature (see, e.g.~\cite{Henriksson:2022rnm}) and Section~\ref{sec:1ptint}. Similarly, as derived in \cite{Dey:2016mcs}, the OPE coefficients receive quantum corrections at this order only for the operators without $SO(d+1,1)$ spin (see the relevant OPE coefficients summarised in Appendix \ref{ap:bulk}).
For these operators, we find the one-point functions to this order
\begin{align}
&A_S=\frac{N_{+-}}{4\sqrt{2N}}
\left[1+\varepsilon\left(\log2+\frac{N+2}{2(N+8)}(H_{-p/2}-1)+\frac{\Npm^2+2N}{2\Npm(N+8)}\frac{H_{-p/2}-H_{-1/2}}{p-1}\right)
\right],
\nonumber\\
&A_T=\frac{1}{2\sqrt{2}N}
\left[1+\varepsilon\left(\log2+\frac{1}{N+8}(H_{-p/2}-1)+\frac{\Npm}{2(N+8)}\frac{H_{-p/2}-H_{-1/2}}{p-1}\right)
\right],
\label{eq:A0}
\end{align}
\begin{figure}[t]
\centering
 \label{fig:BA}
\includegraphics[width=\halfcapwidth]{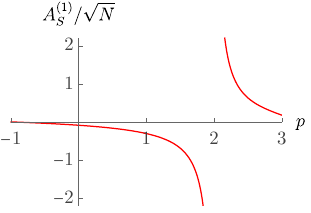}\qquad\includegraphics[width=\halfcapwidth]{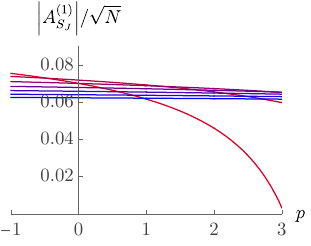}
\caption{The first order correction in $\varepsilon$ to the one-point function coefficient of the scalar operator $S$ \eqref{eq:A0} is plotted on the left, while the corresponding correction for the operator $S_J$ \eqref{eq:ASTJ} with $J=\{2,4,\dots,14\}$ is on the right. Both are plotted as a function of $p$ in the limit $N\ra\infty$ with $N_-$ fixed. The more blue the colour, the larger the spin $J$. We again see a divergence at $p=2$, in this case in the one-point function of $S$ (and for $T$ as well), which should be cancelled by the counterterm \eqref{counter}.}
\end{figure}
where $A_T$ was defined by \eqref{eq:tensornorm}. The result for $A_S$ agrees with the direct computation of the one-point function given in \eqref{eq:S-WF}, written slightly differently (and note that $H_{-1/2}=-\log 4$).

To present the one-point functions of spinning operators, it is useful to again expand the result in a series around $\varepsilon=0$
 \begin{equation}
     A_{S_J}=\sum_{n=0}^\infty \varepsilon^nA^{(0)}_{S_J}\comma A_{T_J}=\sum_{n=0}^\infty \varepsilon^nA^{(0)}_{T_J}\,,
 \end{equation}
where we find that the first two orders are given by
\begin{equation}
    A^{(0)}_{S_J}=\frac{(-1)^{J/2}\Npm}{4\pi^{1/4}\sqrt{2N}}\frac{\Gamma\big(\frac{J+1}{2}\big)}{\Gamma\big(\frac{J+2}{2}\big)}\sqrt{\frac{\Gamma(J+1)}{\Gamma\big(J+\frac{1}{2}\big)}}\comma    A^{(0)}_{T_J}=\frac{(-1)^{J/2}}{2\pi^{1/4}\sqrt{2}N}\frac{\Gamma\big(\frac{J+1}{2}\big)}{\Gamma\big(\frac{J+2}{2}\big)}\sqrt{\frac{\Gamma(J+1)}{\Gamma\big(J+\frac{1}{2}\big)}}\,,
\end{equation}
and
\begin{equation}
\begin{split}
A_{S_{J}}^{(1)}=-\frac{A^{(0)}_{S_J}}{4}\Bigg[&2H_{-1/2}+H_J+H_{(J-1)/2}-H_{J-1/2}-H_{J/2}
    \\
    &-\frac{\Npm^2+2N}{\Npm(N+8)}\frac{\Gamma\big(\frac{p+J}{2}\big)
\Gamma \big(\frac{p-J-1}{2}\big)}
{\Gamma \big(\frac{p-J}{2}\big)
\Gamma \big(\frac{p+J+1}{2}\big)}
\left(H_{-p/2}-\frac{\xi_J(p-1)}{2}\right)\Bigg]\,,
\\
A_{T_{J}}^{(1)}=-\frac{A^{(0)}_{T_J}}{4}\Bigg[&2H_{-1/2}+H_J+H_{(J-1)/2}-H_{J-1/2}-H_{J/2}
    \\
    &-\frac{\Npm}{N+8}\frac{\Gamma\big(\frac{p+J}{2}\big)
\Gamma \big(\frac{p-J-1}{2}\big)}
{\Gamma \big(\frac{p-J}{2}\big)
\Gamma \big(\frac{p+J+1}{2}\big)}
\left(H_{-p/2}-\frac{\xi_J(p-1)}{2}\right)\Bigg]\label{eq:ASTJ}\,.
\end{split}
\end{equation}
Here, $\xi_J(x)$ is a ratio of two even polynomials of degree $J$. The first few, which we obtain explicitly by matching, are
\begin{equation}\label{eq:xi}
\begin{split}
&\xi_2(x)=\frac{3x^2}{x^2-1}-4\log2\comma \xi_4(x)=\frac{5x^2(5x^2-38)}{6(x^2-1)(x^2-9)}-4\log2\\
&\xi_6(x)=\frac{7x^2(7x^4-215x^2+1198)}{10(x^2-1)(x^2-9)(x^2-25)}-4\log2
\end{split}
\end{equation}
In all these examples $\xi_J(x)$ has poles at the odd integers between $-J$ and $J$, which in particular cancels the divergences of the $\Gamma$-functions at $p\in\{-1,1,3\}$. The numerator is then fixed by specifying the value at $J+1$ additional points. In particular, a convenient choice of such additional inputs is to use the values at even integers between $-J$ and $J$, where the expressions above become independent of $J$ and equal to\footnote{We are grateful to Johan Henriksson for pointing this out.}
\begin{equation}
    \xi_J(2n)=H_{-\frac{1+2n}{2}}+H_{-\frac{1-2n}{2}}\,,
    \qquad 
    -J\leq 2n\leq J\,.
\label{eq:BFKL}
\end{equation}
Curiously, the right-hand side is the eigenvalue of the BFKL kernel \cite{Kuraev:1977fs,Balitsky:1978ic}.
We checked these relations to far larger $J$ than reported in \eqref{eq:xi} and conjecture that they hold in general. Note that the function on the right-hand side of \eqref{eq:BFKL}---when extended to $x\neq 2n$---also has poles at odd integers, but not only those between $-J$ and $J$, and even at the shared poles, the residues are different. 

\begin{figure}[t]
\centering
  \includegraphics[width=\capwidth]{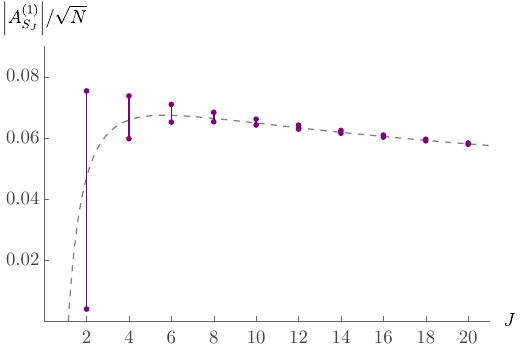}
\caption{The large $J$ asymptotics \eqref{alphaintasymp} are pictured by the dashed grey line in the limit $N\ra\infty$ with $N_-$ fixed. The purple lines are the range of $\left|A^{(1)}_{S_J}\right|$ \eqref{eq:ASTJ} as $p$ varies from $p=-1$ (the top points) to $p=3$ (the bottom points).}
    \label{fig:BaAsymp}
\end{figure}

This mismatch goes away for large $J$, and it is easy to see that $\xi_J(x)$ converges to the right-hand side uniformly in any compact domain excluding the poles, so the right-hand side provides a good approximation of $\xi_J(x)$ in this limit.
This allows us to find the large $J$ asymptotics of the one-point function
\begin{equation}
\begin{split}
&A_{S_{J}}\underset{J\ra\infty}{\sim}\frac{(-1)^J\Npm}{4(\pi J)^{1/4}\sqrt{N}}\left[1-\frac{3}{16J}+\varepsilon\left(\log2+\frac{2-\log8}{16J}-\frac{(\Npm^2+2N)\pi}{4\Npm(N+8)J}\right)+O(\varepsilon^2,1/J^2)\right]\,,
\\
&A_{T_{J}}\underset{J\ra\infty}{\sim}\frac{(-1)^J}{2(\pi J)^{1/4}\sqrt{N}}\left[1-\frac{3}{16J}+\varepsilon\left(\log2+\frac{2-\log8}{16J}-\frac{\Npm\pi}{4(N+8)J}\right)+O(\varepsilon^2,1/J^2)\right]\,.
\end{split}
\label{alphaintasymp}
\end{equation}
A plot of this is given in Figure~\ref{fig:BaAsymp}.

\section{Conclusion}\label{sec:conclusion}
In this work, we initiated the study of crosscap defects (XDCFTs) in conformal field theories in general dimensions. XDCFTs combine structural features of defect CFTs, CFTs at finite temperature and CFTs on $\mathbb{RP}^{d}$: for instance, two-point functions admit three distinct conformal block expansions---bulk, image, and defect channels. The image channel is particularly notable, as it also appears in CFTs on $\mathbb{RP}^{d}$ and CFT at finite temperature, both realizable as quotients of flat space as well. 

We analysed XDCFTs in the $O(N)$ model at the Gaussian and Wilson--Fisher fixed points, computing their CFT data as analytic functions of the fixed-locus dimension $p$. This offers a continuous interpolation between physically distinct setups. 

For $q=1$, XDCFT reduces to BCFT and still shares some properties with BCFT for higher codimension. As an illustration, in the free $O(N)$ model in Section~\ref{sec:free}, an extra deformation localised at the fixed locus \eqref{counter} induces an RG flow from $\iota_p$-even to $\iota_p$-odd boundary conditions, as is familiar from free BCFT flows. In the interacting case, such a counterterm is required at $p=2$ as discussed in Section~\ref{sec:WF-S}. It would be interesting to examine the RG structure in this case in more detail. In particular, it would be natural to explore whether analogues of the ordinary, extraordinary, special and log transitions exist, and whether one can approach the 3d $O(N)$ model with a boundary, by starting from an XDCFT in $d=4-\varepsilon$ at fixed $p=2$ and extrapolating in $d$.

In many respects, XDCFTs closely resemble DCFTs, particularly at the level of kinematics, presented in Section~\ref{sec:crosscap}. Furthermore, we find that the bulk two-point function for the $p=1$ XDCFT \eqref{eq:p=12pt} matches that of the line DCFT \cite{Gimenez-Grau:2022ebb,
Bianchi:2022sbz} in the Wilson--Fisher $O(N)$ model at order $\varepsilon$. A diagrammatic explanation of this similarity is presented in Appendix~\ref{app:comparison}, and likely extends to other values of $p$, although it would presumably break down at higher orders in $\varepsilon$.

Despite these similarities, XDCFTs exhibit important differences from standard DCFTs. Most notably, XDCFTs do not admit displacement and tilt operators. While such operators are universal in DCFTs (and present in BCFTs, corresponding to $p=d-1$), they appear to be absent for other values of $p$, possibly due to the rigidity of the crosscap construction. In standard DCFTs, tilt operators are marginal operators that generate defect conformal manifolds. Hence, XDCFTs that break a global internal symmetry provide interesting examples of conformal manifolds without exactly marginal operators. Note that this does not contradict the recent analysis in \cite{Komatsu:2025cai}, which under certain assumptions shows that the conformal manifolds need to be generated by exactly marginal operators, since the defect spectrum of XDCFT does not contain a stress tensor, whose existence is one of the assumptions of their derivation.

Another notable distinction is the global symmetry preserved by the defects. We constructed defects in the free and critical $O(N)$ models that preserve the full $O(N)$ symmetry for all values of $p$. This contrasts with DCFTs, where no such $O(N)$-invariant line defect ($p=1$) has been found. Known line defects break $O(N)$ to smaller subgroups, such as $O(N-1)$ \cite{Gimenez-Grau:2022ebb,Bianchi:2022sbz,Cuomo:2021kfm,Nishioka:2022qmj,Bianchi:2023gkk} for the case of the magnetic line, or to even smaller subgroups \cite{Pannell:2023pwz}. It may be interesting to try to show that a stable%
\footnote{In $\cN=4$ SYM the non-BPS Wilson loop is a DCFT with $O(6)$ symmetry \cite{Alday:2007he, Polchinski:2011im, Beccaria:2017rbe}, as the bulk, but it has relevant operators.}
$p=1$ DCFT with $O(N)$ symmetry must have a singlet field of dimension close to one. If this or a similar statement is correct, it must also have the exception for XDCFTs.

Our results open several directions for future research. First, it would be interesting to study the crosscap defects of the $O(N)$ model using large $N$ techniques. Given the relationship between crosscap defects and theories on $\bH_{p+1} \times \mathbb{RP}^{q-1}$, recent works on large $N$ theories in AdS \cite{Carmi:2018qzm,Giombi:2020rmc,Dujava:2025php,Copetti:2023sya} may provide useful tools for that purpose. Second, many analytic tools developed for CFTs and DCFTs, including large-spin expansions \cite{Fitzpatrick:2012yx,Komargodski:2012ek,Lemos:2017vnx}, inversion formulas \cite{Caron-Huot:2017vep,Liendo:2019jpu}, dispersion relations and associated sum rules \cite{Carmi:2019cub,Bianchi:2022ppi,Barrat:2022psm}, should extend naturally to XDCFTs. In particular, it would be valuable to develop the harmonic analysis \cite{Isachenkov:2016gim,Schomerus:2016epl,Karateev:2018oml,Buric:2021ywo} associated with crosscap defects, which may provide a simple setting for harmonic analysis on quotient spaces. Such techniques could subsequently be applied to other quotient geometries, such as CFT at finite temperature. Another major step is the analytic and numerical bootstrap study of XDCFT correlators.

It would also be interesting to study CFTs on more general quotient spaces, such as lens spaces. Existing classifications of conformal geometries \cite{Alekseevskii1978} suggest many possible generalisations that could support new classes of crosscap-like defects. In two dimensions, more general crosscaps, which involve non-$\mathbb{Z}_2$ symmetry actions, are studied in \cite{Huiszoon:1999xq,Fuchs:2000cm,Brunner:2002em,Harada:2025uhh}. It would be worth exploring if the construction can be generalised to crosscap defects in higher dimensions.

Finally, developing a holographic description of XDCFTs would be highly valuable. In forthcoming work \cite{toappear}, we study half-BPS examples in $\mathcal{N}=4$ SYM. A complementary ``bottom-up'' holographic treatment, analogous to existing approaches for BCFTs \cite{Takayanagi:2011zk,Fujita:2011fp}, DCFTs \cite{Jensen:2013lxa,Kobayashi:2018lil,Nakayama:2025hqo}, and CFTs on $\mathbb{RP}^{d}$ \cite{Wei:2024zez}, would help clarify the holographic interpretation of crosscap defects and their dynamics.

\section*{Acknowledgements}
We are grateful to J.~Caetano, J.~Henriksson, Z.~Kong, P.~Kravchuk, C.~Kristjansen, M.~Mei\-neri, A.~Stergiou, 
G.~Watts and M.~Wilhelm for useful discussions.
ND would like to acknowledge the hospitality of 
CERN, DESY, EPFL, ICTP-SAIFR, Perimeter 
Institute, the Simons Center for 
Geometry and Physics and The Newton Institute in the course of this work.
ND's research is supported in part by the Science 
Technology \& Facilities council under the
grants ST/P000258/1 and ST/X000753/1.

\appendix

\section{Two-point functions for spinning operators}
\label{app:2ptspinning}
This appendix presents the possible structures of two-point functions of bulk operators in an XDCFT, extending the analysis for scalars in Section~\ref{sec:corfunc}. The analysis is largely equivalent to that in DCFT \cite{Billo:2016cpy}, with certain modifications due to the action of $\iota_p$.

To cut down on the number of explicit indices in the following expressions, we define the dot product such that if $X$ is an $n$-tensor and $Y$ an $m$-tensor with $n\geq m$,
\begin{equation}
	\begin{split}
	&(X\cdot Y)_{M_1\dots M_{n-m}}=X_{M_1\dots M_{n-m}M_{n-m+1}\dots M_n}Y^{M_{n-m+1}\dots M_n}\,,
	\\
	&(Y\cdot X)_{M_{1}\dots M_{n-m}}=Y^{M_{n-m+1}\dots M_n}X_{M_{n-m+1}\dots M_n M_1\dots M_{n-m}}\,.
	\end{split}
\end{equation}
Thus for example, if 
$X,Y$ are vectors and $C,D$ are two-tensors, then $X\cdot Y=X_M Y^M$ and $C\cdot D=C_{MN} D^{MN}$, while the mixed contractions are given by $(C\cdot X)_M=C_{MN} X^N$ and $ (X\cdot C)_M=X^NC_{NM}$. Likewise, we use the tensor product symbol, such that $C\cdot(X\otimes Y)=(C\cdot X)\cdot Y$.

\subsection{bulk--defect two-point functions}\label{sec:bdspin}
In Section~\ref{sec:bd2pt}, we study the bulk--defect two-point function for scalar bulk operators \eqref{bdscalar}. For bulk operators carrying spin, there are additional structures that the bulk--defect two-point function may depend on. Consider the two-point function of a bulk operator $\mathcal{O}_1(X,Z)$ and a defect operator $\hat{\mathcal{O}}_2(\hat{Y}_+,\hat{W}_+,\hat{W}_-)$. The correlator is built from all the possible contractions of the bulk coordinates $X_+$, $X_-$, of $C_{+-}=X_+\otimes Z_--Z_+\otimes X_-$, of the defect coordinates $\hat{Y}_+,\hat{W}_-$, and of $\hat{D}_{++}=\hat{Y}_+\otimes\hat{W}_+-\hat{W}_+\otimes\hat{Y}_+$. We should keep in mind that not all contractions are independent, e.g.~$X_+\cdot X_+=-X_-\cdot X_-$. Similarly, some contractions vanish; either because the two tensors have non-overlapping components, by the transversality of the defect coordinates \eqref{eq:deftrans} or by antisymmetry of $D_{++}$. Thus, combinations like $X_-\cdot\hat{W}_+$, $\hat{Y}_{+}\cdot \hat{Y}_{+}$ and $\hat{D}_{++}\cdot (X_+\otimes X_+)$ all vanish.

In the end, there are seven independent ways to contract the indices, which can be chosen as
\begin{equation}
\begin{gathered}
X_-\cdot X_-\,,
\qquad 
C_{+-}\cdot C_{+-}\,,
\qquad 
X_-\cdot \hat{W}_-\,,
\qquad 
X_+\cdot \hat{Y}_+\,,
\qquad
(\hat{D}_{++}\cdot X_+)\cdot(C_{+-}\cdot X_-)\,,
\\
C_{+-}\cdot(X_+\otimes W_-)\,,
\qquad 
C_{+-}\cdot (\hat{Y}_+\otimes X_-)\,.\label{eq:combis1}
\end{gathered}
\end{equation}
The two-point function has five quantum numbers: $(\Delta,J)$ for the bulk operator and $(\hat{\Delta},\hat{J},\hat{s})$ for the defect operator. We use the five contractions on the first line of \eqref{eq:combis1} to capture the dependence on those quantum numbers and put those in the second line into completely invariant cross ratios
\begin{equation}
\hat{K}_1=\frac{C_{+-}\cdot(X_+\otimes \hat{W}_-)}{X_-\cdot\hat{W}_-\sqrt{C_{+-}\cdot C_{+-}}}
\comma
\hat{K}_2=\frac{C_{+-}\cdot(\hat{Y}_{+}\otimes X_{-})}{X_+\cdot\hat{Y}_+\sqrt{C_{+-}\cdot C_{+-}}}\,.
\label{eq:K1K2}
\end{equation}
With this, the bulk--defect two-point function for general $p$ is
\begin{equation}
\begin{split}
\langle\mathcal{O}_1(X,Z)\hat{\mathcal{O}}_2(\hat{Y}_+,\hat{W}_+,\hat{W}_-)\rangle
&=\frac{(X_-\cdot\hat{W}_-)^{\hat{s}}(C_{+-}\cdot C_{+-})^{\frac{J-\hat{J}}{2}}}{(2X_+\cdot \hat{Y}_+)^{\hat{\Delta}+\hat{J}}(X_-\cdot X_-)^{\frac{\Delta-\hat{\Delta}+\hat{s}+J+\hat{J}}{2}}}
\\&\quad \times
((\hat{D}_{++}\cdot X_+)\cdot(C_{+-}\cdot X_-))^{\hat{J}}\,B_{12}\left(\hat{K}_1,\hat{K}_2\right)\,.
\label{spinBD}
\end{split}
\end{equation}
Note that parity invariance requires $\mathcal{O}_1$ and $\hat{\mathcal{O}}_2$ to have the same parity in order for the two-point function to be non-zero. Similarly, the two-point function is non-zero only when $\hat{s}$ is even and $\mathcal{O}_1$ is $\iota_p$-even or when $\hat{s}$ is odd and $\mathcal{O}_1$ is $\iota_p$-odd.

By construction, the left hand side of \eqref{spinBD} is a polynomial in $Z$, $\hat{W}_+$ and $\hat{W}_-$.
Thus, the right-hand side must also be polynomial in the same coordinates and of the same degree, which implies in turn that $B_{12}(\hat{K}_1,\hat{K}_2)$ is polynomial in both $\hat{K}_{1}$ and $\hat{K}_2$. The square-roots and negative powers of $C_{+-}\cdot C_{+-}$ in \eqref{eq:K1K2} require further care, so $B_{12}$ must take the form
\begin{equation}
B_{12}(\hat{K}_1,\hat{K}_2)=\sum_{\hat{n}_1=0}^{\min(\hat{s},J-\hat{J})}
\sum_{\hat{n}_2=0}^{J-\hat{J}-\hat{n}_1}
\tfrac{1+(-1)^{J-\hat J-\hat n_1-\hat n_2}}{2}B_{12}^{(\hat{n}_1,\hat{n}_2)}\hat{K}_1^{\hat{n}_1}\hat{K}_2^{\hat{n}_2}\,,\label{eq:Bspin}
\end{equation}
In particular for $J=\hat{J}$, there is no dependence on $\hat K_1$ and $\hat K_2$, as in the case of a scalar bulk operator in \eqref{bdscalar}. Equation \eqref{eq:Bspin} matches the structure of bulk--defect two-point functions in defect CFTs \cite[(3.9)]{Billo:2016cpy}.

The above result is true for generic $p$, but for odd $q=n-p$, there are some subtleties and generalisations that we address below. The main reason is that the transverse rotation group $PO(q)=SO(q)$ for odd $q$ (for even $q$ the groups are different). This means that one can incorporate the Levi--Civita tensor with $q$ indices for two-point functions of operators of opposite $\iota_p$ chirality. As we require spacetime chirality, we cannot use the $\epsilon$ tensor in the defect directions.

If $p=d-1$, i.e.\ $q=1$ or BCFT, there is no notion of transverse spin. There is a single negative component $X_-^d$, which is a $d+1$-dimensional scalar and for $C_{+-}$, the components are $C_{+-}^{Md}$ so is a $d+1$-dimensional null vector. 
In this case the cross ratios in \eqref{eq:K1K2} all vanish, thus \eqref{spinBD} just depends on one number $B_{12}$ and we must require $J\geq\hat{J}$ to make the right-hand side a polynomial in $Z$. As usual, this is valid when $J$ is even and $\mathcal{O}_1$ is $\iota_p$-even or when $J$ is odd and $\mathcal{O}_1$ is $\iota_p$-odd.

What is really new in this case is that it is possible to have non-zero two-point functions when the spin and $\iota_p$ chirality of $\cO_1$ are the opposite. They rely on the same odd structure as in the one-point function \eqref{eq:1ptcodd} with explicit $X_-$ dependence
\begin{equation}
\langle\mathcal{O}_1(X,Z)\hat{\mathcal{O}}_2(\hat{Y}_+,\hat{W}_+)\rangle=\frac{(C_{+-}\cdot\hat{Y}_+)^{J-\hat{J}}(\hat{D}_{++}\cdot (X_+\otimes C_{+-}))^{\hat{J}}}{(2(X_+\cdot\hat{Y}_+))^{\hat{\Delta}+J}|X_-|^{\Delta-\hat{\Delta}+\hat{J}-1}X_-}B_{12}\,.
\end{equation}

If $q=3$ ($p=d-3$), in addition to \eqref{spinBD}, there again can be a non-zero two-point function even when the $\iota_p$ charge of $\mathcal{O}_1$ is even and the transverse spin $\hat{s}$ of $\hat{\mathcal{O}}_2$ is odd or vice versa. Its form is given by
\begin{equation}
\begin{split}
\langle\mathcal{O}_1(X,Z)\hat{\mathcal{O}}_2(\hat{Y}_+,\hat{W}_+,\hat{W}_-)\rangle
&=\frac{(X_-\cdot\hat{W}_-)^{\hat{s}-1}(C_{+-}\cdot C_{+-})^{\frac{J-\hat{J}-1}{2}}\,\epsilon\cdot(X_-\otimes Z_-\otimes \hat{W}_-)}{(2X_+\cdot \hat{Y}_+)^{\hat{\Delta}+\hat{J}}(X_-\cdot X_-)^{\frac{\Delta-\hat{\Delta}+\hat{s}+J+\hat{J}-1}{2}}}
\\&\quad\times
((\hat{D}_{++}\cdot X_+)\cdot(C_{+-}\cdot X_-))^{\hat{J}}\,B_{12}\left(\hat{K}_1,\hat{K}_2\right)\,,
\end{split}
\end{equation}
where $\epsilon_{MNK}$ is the 3D Levi--Civita symbol. In this case, $B_{12}(\hat{K}_1,\hat{K}_2)$ satisfies similar constraints to \eqref{eq:Bspin}, with slight modifications to the limits on the sums
\begin{equation}
B_{12}(\hat{K}_1,\hat{K}_2)
=\sum_{\hat{n}_1=0}^{\min(\hat{s}-1,J-\hat{J}-1)}
\sum_{\hat{n}_2=0}^{J-\hat{J}-1-\hat{n}_1}
\tfrac{1+(-1)^{J-\hat J-\hat n_1-\hat n_2-1}}{2}B_{12}^{(\hat{n}_1,\hat{n}_2)}\hat{K}_1^{\hat{n}_1}\hat{K}_2^{\hat{n}_2}\,.
\end{equation}

Lastly, for $p=-1$, i.e.\ the XCFT, $\iota_p$ has no fixed points and there are thus no bulk--defect two-point functions.
\subsection{Bulk--bulk two-point functions}\label{sec:bbspin}
The two-point function of two bulk scalar operators is given in \eqref{eq:2pt}. For spinning operators, there are six building blocks for the bulk two-point function $\langle\mathcal{O}_1(X,Z)\mathcal{O}_2(Y,W)\rangle$, namely $X_+,X_-$ and $C_{+-}=X_+\otimes Z_--Z_+\otimes X_-$ from the first operator and $Y_+,Y_-$ and $D_{+-}=Y_+\otimes W_--W_+\otimes Y_-$ for the second operator.

Playing the same game as for the bulk--defect two-point function, one finds twelve independent non-vanishing contractions
\begin{equation}
\begin{gathered}
X_-\cdot X_-\,,
\qquad 
Y_-\cdot Y_-\,,
\qquad 
C_{+-}\cdot C_{+-}\,,
\qquad 
D_{+-}\cdot D_{+-}\,,
\\
X_+\cdot Y_+\,,
\qquad 
X_-\cdot Y_-\,,
\qquad
C_{+-}\cdot (Y_+\otimes Y_-)\,,
\qquad 
C_{+-}\cdot (X_+\otimes Y_-)\,,
\qquad
D_{+-}\cdot (X_+\otimes X_-)\,,
\\
D_{+-}\cdot (Y_+ \otimes X_-)\,,
\qquad 
(C_{+-}\cdot Y_-)\cdot (D_{+-}\cdot X_-)\,,
\qquad 
(Y_+\cdot C_{+-})\cdot (X_+\cdot D_{+-})\,.
\end{gathered}
\end{equation}
We use the first four combinations to capture the charges $(\Delta_1,J_1)$, $(\Delta_2,J_2)$ and are left with eight fully invariant cross ratios.

To simplify their expression, we define
\begin{equation}
\begin{aligned}
\bar{X}_\pm^M&=\frac{X_\pm^M}{\sqrt{X_-\cdot X_-}}\,,
&\quad
\bar{C}_{+-}^{MN}&=\frac{C_{+-}^{MN}}{\sqrt{C_{+-}\cdot C_{+-}}}\,,
\\
\bar{Y}_\pm^M&=\frac{Y_\pm^M}{\sqrt{Y_-\cdot Y_-}}\,,
&
\bar{D}_{+-}^{MN}&=\frac{D_{+-}^{MN}}{\sqrt{D_{+-}\cdot D_{+-}}}\,,
\end{aligned}
\end{equation}
that are normalised such that $\mp\bar{X}_\pm\cdot \bar{X}_\pm=\mp\bar{Y}_\pm\cdot \bar{Y}_\pm=\bar{C}_{+-}\cdot \bar{C}_{+-}=\bar{D}_{+-}\cdot\bar{D}_{+-}=1$. The cross ratios are then chosen as
\begin{equation}
\begin{gathered}
\label{eq:K6}
\kappa_+=\bar{X}_+\cdot\bar{Y}_+\,,\qquad 
\kappa_-=\bar{X}_-\cdot\bar{Y}_-\,,\qquad
K_1=\bar{C}_{+-}\cdot (\bar{Y}_+\otimes\bar{Y}_-)\,,
\\
K_2=\bar{C}_{+-}\cdot (\bar{X}_+\otimes \bar{Y}_-)\,,\qquad
K_3=\bar{D}_{+-}\cdot (\bar{X}_+\otimes\bar{X}_-)\,,\qquad
K_4=\bar{D}_{+-}\cdot (\bar{Y}_+\otimes\bar{X}_-)\,,
\\
K_5=(\bar{C}_{+-}\cdot \bar{Y}_-)\cdot(\bar{D}_{+-}\cdot\bar{X}_-)\,,\qquad
K_6=(\bar{Y}_+\cdot \bar{C}_{+-})\cdot(\bar{X}_+\cdot \bar{D}_{+-})\,.
\end{gathered}
\end{equation}
The first two are already in \eqref{eq:kappas}.

The bulk--bulk two-point function can then be written as
\begin{equation}
\langle\mathcal{O}_1(X,Z)\mathcal{O}_2(Y,W)\rangle
=\frac{(C_{+-}\cdot C_{+-})^{\frac{J_1}{2}}(D_{+-}\cdot D_{+-})^{\frac{J_2}{2}}}{(X_-\cdot X_-)^{\frac{\Delta_1+J_1}{2}}(Y_-\cdot Y_-)^{\frac{\Delta_2+J_2}{2}}}G_{12}(\kappa_\pm,K_i)\,,\label{BBspin}
\end{equation}
which vanishes if the two operators have different $\iota_p$ charges.
Like the bulk--defect two-point function \eqref{eq:Bspin}, the dependence on the cross ratios $K_i$ is again highly constrained, and $G_{12}$ must take the form
\begin{equation}
\begin{aligned}
G_{12}(\kappa_\pm,K_i)
&=\sum_{n_5+n_6=0}^{\min(J_1,J_2)}\,
\sum_{n_1+n_2=0}^{J_1-n_5-n_6}
\tfrac{1+(-1)^{J_1-n_5-n_6-n_1-n_2}}{2}
\\&\quad\times\sum_{n_3+n_4=0}^{J_2-n_5-n_6}
\tfrac{1+(-1)^{J_2-n_5-n_6-n_3-n_4}}{2}
G_{12}^{(\vec{n})}(\kappa_\pm)\prod_{i=1}^{6}K_i^{n_i}\,,
\end{aligned}
\end{equation}
with all $n_i\geq0$. The dependence on the additional cross ratios $K_i$ thus drops out if $J_1=J_2=0$, as in \eqref{eq:2pt}. This matches the structure of bulk--defect two-point functions in defect CFTs \cite[(3.17)]{Billo:2016cpy}.

Again, there are a number of extra structures for odd values of $q=d-p$ that we outline below.

If $p=d-1$, both $C$ and $D$ can be viewed as $d+1$-dimensional null vectors. Then the only non-vanishing cross ratios from \eqref{eq:K6} are $\kappa_\pm$ and
\begin{equation}
K'
=\frac{\kappa_+K_5}{\kappa_-K_6}
=\frac{(C_{+-}\cdot D_{+-})(X_+\cdot Y_+)}{(C_{+-}\cdot Y_+)\cdot(D_{+-}\cdot X_+)}\,.
\end{equation}
This simplifies the expression for $G_{12}$ in \eqref{BBspin}
and it holds, as usual, when the combined $\iota_p$ charge and total spin $J_1+J_2$ are both even or both odd.

The new possibility is when the combined $\iota_p$ charge is even and total spin is odd or vice versa, in which case the new form is (cf.\ \eqref{eq:1ptcodd})
\begin{equation}
    \left<\mathcal{O}_1(X,Z)\mathcal{O}_2(Y,W)\right>=\frac{(C_{+-}\cdot Y_+)^{J_1}(D_{+-}\cdot X_+)^{J_2}}{|X_-|^{\Delta_1+J_1+J_2}|Y_-|^{\Delta_2+J_1+J_2-1}Y_-}G_{12}(\kappa_+,K')\,.
\end{equation}
In either case, the function $G_{12}(\kappa_+,K')$ must be a polynomial in $K'$ of order at most $\min(J_1,J_2)$.

If $q=3$, i.e.\ $p=d-3$, the two-point function of operators with the same $\iota_p$ charge again takes the form \eqref{BBspin}. However, for this specific value of $p$, the two-point function can be non-zero even if the two operators have different $\iota_p$ charges, and can be given as
\begin{equation}
\langle\mathcal{O}_1(X,Z)\mathcal{O}_2(Y,W)\rangle
=\frac{(C_{+-}\cdot C_{+-})^{\frac{J_1-1}{2}}(D_{+-}\cdot D_{+-})^{\frac{J_2}{2}}\epsilon\cdot(X_-\otimes Y_-\otimes Z_-)}{(X_-\cdot X_-)^{\frac{\Delta_1+J_1}{2}}(Y_-\cdot Y_-)^{\frac{\Delta_2+J_2+1}{2}}}G_{12}(\kappa_\pm,K_i)\,,
\end{equation}
with
\begin{equation}
\begin{aligned}
    G_{12}(\kappa_\pm,K_i)
&=\sum_{n_5+n_6=0}^{\min(J_1-1,J_2)}\,
\sum_{n_1+n_2=0}^{J_1-n_5-n_6-1}\,
\tfrac{1+(-1)^{J_1-n_5-n_6-n_1-n_2-1}}{2}
\\&\quad\times
\sum_{n_3+n_4=0}^{J_2-n_5-n_6}\,
\tfrac{1+(-1)^{J_1-n_5-n_6-n_3-n_4-1}}{2}
	G_{12}^{(\vec{n})}(\kappa_\pm)\prod_{i=1}^{6}K_i^{n_i}\,,
\end{aligned}
\end{equation}
with all $n_i\geq0$.

Lastly, for $p=-1$, $C$ and $D$ can again be considered as $d+1$-dimensional null vectors and the bulk--bulk two-point function takes the form
\begin{equation}
    \left<\mathcal{O}_1(X,Z)\mathcal{O}_2(Y,W)\right>=\frac{(C_{+-}\cdot Y_-)^{J_1}(D_{+-}\cdot X_-)^{J_2}}{(X_-\cdot X_-)^{\frac{\Delta_1+J_1+J_2}{2}}(Y_-\cdot Y_-)^{\frac{\Delta_2+J_1+J_2}{2}}}G_{12}(\kappa_+,K')\,.
\end{equation}
Like the two-point function for arbitrary $p$, this two-point function vanishes for two operators with different $\iota_p$ charges.

\section{Feynman integrals}\label{sec:Feynman}
We compute here the Feynman integrals in Figure~\ref{fig:Feynman}. As mentioned, they rely on the evaluation of the integral
\begin{equation}
I(d,p|x,y)=-\lambda\int\dd^duK(x,u)K(y,u)K(u,u')\,,
\end{equation}
with the $d$-dimensional propagator
\begin{equation}
K(x,y)=\frac{\Gamma(d/2-1)}{4\pi^{d/2}(x-y)^{d-2}}\,.
\end{equation}
Writing out the integral explicitly, we find
\begin{align}
I(d,p|x,y)&=-\frac{\lambda\Gamma^3(d/2-1)}{64\pi^{3d/2}}\int \frac{\dd^qu_\perp}{(4u_\perp^2)^{d/2-1}}
\\&\quad\nonumber
\times\int\dd^p u_\parallel\frac{1}{\big((x_\parallel-u_\parallel)^2+(x_\perp-u_\perp)^2\big)^{d/2-1}
\big((y_\parallel-u_\parallel)^2+(y_\perp-u_\perp)^2\big)^{d/2-1}}
\,.
\end{align}
To integrate over $u_\parallel$, we use Feynman parameters $a_i$
\begin{align}
I(d,p|x,y)&=-\frac{\lambda\Gamma(d/2-1)\Gamma(d-2)}{64\pi^{3d/2}} \int\frac{\dd^qu_\perp}{(4u_\perp^2)^{d/2-1}}
    \int_0^1\dd a_1\dd a_2
\\&\qquad
    \times\int\dd^pu_\parallel\frac{\delta(a_1+a_2-1)(a_1a_2)^{d/2-2}}{\big(u_\parallel^2+a_1a_2(x_\parallel-y_\parallel)^2+a_1(x_\perp-u_\perp)^2+a_2(y_\perp-u_\perp)^2\big)^{d-2}}\,.
\nonumber
\end{align}
The integral over $u_\parallel$ then leads to
\begin{align}
I(d,p|x,y)&=-\frac{\lambda\Gamma\left(d-2-p/2\right)\Gamma(d/2-1)}{64\pi^{3d/2-p/2}}\int_0^1\dd a_1\dd a_2
\\&\qquad
\times
\int\frac{\dd^qu_\perp}{(4u_\perp^2)^{d/2-1}}\frac{\delta(a_1+a_2-1)(a_1a_2)^{d/2-2}}{\big(a_1a_2(x_\parallel-y_\parallel)^2+a_1(x_\perp-u_\perp)^2+a_2(y_\perp-u_\perp)^2\big)^{d-2-p/2}}\,.
\nonumber
\end{align}

For the integral over $u_\perp$ we introduce further Feynman parameters $b_i$
\begin{align}
&I(d,p|x,y)=-\frac{\lambda\Gamma\left(3d/2-3-p/2\right)}{2^{d+4}\pi^{3d/2-p/2}}\int_0^1\dd a_1\dd a_2\int_0^1\dd b_1\dd b_2
\int\dd^qu_\perp
\\&\quad\nonumber
\times\frac{\delta(a_1+a_2-1)\delta(b_1+b_2-1)(a_1a_2b_2)^{d/2-2}b_1^{d-3-p/2}}{\big(u_\perp^2+a_1^2b_1b_2x_\perp^2+a_2^2b_1b_2y_\perp^2 -2a_1a_2b_1\big(x_\para\cdot y_\para-\frac{x^2+y^2}{2}\big)-2a_1a_2b_1^2x_\perp\cdot y_\perp\big)^{3d/2-3-p/2}}\,.
\end{align}
In the denominator we recognise the expressions of $\kappa_\pm$ from \eqref{eq:flatkappa}. The $u_\perp$ integral then gives
\bal
I(d,p|x,y)&=-\frac{\lambda\Gamma(d-3)}{2^{d+5}\pi^{d}|x_\perp|^{d-3}|y_\perp|^{d-3}}\int_0^1\dd a_1\dd a_2\int_0^1\dd b_1\dd b_2
\\&\hskip2cm
\times\frac{\delta(a_1+a_2-1)\delta(b_1+b_2-1)(a_1a_2b_2)^{d/2-2}b_1^{-p/2}}{\big(a_1^2b_2\cM+a_2^2b_2\cM^{-1}+2a_1a_2\kappa_+-2a_1a_2b_1\kappa_-\big)^{d-3}}\,,\label{eq:Fullint}
\eal
where we introduced $\cM=\frac{|x_\perp|}{|y_\perp|}$. In $d=4$ the integral is secretly independent of $\cM$ in agreement with \eqref{eq:2pt}. When working to order $\varepsilon$, we can set $d=4$, so we take $\mathcal{M}=1$ below.

\subsection{One-point integral}\label{app:1pt}
To get the one-point function of the operators $S$ and $T$, we take 
the limits $y\to x$ and $y\to x'$, corresponding to $\kappa_+\ra 1$ and $\kappa_-\ra \pm 1$. Focusing first on $y\to x$, the integral \eqref{eq:Fullint} simplifies to
\bal
I(d,p|x,x)=-\frac{\lambda\Gamma(d-3)}{2^{d+5}\pi^{d}|x_\perp|^{2d-6}}
&\int_0^1\dd a_1\dd a_2\delta(a_1+a_2-1)(a_1a_2)^{d/2-2}
\\&
\times\int_0^1\dd b_1\dd b_2\delta(b_1+b_2-1)b_2^{1-d/2}b_1^{-p/2}\,,\label{eq:-11}
\eal
It is then straightforward to compute the remaining integrals, resulting in
\begin{equation}
I(d,p|x,x)=-\frac{\lambda }{2^{d+5}\pi^d|x_\perp|^{2d-6}}\frac{\Gamma(d/2-1)^2\Gamma(2-d/2)\Gamma(1-p/2)}{(d-3)\Gamma(3-d/2-p/2)}\,.\label{eq:-11e}
\end{equation}
This expression is divergent both in the limit $d\ra 4$ and the limit $p\ra2$. Focusing on the former limit by setting $d=4-\varepsilon$, we obtain
\begin{equation}
I(4-\varepsilon,p|x,x)
=-\frac{\lambda}{(4\pi)^4x_\perp^2}
\left(\frac{1}{\varepsilon}+1+{\gamma_E}+\log(2\pi x_\perp^2)-\frac12H_{-p/2}+O(\varepsilon)\right).
\end{equation}

Next, for the contribution from the antipodal point $x'=\iota_p(x)$, \eqref{eq:Fullint} becomes
\bal
I(d,p|x,x')&=-\frac{\lambda\Gamma(d-3)}{2^{d+5}\pi^{d}|x_\perp|^{d-3}|y_\perp|^{d-3}}\int_0^1\dd a_1\dd a_2\int_0^1\dd b_1\dd b_2
\\&\quad\times
    \frac{\delta(a_1+a_2-1)\delta(b_1+b_2-1)(a_1a_2b_2)^{d/2-2}b_1^{-p/2}}{\big(a_1^2b_2+a_2^2b_2+2(1+b_1)a_1a_2\big)^{d-3}}\,,\label{eq:-1-1}
\eal
In contrast to \eqref{eq:-11}, this integral does not have a pole in four dimensions, so we can readily set $d=4$. Integrating over Feynman parameters, we then obtain (note that $H_{-1/2}=-\log 4$)
\begin{equation}
I(4,p|x,x')=\frac{\lambda}{(4\pi)^4x_\perp^2}\frac{H_{-p/2}-H_{-1/2}}{2(p-1)}\,.
\end{equation}

\subsection{Two-point integral}\label{app:2pt}
We now return to \eqref{eq:Fullint} for $y\neq x, x'$. We see that we can set $d=4$ without any obstructions and it is then straightforward to express the integral over $b_i$ in terms of a hypergeometric function
\begin{equation}
\begin{split}
I(4,p|x,y)&=-\frac{\lambda}{(4\pi)^4|x_\perp||y_\perp|}\frac{1}{2-p}\int_0^1\dd a_1\dd a_2\frac{\delta(a_1+a_2-1)}{a_1^2+a_2^2+2a_1a_2\kappa_+}
\\&\quad
\times{}_2F_1\left(1,1-p/2;2-p/2;\frac{a_1^2+a_2^2+2a_1a_2\kappa_-}{a_1^2+a_2^2+2a_1a_2\kappa_+}\right)\,.
\label{eq:aintegral}
\end{split}
\end{equation}
Switching to light-cone coordinates \eqref{eq:LCcoord} and parametrizing $a_1$ as
\begin{equation}
    a_1=\frac{w}{w+|z|}\,,
\end{equation}
the integral becomes
\begin{equation}
I(4,p|x,y)=-\frac{\lambda|z|}{(4\pi)^4(2-p)|x_\perp||y_\perp|}\int_0^\infty \dd w\frac{{}_2F_1\left(1,1-p/2;2-p/2;\frac{(w+z)(w+\bar{z})}{(w+1)(w+|z|^2)}\right)}{(w+1)(w+|z|^2)}\,.
\label{eq:wintegral}
\end{equation}

The hypergeometric function simplifies when $p$ is an integer, in which case the remaining integral can be performed. This is done in Section~\ref{sec:int2pt}.

\subsection{\texorpdfstring{${p=2}$}{p=2} counterterm}\label{sec:counter}
For $p=2$, the expression for the one-point function in Section~\ref{sec:1ptint} diverges. We then need to add an additional defect-localised counterterm \eqref{counter}
\begin{equation}
S^{D}=\frac{1}{2}h_0\int\dd^{p}x_{\para}\,\big(\phi_+(x_\para)\big)^2\,,
\end{equation}
where $p=2-\delta$. This leads to a new Feynman diagram contributing to the two point function. The integral takes the form
\begin{align}
D(d,p|x,y)&\equiv -h_0\int\dd^pu_\parallel K(x,z)K(z,y)
\\&\nonumber
=-\frac{h_0\Gamma^2(d/2-1)}{16\pi^{d}}\int \dd^pu_\para \frac{1}{\big((x_\para-u_\para)^{2}+x_\perp^2\big)^{d/2-1}\big((y_\para-u_\para)^2+y_\perp^2\big)^{d/2-1}}\,.
\end{align}
We once again introduce Feynman parameters to write the integral as
\begin{equation}
D(d,p|x,y)=-\frac{h_0\Gamma(d-2)}{16\pi^d}\int\dd^pu_\para\int_0^1\dd a_1\dd a_2\frac{\delta(a_1+a_2-1)(a_1a_2)^{d/2-2}}{\big(u_\para^2+a_1a_2(x_\para-y_\para)^2+a_1x_\perp^2+a_2y_\perp^2\big)^{d-2}}\,.
\end{equation}
Integrating over $u_\para$ gives
\begin{equation}
D(d,p|x,y)=-\frac{h_0\Gamma(d-2)}{16\pi^{d-p/2}(|x_\perp||y_\perp|)^{d-2-p/2}}\int_0^1\dd a_1\dd a_2\frac{\delta(a_1+a_2-1)(a_1a_2)^{d/2-2}}{\big(a_1^2\cM+a_2^2\cM^{-1}+2a_1a_2\kappa_+\big)^{d-2-p/2}}\,.
\end{equation}
\paragraph{One-point function:}
We compute the one-point function by setting $\kappa_+=1$. Like for the bulk interaction term, the denominator simplifies and we get
\begin{equation}
D(d,p|x,x)=-\frac{h_0\Gamma(d/2-1)^2}{16\pi^{d-p/2}|x_\perp|^{2d-4-p}}\,.\label{eq:D-1}
\end{equation}
We are then free to set $d=4-\varepsilon$, $p=2-\delta$ and expand around $\varepsilon=\delta=0$. We obtain
\begin{equation}
D(4-\varepsilon,2-\delta|x,x)=-\frac{2h_0}{(4\pi)^3x_\perp^2}
\big(2-\delta\log(\pi x_\perp^2)\big)
\big(1+\varepsilon(\gamma_E+\log(\pi x_\perp^2))\big)+O(\varepsilon^2,\delta^2)\,.
\label{eq:Dexp}
\end{equation} 
\paragraph{Two-point function:}
Following the same steps as in Appendix~\ref{app:2pt}, the remaining integral for $p=2-\delta$ is
\begin{equation}
D(4,2-\delta|x,y)=\frac{2h_0}{(4\pi)^3|x_\perp||y_\perp|}\frac{|z| \log |z|}{1- |z|^2}\left(2+ \delta  \left(\frac{3}{4}\log \frac{|z|}{(1+|z|)^2}-\gamma_E-\log\pi\right)\right)+O(\delta^2)
\,.\label{eq:D2}
\end{equation}

\section{Comparison to regular defects}
\label{app:comparison}

In this appendix we explain the similarity between the bulk two-point function for a crosscap defect and for a defect given by integrating an operator over a submanifold. This can be seen when comparing our $p=1$ result in Section~\ref{sec:p=1} to the calculation in~\cite{Bianchi:2022sbz, Gimenez-Grau:2022ebb}, which both give the same function at order $\varepsilon$.

The reason is easily illustrated in an example, so take $p=3$ where defects involving the integration of cubic operators $h_{ijk}\phi^i\phi^j\phi^k$ have been constructed in \cite{Harribey:2023xyv,Harribey:2024gjn}. For such a defect, the two-point function at order $\varepsilon$ is given by the Feynman diagram
\begin{align}
\label{phi^3}
\begin{tikzpicture}[scale=0.5, baseline=(vert_cent.base)]
\node[inner sep=0pt,outer sep=0pt] (vert_cent) at (3,2.5) {$\phantom{\cdot}$};
\draw[very thick,black] (-3,0)--(3,0);
\draw (-2,0) to[out=60,in=120] (2,0);
\draw (-2,0) to[out=50,in=130](2,0);
\draw (-2,0) to[out=70,in=-140](0,2.5);
\draw (2,0) to[out=110,in=-40](0,2.5);
\draw (0,2.5) -- (2,4);
\draw (0,2.5) -- (-2,4);
\fill[black] (-2,4) circle (4pt);
\fill[black] (2,4) circle (4pt);
\fill[black] (0,2.5) circle (3pt);
\fill[black] (-2,0) circle (3pt);
\fill[black] (2,0) circle (3pt);
\node[below] at (-2,0) {$\tau_1$};
\node[below] at (2,0) {$\tau_2$};
\node[right] at (0.2,2.5) {$\lambda$};
\node[above] at (-2,4) {$\phi(x_1)$};
\node[above] at (2,4) {$\phi(x_2)$};
\end{tikzpicture}
&\begin{aligned}
&=-\lambda h^2\int \dd^dy\,K(x_1,y)K(x_2,y)
\\&\quad\times\int \dd^p\tau_1\,\dd^p\tau_2\,
K(\tau_1,\tau_2)^2K(\tau_1,y)K(\tau_2,y)
\\&=\#\int \dd^dy\,K(x_1,y)K(x_2,y)\int \dd\tau_1\,\dd\tau_2\,d\cos\theta
\\&\hskip2cm\times\frac{\tau_1^{p-1}\tau_2^{p-1}}
{(\tau_1^2+y_\perp^2)^\frac{d-2}{2}(\tau_2^2+y_\perp^2)^\frac{d-2}{2}(\tau_1^2+\tau_2^2-2\tau_1\tau_2\cos\theta)^{d-2}}
\end{aligned}
\nonumber\\
&=\#\int \dd^dy\,K(x_1,y)K(x_2,y)|y_\perp|^{2p+8-4d}\,.
\end{align}
This expression is for a defect of dimension $p\sim3$ and we did not keep track of numerical and tensorial factors. For $d\sim4$ and $p\sim3$ the power of $y_\perp$ is roughly $K(y_\perp,0)$, so similar to the propagator $K(u,\iota_p(u))$ in the expression for the two-point function for the crosscap \eqref{IpInt}.

More generally, for an arbitrary dimension $p$ defect defined by the contribution to the action
\begin{equation}
S_4=h\int \dd^p\tau\, O(\tau)\,,
\end{equation}
where schematically we can regard $O(\tau)\sim\phi O'(\tau)$ with $O'$ of dimension $\Delta\sim p-1$, on dimensional grounds we find the same expression as above with $|y_\perp|^{2p+4-2d-2\Delta}\sim|y_\perp|^{4-2d}$ as in the crosscap propagator.

\section{Summary of bulk CFT data}\label{ap:bulk}
For readers' convenience, we review here the bulk CFT data for the free and the WF $O(N)$ theories. Those are used in the main text for comparison, and for extraction of $1$-point functions. The results below can be found in \cite{Henriksson:2022rnm, Dey:2016mcs}.

In the free $O(N)$ theory, the conformal dimensions of operators quadratic in $\phi^{i}$ are (here and below we assume $J\geq 1$)
\begin{align}
\Delta_S^{\rm free}=\Delta_{T}^{\rm free}=d-2\,,\qquad \Delta^{\rm free}_{S_J}=\Delta^{\rm free}_{T_J}=J+d-2\,,
\end{align}
while their OPE coefficients with two fundamental fields are
\begin{align}
\begin{aligned}
\sqrt{N}\lambda_S^{\rm free}=
\lambda_{T}^{\rm free}=\sqrt{2}\,,\quad 
\sqrt{N}\lambda^{\rm free}_{S_J}=\lambda^{\rm free}_{T_J}=\frac{\Gamma\left(\frac{d}{2}+J-1\right)}{\Gamma\left(\frac{d}{2}-1\right)}
\sqrt{\frac{2\,\Gamma(J+d-3)}{\Gamma(2J+d-3)\Gamma(J+1)}}\,.
\end{aligned}
\end{align}
The three-point functions of $O(N)$ tensors from \cite{Dey:2016mcs} are defined in the same normalisation convention as in \eqref{eq:normalisations}. Upon setting $d=4-\varepsilon$, these expressions can be expanded in $\varepsilon$.

In the WF $O(N)$ theory at $d=4-\varepsilon$, their dimensions up to order $\varepsilon$ are given by
\begin{align}
\begin{aligned}
&\Delta_{S}=\Delta_{S}^{\rm free}+\frac{N+2}{N+8}\varepsilon\,,\quad  
\Delta_{T}=\Delta_{T}^{\rm free}+\frac{2}{N+8}\varepsilon\,, \quad \Delta_{S_J}=\Delta_{T_J}=\Delta_{S_J}^{\rm free}=\Delta_{T_J}^{\rm free}\,,
\end{aligned}
\end{align}
while their OPE coefficients up to order $\varepsilon$ are
\begin{align}
\begin{aligned}
&\frac{\lambda_S}{\lambda_S^{\rm free}}=1-\frac{N+2}{2(N+8)}\varepsilon\,,\quad \frac{\lambda_T}{\lambda_T^{\rm free}}=1-\frac{1}{N+8}\varepsilon\,,\quad\frac{\lambda_{S_J}}{\lambda_{S_J}^{\rm free}}=\frac{\lambda_{T_J}}{\lambda_{T_J}^{\rm free}}=1\,.
\end{aligned}
\end{align}

\bibliographystyle{utphys2}
\bibliography{refs}
\end{document}